%% file: 3pi.tex
\documentclass[prd,showpacs,showkeys,preprintnumbers,floatfix,
nofootinbib,superscriptaddress]{revtex4-1}
\usepackage{float}
\usepackage{nicefrac}
\usepackage{mathtools}
\usepackage{amsfonts} 
\usepackage{amssymb} 
\usepackage{amsmath} 
\usepackage{graphicx} 
\usepackage{subfigure} 
\usepackage{array} 
\usepackage{dcolumn} 
\usepackage{bm} 
\usepackage{latexsym} 
\usepackage{longtable} 
\usepackage{hyperref} 
\usepackage{verbatim}
\usepackage{epsfig}
\usepackage{color}
\newcommand{\dl}{| \hspace{-0.5pt} |}

\newcommand{\CL}[1]{C_{L}^{(#1)}}
\newcommand{\CI}[1]{C_{\infty}^{(#1)}}

\newcommand{\CIeps}[1]{C_{\infty}^{(#1),i \epsilon}}
\newcommand{\CLF}[2]{C_{L, #1 F}^{(#2)}}

\newcommand{\K}[0]{\mathcal K_2}
\newcommand{\Kin}[1]{\mathcal K_{2; #1}}

\newcommand{\Kdfth}[0]{\mathcal K_{\mathrm{df},3}}
\newcommand{\Mdfth}[0]{\mathcal M_{\mathrm{df},3}}
\newcommand{\Kdfthin}[1]{\mathcal K_{\mathrm{df},3; #1}}
\newcommand{\Mdfthin}[1]{\mathcal M_{\mathrm{df},3; #1}}
\newcommand{\M}[0]{\mathcal M_2}
\newcommand{\Mth}[0]{\mathcal M_{ 3}}
\newcommand{\Kth}[0]{\mathcal K_{ 3}}
\newcommand{\Kthin}[1]{\mathcal K_{ 3 ; #1}}
\newcommand{\smi}[1]{\bigg [\frac{1}{L^3} \sum_{\vec {#1}} - \int_{\vec {#1}} \bigg ]}
\newcommand{\PV}[0]{\widetilde{\mathrm{PV}}}

\newcommand{\KthnL}[1]{\mathcal K_{3, L}^{(#1,u,u)}}
\newcommand{\KthinnL}[2]{{\cal K}_{3, L;#2}^{(#1,u,u)}}

\newcommand{\tr}[0]{\mathrm{Tr\,}}

\newcommand{\AnL}[2]{    A^{(#1,u)}_{ L  #2}    }
\newcommand{\ApnL}[2]{    A'^{(#1,u)}_{ L #2}    }

\begin{document}
\title{Relativistic, model-independent, three-particle quantization condition}
\author{Maxwell T. Hansen}
\email[e-mail: ]{mth28@uw.edu}
\affiliation{
 Physics Department, University of Washington, 
 Seattle, WA 98195-1560, USA \\
}
\author{Stephen R. Sharpe}
\email[e-mail: ]{srsharpe@uw.edu}
\affiliation{
 Physics Department, University of Washington, 
 Seattle, WA 98195-1560, USA \\
}
\date{\today}
\begin{abstract}
We present a generalization of L\"uscher's relation between the 
finite-volume spectrum and scattering amplitudes to the case of three
particles. We consider a relativistic 
scalar field theory in which the couplings
are arbitrary aside from a \(\mathbb{Z}_2\) symmetry that removes vertices
with an odd number of particles. 
The theory is assumed to have two-particle phase shifts that are bounded by 
\(\pi/2\) in the regime of elastic scattering. 
We determine the spectrum of the finite-volume theory from the poles 
in the odd-particle-number 
finite-volume correlator, which we analyze 
to all orders in perturbation theory.
We show that it depends on the infinite-volume two-to-two K-matrix as well as a nonstandard infinite-volume
three-to-three K-matrix. A key feature of our result
is the need to subtract physical singularities in the
three-to-three amplitude and thus deal with a divergence-free quantity.
This allows our initial, formal result to be truncated to a finite
dimensional determinant equation.
At present, the relation of the three-to-three K-matrix to the
corresponding scattering amplitude is not known, although previous
results in the non-relativistic limit suggest that such a relation exists.

\end{abstract}
%
\maketitle

\input{introres.tex}

\input{isotropic.tex}
\input{derivation.tex}

\input{noswitches.tex}

\input{oneswitch.tex}

\input{twoswitches.tex}

\input{remdiagrams.tex}

\input{B3.tex}

\input{conclusions.tex}

\section*{Acknowledgements}
We thank Ra\'ul Brice\~no, Zohreh Davoudi and Akaki Rusetsky for discussions.
This work was supported in part by the United States Department of Energy 
grants DE-FG02-96ER40956 and DE-SC0011637.
MTH was supported in part by the Fermilab Fellowship in Theoretical Physics.
Fermilab is operated by Fermi Research Alliance, LLC, under Contract
No.~DE-AC02-07CH11359 with the United States Department of Energy.


\input{apps.tex}

\bibliography{ref} 

\end{document}

%% file: introres.tex
\section{Introduction}

\indent
In the last few years, lattice QCD calculations of
the properties of resonances have become widespread.\footnote{%
For recent reviews see 
Refs.~\cite{Prelovsek:2013cta,Thomas:2014dpa,Doering:2014fpa}.}
Most use a method first proposed by L\"uscher in
Refs.~\cite{Luscher:1986n2,Luscher:1991n1,Luscher:1991n2},
in which the finite-volume spectrum (obtained using lattice simulations)
can be related to infinite-volume scattering amplitudes.
This method initially applied to two-particle systems below the
inelastic threshold, but has since been extended to systems with
multiple two-particle channels~\cite{Lage:2009,Bernard:2010,Doering:2011,Hansen:2012tf,Briceno:2012yi}.
A striking example of the practical implementation of this
multi-channel formalism is the recent
lattice study of the properties of kaon resonances~\cite{Dudek:2014qha}.

Lattice calculations can now determine
many spectral levels for a given set of total quantum numbers, and can do
so for quark masses approaching physical values.
This means that channels involving three or more particles
are opening up and must be incorporated into the formalism.
Examples include $\omega\to3\pi$, $K^*\to K\pi\pi$, and
$N^*\to N\pi\pi$. Indeed, the study of Ref.~\cite{Dudek:2014qha}, although
using an unphysically heavy pion mass of 390 MeV, was limited
by the opening of the $K\pi\pi$ channel.
Thus there is strong motivation to extend the finite-volume
formalism to include three (or more) particles.

First steps in this direction have been taken in 
Refs.~\cite{Polejaeva:2012ut} and \cite{Briceno:2012rv}.
The former work considers the problem in a non-relativistic context,
and shows that the finite-volume spectrum 
is determined (via integral equations)
by infinite-volume scattering amplitudes.
The latter work reaches the same conclusion in
the case in which pairs of particles interact only in the s-wave. Related problems have also been considered in Refs.~\cite{Guo:2013qla} and \cite{Aoki:2013cra}.
We attempt here to go beyond these works by considering 
a relativistic theory in which we make no approximation concerning
the nature of the two-particle interactions.

Our approach is a generalization of the diagrammatic, field-theoretic
method introduced for two particles in Ref.~\cite{Kim:2005}.
The finite-volume spectrum is determined by the poles in an appropriate
finite-volume correlation function. The method consists of rewriting
this correlation function, diagram by diagram, in terms of infinite-volume
contributions and kinematic functions which depend on the volume.
Summing all diagrams then leads to the desired quantization condition.
This approach is straightforward in the two-particle case, but several
complications arise with three particles. 
In the end, however, we are able to
obtain a simple-looking quantization condition [Eq.~(\ref{eq:mainres})],
which succeeds in
separating finite-volume dependence into kinematical functions.

As in the two-particle quantization conditions, our result is formal
in that it involves a determinant over an infinite-dimensional space.
Practical applications require truncation of this space.
It turns out that such a truncation can be justified for three particles
by a simple extension of the arguments used for two particles.

The main drawback of our result is that it depends on a non-standard infinite-volume three-to-three scattering quantity, a modified three-particle K-matrix. The relation of this quantity to
physical scattering amplitudes is as yet unclear.
Nevertheless, given the results
of Refs.~\cite{Polejaeva:2012ut,Briceno:2012rv} in the non-relativistic
context, we think it very likely that such a relation exists. 

The remainder of this article is organized as follows.
We begin, in Sec.~\ref{sec:res}, by presenting our main result.
This in itself requires a fairly lengthy introduction and
explanation of notation.
Next, in Sec.~\ref{sec:iso}, we describe briefly how the result might
be used in practice.
The core of the paper is Sec.~\ref{sec:der},
in which we derive our main result.
We conclude and discuss the future outlook in Sec.~\ref{sec:con}.

Three appendices discuss technical details.
Appendix~\ref{app:sumintegral}
derives the key sum-integral difference
identity used throughout the derivation.
Appendix~\ref{app:PVtilde}
describes the properties of the modified principal-value pole prescription that we use.
Finally, Appendix~\ref{app:iso} discusses in detail an example
of using our quantization condition in the isotropic approximation.

A sketch of the result has been given previously in Ref.~\cite{Hansen:2013dla},
although some of the technical remarks in that work are incorrect
and have been corrected here.

\section{Quantization Condition}

\label{sec:res}

\indent

In this section we present the three-particle 
quantization condition.
To explain the result requires some preliminary discussion,
particularly about the three-particle scattering amplitude. It also
requires the introduction of some rather involved notation. We have
attempted to make this section self-contained so that the reader can
skip the subsequent lengthy derivation if desired.  

Lattice calculations can determine the spectrum of QCD in finite 
spatial volumes.
We assume here a cubic spatial volume of extent \(L\) 
with fields satisfying periodic boundary conditions. 
We take \(L\) large enough to allow neglect of exponentially suppressed
corrections of the form \(e^{- mL}\), where \(m\) is the particle mass. 
We also assume that discretization errors are small and can be ignored, 
and so work throughout with continuum field theory (zero lattice spacing). 

We work in general in a ``moving frame''. That is, we consider states
with non-zero total three-momentum $\vec P$.  This three-momentum is
constrained by the boundary conditions to satisfy $\vec P=2\pi \vec n_P/L$, 
with $\vec n_P$ a vector of integers.  The total moving-frame
energy is denoted $E$, while $E^*$ is the energy in the center-of-mass
(CM) frame: \(E^{*2}=E^2 - \vec P^2\). 
(The superscript \(*\) is used throughout this work to indicate a
quantity boosted to an appropriate CM frame.)  
The goal of this section is, at fixed fixed \(\{L, \vec P\}\),
to determine the spectrum of the finite-volume system
in terms of infinite-volume scattering quantities.

We choose a simple theory for this study: a single real scalar field
$\phi$ describing particles of physical mass \(m\).  
Thus all results in this work hold for identical particles.  
For simplicity, we assume the Lagrangian has a \(\mathbb Z_2\) symmetry
that prevents vertices having an odd number of particles. (For pions
in QCD this is G-parity.) We otherwise include all vertices, with any
even number of fields, and make no assumptions about relative coupling
strengths.

Given the $\mathbb Z_2$ symmetry, the Hilbert space splits into even-
and odd-particle states. We are interested here in the latter, which
are those created from the vacuum by the field $\phi$ (or by $\phi^3$,
$\phi^5$, etc.). The spectrum in this sector consists of an isolated
single-particle state with $E^*=m$, followed by a tower of states that
lies close (for large $L$) to the energies of three free particles in
the finite volume.  Such states begin at $E^*\approx 3 m$, and it is
these that we focus on.  Their energies typically are shifted from
those of three free particles by a difference $\Delta E$ which scales
as an inverse power of $L$.  Once $E^*$ reaches $5m$, one also has
states which lie close to the energies of five free finite-volume
particles. Our derivation breaks down at this point. Thus we focus on
the range $m < E^* < 5 m$, within which it turns out that the only
infinite-volume observables that enter are quantities related
to two-to-two and three-to-three scattering.\footnote{%
  Were we to remove the $\mathbb Z_2$ symmetry, we would also need to include
  two-to-three amplitudes, as has been done in Ref.~\cite{Polejaeva:2012ut}.}

An additional technical requirement is that the two-particle
K-matrix remain finite in the kinematical range of interest.
This range runs from $0 < E_{2}^* < 4 m$,
where $E_{2}^*$ is the two-particle CM energy. 
This requirement means that the phase shifts must satisfy
$|\delta_\ell| <\pi/2$ below the four-particle threshold
for all $\ell$. In other words, two-particle interactions can be neither
attractive enough to produce a resonance nor overly repulsive.

\bigskip

We begin by establishing our notation for three-particle kinematics,
considering first the case where all particles are on shell.
If the momenta of two of these particles are $\vec k$ and $\vec a$,
then that of the third is fixed to be
\(\vec b_{ka} \equiv \vec P - \vec k - \vec a\) by momentum conservation.
The corresponding energies are denoted 
\begin{equation}
\omega_k=\sqrt{\vec k^2+m^2}\,,\ \
\omega_a=\sqrt{\vec a^2+m^2}\,,\ \ {\rm and}\ \ 
\omega_{ka} =\sqrt{(\vec P - \vec k - \vec a)^2 + m^2} \,, \ \
{\rm respectively}.
\label{eq:omegadef}
\end{equation}
The momenta \(\vec k\) and \(\vec a\) cannot be chosen freely: on-shell 
and total energy constraints require 
\begin{equation}
\label{eq:Eeqthree}
E = \omega_k + \omega_a + \omega_{ka} \,.
\end{equation}
It is convenient to separate the three particles into a
``spectator'',  which we take to be that with momentum $\vec k$,
and the remaining two-particle pair,
with four-momentum $P_2=(E-\omega_k,\vec P-\vec k)$.
The energy of this pair in its CM-frame 
(which we stress is different, in general, from the
CM-frame of all three particles) is labeled $E_{2,k}^*$, where
\begin{equation}
\label{eq:Etwostar}
E_{2,k}^{*2} = (P_2)^2 = (E - \omega_k)^2 - (\vec P - \vec k)^2 \,.
\end{equation}
For Eq.~(\ref{eq:Eeqthree}) to hold, we must have that
$E_{2,k}^*\ge 2m$. For fixed total energy-momentum, this condition
holds only for a finite region of $\vec k$.

We now boost to the two-particle CM frame, which requires a boost velocity of
\begin{equation}
\vec \beta_k \equiv - \frac{\vec P - \vec k}{E - \omega_k} \,.
\end{equation}
We denote by \((\omega_a^*, \vec a^*)\) and \((\omega_{ka}^*, \vec b^*_{ka})\) 
the four vectors reached by boosting \((\omega_a, \vec a)\) 
and \((\omega_{ka}, \vec b_{ka})\), respectively. 
If Eq.~(\ref{eq:Eeqthree}) holds, then we have
\begin{equation}
\omega_a^* = \omega_{ka}^* = \frac{E^*_{2,k}}2  \ \ {\rm and}\ \ 
\vec a^* = - \vec b^*_{ka} \,,
\label{eq:onshellkin}
\end{equation}
while the magnitudes of the momenta in the two-particle CM frame
satisfy
\begin{equation}
a^*=b_{ka}^* = q_k^* \equiv \sqrt{E_{2,k}^{*2}/4 - m^2} \,.
\label{eq:qkstardef}
\end{equation}
Thus, once \((E, \vec P)\) and \(\vec k\) are fixed, the
remaining degrees of freedom for three on-shell particles can
be labeled by a single unit vector, \(\hat a^*\). 
This is simply the direction of motion for one of the two non-spectator
particles in their two-particle CM frame. We will often parametrize
the dependence on this direction in terms of spherical harmonics.

We can also interchange the roles of $\vec k$ and $\vec a$, treating
the latter as the spectator. In this case the CM energy of the
non-spectator pair is $E_{2,a}^*$ where
\begin{equation}
E_{2,a}^{*2} \equiv (E-\omega_a)^2-(\vec P-\vec a)^2 \,,
\end{equation} 
while the required boost has velocity
\begin{equation}
\vec \beta_a = - \frac{\vec P-\vec a}{E-\omega_a} \,.
\end{equation}
This boost leads to $(\omega_k,\vec k) \to (\omega_k^*,\vec k^*)$,
and the on-shell condition implies
\begin{equation}
k^*=q_a^*\equiv \sqrt{E_{2,a}^{*2}/4-m^2} \,,
\end{equation}
so that the three on-shell particles [with fixed $(E, \vec P)$] are parametrized by $\vec a, \hat k^*$.
This discussion exemplifies the notation that we will use repeatedly
below, wherein the subscripts denote which momentum is that of the
spectator, and it is clear from the context in which two-particle CM
frame starred quantities are defined.

Also relevant are situations in which two of the particles,
say those with momenta $\vec k$ and $\vec a$, are on shell, 
while the third is not. The energy-momentum of the third particle
is then $(E-\omega_k-\omega_a,\vec b_{ka})$.
As long as $E_{2,k}^{*2}>0$, we can still boost to the two-particle
CM-frame (with boost velocity $\vec \beta_k$), leading to
\begin{equation}
(\omega_a, \vec a) \longrightarrow (\omega_a^*, \vec a^*)\,,
\ \ {\rm and} \ \
(E-\omega_k-\omega_a,\vec b_{ka}) \longrightarrow
(E_{2,k}^*-\omega_a^*,-\vec a^*)
\,.
\label{eq:boost2}
\end{equation}
In this case, however, $a^*\ne q_k^*$, so the degrees of freedom
are now parametrized by $\vec k$ and the {\em vector} $\vec a^*$. As in the on-shell case, also here we can exchange the roles of $\vec k$ and $\vec a$. As long as $E^{*2}_{2,a}>0$, we can boost $(\omega_k, \vec k)$ by $\vec \beta_a$ to define $(\omega_k^*, \vec k^*)$, with $k^*$ now unconstrained.

\bigskip

We use these coordinates to express the momentum dependence of the
on-shell quantities appearing in the final result.
We start with two-to-two scattering, which occurs as a subprocess within
the larger three-to-three process.
We denote the two-to-two scattering amplitude by $\M$ and the corresponding
K-matrix\footnote{%
Our K-matrix $\K$ is standard above threshold, 
while below threshold it is defined by analytic continuation. 
This is discussed further below 
[see Eqs.~(\ref{eq:MKident}), (\ref{eq:Krhoseries}) and (\ref{eq:MKdif})].}
by $\K$. Assuming the particle with momentum $\vec k$ is the
unscattered spectator, an appropriate functional dependence is
$\M(\vec k, \hat a'^*, \hat a^*)$ and $\K(\vec k, \hat a'^*, \hat a^*)$.
In each quantity, the role of the first argument is to
specify the energy-momentum of the two scattering particles.
Knowing the spectator momentum $\vec k$, as well as the total energy-momentum, 
one can determine the lab-frame total momentum of the scattering pair
[$(E-\omega_k,\vec P-\vec k)$]
as well as the boost velocity $\vec \beta_k$ needed to move to the 
scattering CM-frame.
In the latter frame, $\hat a^*$ and $\hat a'^*$ are, respectively,
the initial and final directions of one of the scattered particles.
Decomposing the dependence on these directions into spherical harmonics,
we write\footnote{%
  This is the only exception to our notation
  involving superscript \(*\). The \(*\) on $Y^*_{\ell m}$ indicates
  complex conjugation.}
\begin{equation}
\K(\vec k, \hat a'^*, \hat a^*) =
4 \pi Y^*_{\ell',m'}(\hat a'^*)\Kin {\ell', m'; \ell, m} (\vec k) 
Y_{\ell,m}(\hat a^*)\,,
\label{eq:K2def}
\end{equation}
and similarly for $\M$. Here and in the following there is an implicit
sum over repeated indices. The factor of $4\pi$ is 
conventional~\cite{Kim:2005}.
Rotational invariance implies that
$\Kin {\ell', m'; \ell, m} (\vec k) \propto \delta_{\ell',\ell} \delta_{m',m}$,
and that for each $\ell$ there is only one independent physical quantity, the 
scattering phase-shift in the given partial wave. 

Now we turn to three-to-three scattering. Although our final quantization
condition contains a three-particle K-matrix, we first discuss the
standard three-to-three scattering amplitude, $\Mth$. This allows us
to describe a new issue that arises with three particles in a more
familiar context. As usual, $\Mth$ is the sum of all connected
six-point diagrams with external legs amputated and on shell.
We write its functional dependence as
\(\Mth(\vec k', \hat a'^*, \vec k, \hat a^*)\), 
where now the ``spectator'' momentum changes from the initial
($\vec k$) to the final ($\vec k'$) state.
The two direction vectors $\hat a^*$ and $\hat a'^*$ are defined in
the corresponding two-particle CM frames, which are different for the
initial and final states.
We stress that $\Mth$ is symmetric under particle interchange 
separately in the initial and final states, so that the choice of
spectator is arbitrary. We use asymmetric coordinates because
of the presence of two-to-two scatterings.

\begin{figure}
\begin{center}
\includegraphics{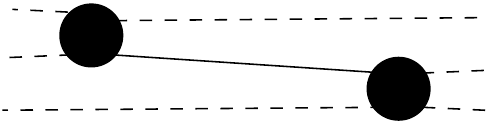}
\caption{Example of singular contribution to the
  on-shell three-to-three scattering amplitude.  
   Dashed lines are on-shell, amputated, external propagators, while the solid line is a fully dressed propagator, which can in general be off-shell.
  Filled circles represent two-to-two
  scattering amplitudes.  The internal (solid) line can become
  on shell for physical external momenta, corresponding to two
  isolated two-to-two scattering events.}
\label{fig:singdiagram}
\end{center}
\end{figure}

We would like to decompose $\Mth$ into spherical harmonics,
as in Eq.~(\ref{eq:K2def}). Although we can do this formally,
we do not expect the sum over angular momenta to converge uniformly.
This is because of a complication not present
in the two-to-two case: the three-to-three scattering amplitude has
physical singularities above threshold.\footnote{%
  The properties and physical consequences of
  these singularities are discussed, for example, in
  Refs.~\cite{Rubin:1966zz,Brayshaw:1969ab,Taylor:1977A,Taylor:1977B}.}
These singularities have nothing to do with bound states, but are
instead due to the possibility of two particles scattering and then
traveling arbitrarily far before one of them scatters off the third
particle (see Fig.~\ref{fig:singdiagram}). 
The three-particle interaction can thus become arbitrarily non-local.  
This means that, even at low energies, a truncation of the 
angular momentum sum is not justified, since a truncated
expansion will give a function that is everywhere finite.
Because truncation is crucial for practical applications of
the quantization condition, we must find a way around this problem.

\begin{figure}
\begin{center}

\includegraphics[scale=0.46]{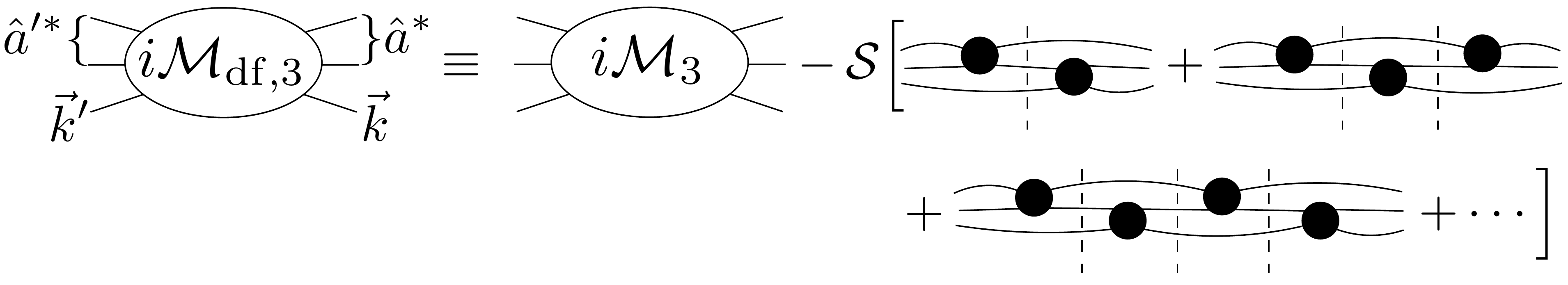}

\caption{Diagrammatic definition of the divergence-free three-to-three
  amplitude, \(\Mdfth {}\). In the subtracted term, filled circles
  represent on-shell two-to-two scattering amplitudes \(\M\). Dashed
  cuts stand for simple kinematic factors that appear between adjacent $\M$. These factors have the requisite poles so that the subtracted terms cancel the singularities in $\Mth{}$.
 The \(\mathcal S\) outside the square
  brackets indicates that the subtracted terms are symmetrized.}
\label{fig:Mdfdef}
\end{center}
\end{figure}

Our solution is to introduce an intermediate quantity that has
the same singularities as the three-to-three scattering amplitude but
depends only on the on-shell two-to-two amplitude \(\M\). This is
possible because divergences in the three-to-three scattering
amplitude are always due to diagrams with only pairwise scatterings,
with all intermediate states on shell.\footnote{%
  Indeed, a diagram
  with \(n\) two-to-two scatterings is divergent if and only if it is
  kinematically possible to have \(n\) classical pairwise scatterings
  (not counting events with zero momentum transfer). For degenerate
  particles only three scatterings are possible so there are two
  divergent diagrams.  For non-degenerate particles further
  scatterings are possible. 
  This is explained in Ref.~\cite{Rubin:1966zz}.  As we will find, our
  derivation requires that we subtract all the diagrams that are
  needed to render the non-degenerate $\Mdfth$ finite, even though all
  but two of these are finite for the degenerate case we study.}  
Labeling this intermediate quantity \(\mathcal M_{\mathrm{sing},3}\),
we define the ``divergence-free'' amplitude by
\begin{equation}
\label{eq:Mdfdef1}
\Mdfth(\vec k', \hat a'^*, \vec k, \hat a^*)
\equiv
\Mth (\vec k', \hat a'^*, \vec k, \hat a^*) 
- \mathcal M_{\mathrm{sing},3}(\vec k', \hat a'^*, \vec k, \hat a^*)
\,.
\end{equation}
This is shown diagrammatically in Fig.~\ref{fig:Mdfdef}.
By construction, $\Mdfth$ is a smooth function, 
and therefore has a uniformly convergent partial-wave expansion:  
\begin{equation}
\Mdfth(\vec k', \hat a'^*, \vec k, \hat a^*) =
4 \pi Y^*_{\ell',m'}(\hat a'^*) 
\Mdfthin {\ell', m'; \ell, m} (\vec k', \vec k) Y_{\ell,m}(\hat a^*) 
\,.
\label{eq:Mdf3def}
\end{equation}
The singular part, $\mathcal M_{\mathrm{sing},3}$, must be included without
partial-wave decomposition. A diagrammatic definition of
$\mathcal M_{\mathrm{sing},3}$ in sketched in Fig.~\ref{fig:Mdfdef};
it can be defined formally as the solution to an integral equation. 
Since we do not need this quantity in this work we do not
go into the details here.

As already noted above, our quantization condition depends not on $\Mdfth$
but rather on a closely related K-matrix-like quantity $\Kdfth$.
Roughly speaking,
this is built up of the same Feynman diagrams as $\Mdfth$, 
and has the above-threshold divergence removed in a similar way. However, 
to define $\Kdfth$,
a modified principal-value (PV) pole prescription is used instead of 
the $i\epsilon$ prescription, and there are some additional subtleties.
Thus we delay a full definition until we present the derivation
of the quantization condition.
What matters here is that $\Kdfth$ is a non-singular, infinite-volume
quantity, closely related to the scattering amplitude.
It is also separately symmetric under initial and final particle interchange.
Its functional dependence and harmonic decomposition is as for $\Mdfth$:
\begin{equation}
\Kdfth(\vec k', \hat a'^*, \vec k, \hat a^*) =
4 \pi Y^*_{\ell',m'}(\hat a'^*) \Kdfthin{\ell', m'; \ell, m} (\vec k', \vec k) 
Y_{\ell,m}(\hat a^*) 
\,.
\label{eq:Kdf3def}
\end{equation}
We stress that \(\Kdfthin {\ell', m'; \ell, m} (\vec k', \vec k)\) 
is not diagonal in $\ell$ or $m$,
since two-particle angular momentum is not a good quantum
number in three-to-three scattering.
It is also noteworthy that our derivation of the quantization condition
automatically leads to removal of the divergent part from $\Kdfth$.
Thus not only is the subtraction reasonable from the
perspective of defining useful infinite-volume observables
(i.e. allowing a convergent partial-wave expansion)
it also arises naturally in our investigation of the finite-volume theory.

\bigskip
We are now in a position to present the quantization condition:
a relation between \(\K\), \(\Kdfth\) and the finite-volume spectrum. 
This relation involves three-particle phase space restricted by the
constraint of finite-volume. In particular, we need
\(\Kin {\ell',m';\ell,m}(\vec k)\) and 
\(\Kdfthin{\ell',m';\ell,m}(\vec k', k)\) 
only for \(\vec k, \vec k' \in (2 \pi/L) \mathbb Z^3\). We
therefore define the finite-volume restrictions of these amplitudes
\begin{align}
\Kin {k',\ell',m';k,\ell,m} & 
\equiv \delta_{k',k} \Kin {\ell',m';  \ell,m}(\vec k) 
\ \ \ \ \mathrm{for} \ \ \ \ \vec k \in (2 \pi/L) \mathbb Z^3 \,, 
\label{eq:K2deflong}
\\ 
\Kdfthin {k',\ell',m';k,\ell,m} & \equiv
\Kdfthin {\ell',m';\ell,m}(\vec k', \vec k)
\ \ \ \ \mathrm{for} \ \ \ \ \vec k', \vec k \in (2 \pi/L) \mathbb Z^3
\,.
\end{align}
The left-hand sides of these equations are to be viewed as
matrices in an extended space with indices\footnote{%
Our notation for the momentum indices, $k$ and $k'$, is somewhat imprecise.
These each are stand-ins for three-dimensional integer vectors labeling
the allowed finite-volume momenta. 
In other words, whenever a spectator momentum occurs as an index,
it indicates implicitly that the corresponding three-vector momentum
is one of those allowed in finite volume.}
\begin{equation}
\label{eq:matspace}
[\mathrm{finite\ volume\ momentum\ } \vec k \in (2 \pi/L) \mathbb Z^3] 
\times [\mathrm{two\ particle\ angular\ momentum}] \,.
\end{equation}
All other quantities entering our final result will also be matrices
acting on this space.

The finite-volume spectrum is determined by 
\begin{equation}
\label{eq:mainres}
\det \! \big [1 +F_{3} \Kdfth \big ] = 0 \,,
\end{equation}
where the determinant is over the direct product space just introduced.  
The matrix $F_3$ is 
\begin{equation}
\label{eq:F3def}
F_{3} \equiv \frac{F}{2 \omega L^3} 
\left[-\frac23 + \frac{1}{1 + [1 + \K G ]^{-1} \K F}\right ] \,,
\end{equation}
where
\begin{align}
\label{eq:omegamatdef}
\left[ \frac{1}{2 \omega L^3} \right]_{k',\ell',m';k,\ell,m} & \equiv
\delta_{k',k} \delta_{\ell',\ell} \delta_{m',m} \frac{1}{2 \omega_k L^3} 
\,,\\ 
\label{eq:Gdef}
G_{p, \ell', m' ; k, \ell, m} 
& \equiv
\left(\frac{k^*}{q_p^*}\right)^{\ell'} 
\frac{4 \pi Y_{\ell',m'}(\hat  k^*) 
H(\vec p\,) H(\vec k\,) Y_{\ell,m}^*(\hat p^*)} 
{2 \omega_{kp} (E - \omega_k - \omega_p - \omega_{kp})}
\left(\frac{p^*}{q_k^*}\right)^\ell 
\frac{1}{2 \omega_k L^3} \,, 
\\
\label{eq:Fdef1}
F_{k', \ell',m';k,\ell,m} 
& \equiv 
\delta_{k',k} F_{\ell',m';\ell,m}(\vec k)\,,
\\
\label{eq:Fdef2}
F_{\ell',m';\ell,m}(\vec k)
&=
F^{i\epsilon}_{\ell',m';\ell,m}(\vec k)
+
\rho_{\ell', m'; \ell, m}(\vec k) \,,
\\
\label{eq:Fdef3}
F^{i\epsilon}_{\ell',m';\ell,m}(\vec k)
&=
\frac12 \left[\frac{1}{L^3} \sum_{\vec a} - \int_{\vec a} \right] 
\frac{{4 \pi} Y_{\ell',m'}(\hat a^*) Y_{\ell,m}^*(\hat a^*) H(\vec k)
  H(\vec a\,)H(\vec b_{ka})}
{2 \omega_a 2 \omega_{ka}(E - \omega_k - \omega_a -  \omega_{ka} + i \epsilon)}
\left(\frac{a^*}{q_k^*}\right)^{\ell+\ell'} \,,
\end{align}
where $\int_{\vec a} \equiv \int d^3 a/(2 \pi)^3$ and
the sum over $\vec a$ in $F^{i\epsilon}$  runs over all finite-volume momenta. 
Here $\rho$ is a phase-space factor defined by
\begin{align}
\label{eq:rhodef}
\rho_{\ell',m';\ell,m}(\vec k)& \equiv 
\delta_{\ell',\ell} \delta_{m',m} H(\vec k) \tilde\rho(P_2)\,,
\\
\label{eq:rhodef2}
\tilde\rho(P_2) &\equiv \frac{1}{16 \pi  \sqrt{P_2^2}} \times
\begin{cases} 
-  i \sqrt{P_2^2/4-m^2} & (2m)^2< E_{2,k}^{*2} \,, 
\\ 
\vert \sqrt{P_2^2/4-m^2} \vert &   0<E_{2,k}^{*2} \leq (2m)^2 \,,
\end{cases}
\end{align}
where we recall that $P_2$ is the four-momentum of the non-spectator pair.
Finally, $H$ is a smooth cut-off function to be defined shortly. 

The quantization condition Eq.~(\ref{eq:mainres}) is our main
result, and will be derived in Sec.~\ref{sec:der}.
Here we work our way through the definitions,
explaining the origin and meaning of each contribution.
As noted above, $\Kdfth$ is closely related to the divergence-free
part of the full three-to-three scattering amplitude.
The singular parts of this amplitude end up in the quantity $F_3$,
where they lead to chains of the form $\dots \K G\K G \K\dots$ which
are obtained by expanding out $[1+\K G]^{-1}\K$. 
These chains arise from subtraction terms like those in Fig.~\ref{fig:Mdfdef},
with the filled circles now representing on-shell K-matrices $\K$
(rather than $\M$). The singular ``cuts'' between K-matrices give rise to the kinematical factors $G$.

In the definition of $G$, Eq.~(\ref{eq:Gdef}), we are using the notation
described in Eqs.~(\ref{eq:omegadef})-(\ref{eq:boost2}), with $\vec p$ in place of $\vec a$.
 Observe in particular that $G$ makes use of the off-shell phase-space described in the paragraph containing Eq.~(\ref{eq:boost2}). Since both $\vec k$ and $\vec p$ can equal any finite-volume three momentum, $(E-\omega_k-\omega_p,\vec b_{pk})$ will generally not be on shell. For this reason the magnitude of $\vec k^*$ (defined via a boost with velocity $\vec \beta_p$) and that of $\vec p^*$ (boost velocity $\vec \beta_k$) are unconstrained. These magnitudes appear in the factors $(k^*/q_p^*)^{\ell'}$ and $(p^*/q_k^*)^\ell$, which
remove singularities due to the spherical harmonics and so ensure that $G$
is non-singular for $\vec k^*$ or $\vec p^*$ equal to zero. (A similar factor $(a^*/q_k^*)^{\ell+\ell'}$ appears in $F$ for the same reason.)

\begin{figure}
\begin{center}
\includegraphics[scale=0.8]{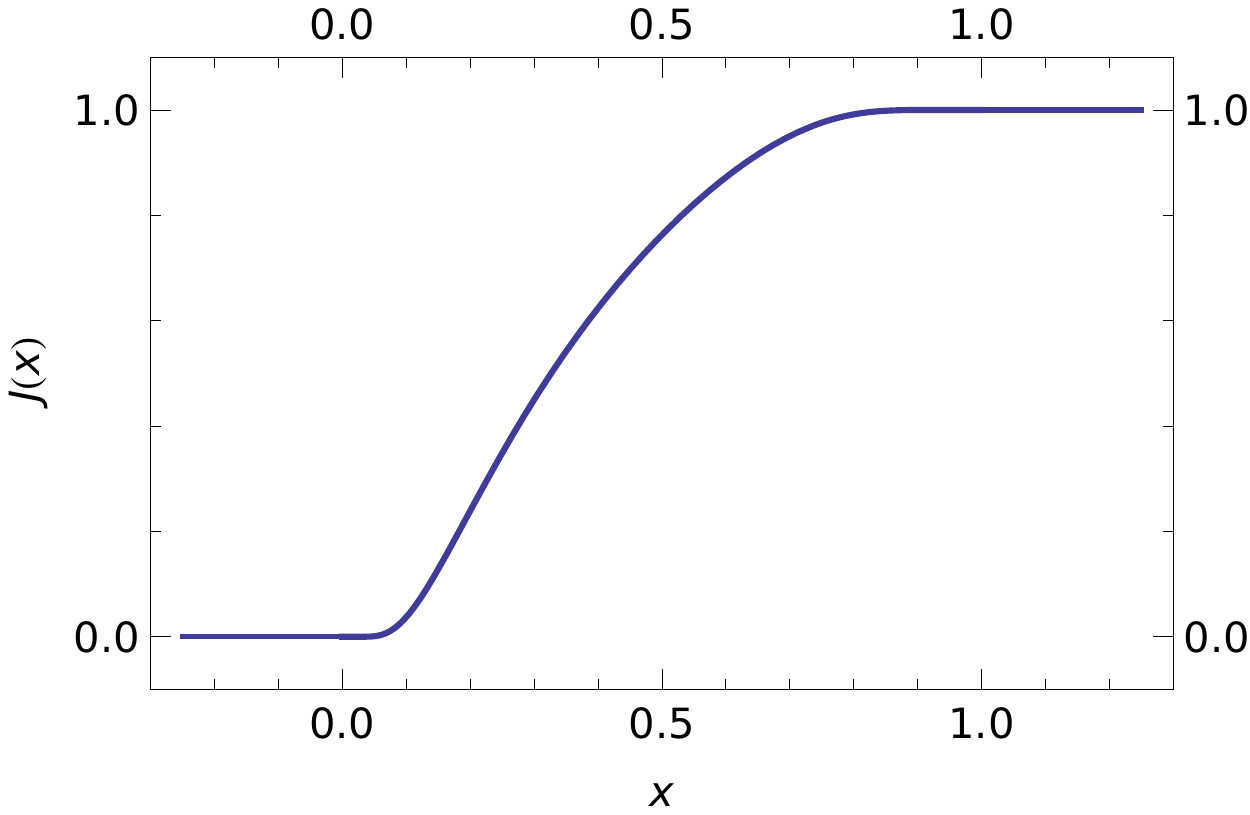}

\caption{The smooth cutoff function $H(\vec k)\equiv J(x)$
with $x=E_{2,k}^{*2}/[4m^2]$. The function varies from 0 to 1 as
\\$E^{*2}_{2,k} \equiv (E - \omega_k)^2 -(\vec P - \vec k)^2 $ varies from $0$ to $4 m^2$. Using this range of variation ensures that the function has width $\Delta \approx m$ in the space of spectator momentum $\vec k$. }
\label{fig:hfunction}
\end{center}
\end{figure}

The final ingredient in $G$ is the function $H$
(which appears also in $F$).
The role of $H$ is to provide a smooth ultraviolet cut-off
on the sum over spectator momentum.
There are two cut-off functions, $H(\vec p)$ and $H(\vec k)$,
because $G$ has different spectator momenta in its
left- and right-handed indices
($\vec p$ and $\vec k$, respectively). 
To understand the need for the cut-off we note that, for fixed
$(E,\vec P)$, as the spectator momentum (say $\vec k$) increases 
in magnitude, the energy-momentum of the other two particles falls
below threshold, $E_{2,k}^*< 2m$.
Now, in the quantization condition (\ref{eq:mainres}),
the determinant runs over {\em all} values of spectator momentum, which
leads to values of $E_{2,k}^{*2}$ arbitrarily far below threshold.
Once $E_{2,k}^{*2}\le 0$, however,
the boost needed to define $p^*$ becomes unphysical
($|\beta_k|\ge 1$).
The cut-off function $H(\vec k)$ resolves this issue.
It has the properties
\begin{equation}
\label{eq:HvalA}
 H(\vec k) = 
\begin{cases}
0 \,, & E_{2,k}^{*2} \le 0 \,;
\\ 
1 \,, & (2m)^2 < E_{2,k}^{*2} \,.
\end{cases}
\end{equation}
where the first condition removes unphysical boosts
and the second ensures that the cut-off does not change the
contributions from on-shell intermediate states.
In the intermediate region, \(0 < E_{2,k}^{*2} < (2m)^2\), 
$H(\vec k)$ interpolates between 0 and 1.
For reasons that will become clear in the derivation below,
this interpolation must be smooth. An example of a function which does the job is
\begin{equation}
\label{eq:Hdef}
H(\vec k) \equiv J(E_{2,k}^{*2}/[4m^2]) \,,
\end{equation}
with
\begin{equation}
\label{eq:Jdef}
J(x) \equiv
\begin{cases}
0 \,, & x \le 0 \,; 
\\ 
\exp \left( - \frac{1}{x} \exp \left [-\frac{1}{1-x} \right] \right ) \,, 
& 0<x \le 1 \,; 
\\ 
1 \,, & 1<x \,.
\end{cases}
\end{equation}
This function is plotted in Fig.~\ref{fig:hfunction}.  

It would also be consistent with the requirements stated so far 
to have $H$ remain smooth but transition more rapidly from 0 to 1.
In that case, however, the difference between a sum and an integral over $H$ will be enhanced
\begin{equation}
\smi k H(\vec k) = \mathcal O(e^{-\Delta L}) \,,
\end{equation}
with $\Delta$ the width
of the drop-off region. Since these
corrections are neglected, an enhancement from using too small a width
would invalidate our final result. We must thus additionally require
\begin{equation}
\label{eq:Hsmooth}
\smi k H(\vec k) = \mathcal O(e^{-mL}) \,.
\end{equation}
In other words we must ensure that
\(m\) is the smallest energy scale in the problem, and thus take
$\Delta\approx m$.  The form sketched in Fig.~\ref{fig:hfunction}
satisfies this requirement.

The appearance of sub-threshold momenta is a general feature
of the three-particle quantization condition, as first pointed out
in Ref.~\cite{Polejaeva:2012ut}.
Indeed, for spectator momenta such that
$0< E_{2,k}^{*2}< (2m)^2$,
the two-particle K-matrices in $F_3$ are evaluated below threshold.
Our modified PV prescription 
[denoted $\PV$ and defined in Eqs.~(\ref{eq:PVtildedef}) and (\ref{eq:PVtildedef2}) below]
ensures that this is achieved by analytic continuation.\footnote{%
  This is in distinction to the standard PV prescription,
  which leads to a cusp in $\K$ at threshold.
  Our definition is the same as that used in studies of 
  bound-state energies using L\"uscher's two-particle quantization
  condition (see, e.g., Refs.~\cite{Beane:2003da,Sasaki:2006jn}). 
  In particular, the quantity
  \((a^*)^{2\ell+1} \cot \delta_\ell(a^*)\) has
  a Taylor expansion in $(a^*)^2$ that can be 
  analytically continued to $(a^*)^2<0$.}
The cut-off functions in $G$ (and in $F$) ensure that these
sub-threshold contributions are absent for $E_{2,k}^{*2}\le 0$.
The three-particle amplitude $\Kdfth$ must also be evaluated 
for sub-threshold momenta, which is achieved by analytic continuation. 

The final matrix that enters the quantization condition is $F$.
This is the kinematic factor that brings in finite-volume effects.  
Its definition uses the notation introduced
in Eqs.~(\ref{eq:omegadef})-(\ref{eq:boost2}). 
As shown in Eq.~(\ref{eq:Fdef1}), it is diagonal in spectator momentum,
and is thus a two-particle quantity.
Indeed, the matrix $F^{i\epsilon}(\vec k)$ defined in Eq.~(\ref{eq:Fdef3})
is essentially the same as the kinematic quantity of the same name
introduced in Ref.~\cite{Kim:2005} in the formulation of the two-particle
quantization condition in a moving frame.  The precise relation,
given in Eq.~(\ref{eq:comparetoKim}) in Appendix~\ref{app:sumintegral}, allows $F^{i\epsilon}$ to be written in terms of
generalizations of the zeta-functions introduced in
Refs.~\cite{Luscher:1986n2,Luscher:1991n1}. 
The only difference between our $F^{i\epsilon}$ 
and that of Ref.~\cite{Kim:2005} is
that we use a different ultraviolet cut-off---our cut-off is provided
by the product of $H$ functions. This change in cut-off leads, however,
to differences proportional to $e^{-mL}$,
which are exponentially suppressed as $L\to\infty$.

The kinematic factor which enters the quantization condition is $F$
rather than $F^{i\epsilon}$. The difference between these two quantities, 
given by Eq.~(\ref{eq:Fdef2}),
is the phase space factor $\rho$, a quantity that appears
repeatedly in the derivation of Sec.~\ref{sec:der} and which is diagonal in angular momentum.
For example, the relation between the two-particle scattering amplitude
and K-matrix is [see Eq.~(\ref{eq:MKdif})] 
\begin{equation}
\label{eq:MKident}
\M^{-1} = \K^{-1} + \rho \,,
\end{equation}
The $\rho$ term in $F$ 
arises because of our use of a modified PV pole prescription.
As can be seen from Eq.~(\ref{eq:Fdef3}), $F^{i\epsilon}$ is the
difference between a sum and integral of three-particle cut,
with the integral defined using the $i\epsilon$ prescription.
The $\rho$ term in Eq.~(\ref{eq:Fdef2}) is exactly what is needed
so that $F$ itself is the sum-integral difference with the integral
defined by the $\PV$ prescription. The latter is the quantity that
appears naturally in our derivation. This means that $F$ is real.
In addition, as we move below threshold [$E_{2,k}^{* 2}< (2m)^2$],
while $F^{i\epsilon}$ drops to zero rapidly, since the 
summand/integrand is no longer singular, $\rho$ (and thus $F$) grows
since $|q_k^*|$ is increasing.
Eventually, however, as $E_{2,k}^{* 2}$ approaches zero, this growth
is overcome by the decrease in the cut-off function $H$, 
such that $\rho$ vanishes for $E_{2,k}^{* 2}\le 0$.

\bigskip
The quantization condition (\ref{eq:mainres}) is similar in
form to that for two particles [see Eq.~(\ref{eq:twoquant}) below,
as well as Refs.~\cite{Kari:1995,Kim:2005,Hansen:2012tf,Briceno:2012yi}].
In principle, they are both to be used in the same way:
if one knows the scattering amplitudes $\K$ and $\Kdfth$ then,
for a given choice of $\{L,n_{\vec P}\}$, the quantization conditions
predict the finite-volume energy levels.
Of course, what we are really interested in is inverting this
prediction, i.e. using 
numerically determined energy levels to extract
information about infinite-volume scattering amplitudes.
This more challenging task is discussed in the following section.

%% file: isotropic.tex
\section{Truncating the quantization condition}

\indent

\label{sec:iso}

In this section we discuss how one might use the 
three-particle quantization condition, Eq.~(\ref{eq:mainres}),
in practice. 
Specifically, we assume that, using lattice simulations,
one has determined some number of three-particle
energy levels for a set of choices of $\{L, \vec P\}$.
From this information, we want to learn as much as possible
about $\Kdfth$.

The first step is to assume that, using L\"uscher's two-particle
quantization condition and its generalizations, the two-particle
K-matrix \(\Kin {\ell', m'; \ell, m}(\vec k)\) has been determined.
To do so in practice one must assume that $\K$ is negligible for large
enough angular momenta, which is a generally a 
good approximation for any fixed two-particle energy. 
Specifically, we assume $\K=0$
for $\ell>\ell_{{\rm max},2}$. In this case the
two-particle quantization condition is truncated to a 
solvable finite matrix condition.
In addition, since lattice results only determine values of $\K$
(or, equivalently, the phase shifts) 
for a discrete set of kinematical points, 
we assume that these have been suitably interpolated and/or extrapolated
to obtain continuous functions.

In the three-particle case, we are dealing with a larger index space,
containing the additional label for finite-volume spectator momenta.
However, the regulator function $H$ provides an automatic truncation
of this space. This occurs because, for fixed $(E,\vec P)$, there are a finite
number of values of $\vec k$ for which $H(\vec k)$ is non-vanishing. 
We call this number of values $N$.
This automatically truncates $G$ and $F$ 
(which contain $H(\vec k)$) to be $N\times N$ matrices
in spectator-momentum space, with all other entries vanishing.
Since $\K$ always sits between factors of
$F$ and $G$ [as can be seen by expanding out the nested geometric
  series in Eq.~(\ref{eq:F3def})], $\K$ is also effectively
truncated (in the sense that the terms in $\K$ lying outside the
$N\times N$ block do not contribute).  Since $F_3$ always has an $F$
at both ends (again after expanding out), it also is truncated.
Finally, expanding out the determinant
(e.g. using $\det Z =\exp\tr\ln Z$)
one sees that $\Kdfth$ always has an $F_3$ on both sides and so it
also is effectively truncated.\footnote{%
  The fact that the sum over $\vec k$ is truncated makes sense in the
  limit of weak interactions. If all interactions vanish, then,
  for given $\vec P$, there will only be a finite number of free three-particle
  states with energies below, or in the vicinity of, 
  any given choice for $E$. It is   primarily these states
  which are mixed by interactions to form the finite-volume eigenstates 
  with $E_i < E$.}

Next we consider the spherical harmonic indices. As already noted, we
assume $\K$ is truncated in these indices at $\ell_{{\rm max},2}$. To reduce the determinant condition to a finite-dimensional space, we must further assume that \(\Kdfth\) is truncated, in both
\(\ell\) and \(\ell'\), at \(\ell_{\mathrm{max,3}}\).  
This is reasonable because $\Kdfth$ is a smooth function, as is made clear in the course of defining it, in Sec.~\ref{sec:der} below. Defining $\ell_{\rm max}$ as the larger of 
$\ell_{{\rm max},2}$ and $\ell_{{\rm max},3}$, we find that all factors of $F$ and $G$ appearing in the quantization-condition are projected onto a \((2 \ell_{\mathrm{max}}+1)  \times (2 \ell_{\mathrm{max}}+1) \) subspace of the angular-momentum space. This follows from the argument already given above: expanding in $F$ and $G$, one finds that every factor of these two kinematic matrices sits between (and is thus truncated by) factors of either $\K$ or $\Kdfth$. The net result is that the
quantization condition collapses to that for truncated matrices of size 
\((2 \ell_{\mathrm{max}}+1) N \times (2 \ell_{\mathrm{max}}+1) N\). In this way the formal result has been turned into something more 
practical.\footnote{%
We suspect that it is inconsistent to 
choose $\ell_{{\rm max},3}<\ell_{{\rm max},2}$, because three-particle
scattering involves two-to-two subprocesses. Indeed the latter are the
leading cause of complications in the derivation presented below.
The most natural choice appears to us to be
$\ell_{{\rm max},3}=\ell_{{\rm max},2}$, although we do not know
how to demonstrate that this is a rigorous requirement.}

The final step is to assume a parametrization of the $\vec k'$ and $\vec k$
dependence of the non-zero angular-momentum components of $\Kdfth$.
We stress again that $\Kdfth$ is not diagonal in its angular-momentum
indices (unlike $\K$) so that there will be a larger number of components
to parametrize. 
Nonetheless, given knowledge of $\K$ (including analytic continuation below threshold),
each of the measured three-particle energy levels gives a relation between the parameters characterizing $\Kdfth$.
Thus, given enough energy levels one can solve for any finite set of parameters.
Although this sounds complicated, we note that the recent kaon 
resonance study of Ref.~\cite{Dudek:2014qha} was able to deal with
multiple (two-particle)
channels using a suitable parametrization and many energy levels.

We close this section by working out the simplest possible case of
the above-described program. We assume that
both $\K$ and $\Kdfth$ are s-wave dominated
(i.e. $\ell_{{\rm max},2}=\ell_{{\rm max},3}=0$),
and that $\Kdfth$ is a function only of the total three-particle
CM energy. These assumptions are summarized by
\begin{equation}
\K(\vec k, \hat a'^*, \hat a^*) = \K^s(E_{2,k}^*)
\ \ {\rm and}\ \ 
\Kdfth (\vec k', \hat a'^*, \vec k, \hat a^*) = \Kdfth^{\rm iso}(E^*)
\,.
\label{eq:isoapprox}
\end{equation}
All matrices entering the quantization condition thus collapse to
$N\times N$ matrices in spectator-momentum space, 
and have the explicit forms
\begin{align}
{\cal K}_{2;k',k}^s & \equiv \delta_{k',k}\, \K^{s}(E_{2,k}^*) \,,
\label{eq:K2sdef}
\\
{\cal K}_{{\rm df},3;k',k}^s & \equiv \Kdfth^{\rm iso}(E^*)\,,
\label{eq:Kdf3sdef}
\\
G_{p,k}^s & \equiv \frac{H(\vec p\,) H(\vec k)}
{2 \omega_{kp} (E - \omega_k -   \omega_p - \omega_{kp})} 
\frac{1}{2 \omega_k L^3} \,, 
\label{eq:Gsdef}
\\ F_{k',k}^s & \equiv \delta_{k',k}
\frac{1}{2} \left[\frac{1}{L^3} \sum_{\vec a} - \int_{\vec a} \right]
\frac{H(\vec k\,) H(\vec a\,)H(\vec b_{ka})}
{2 \omega_a 2 \omega_{ka}(E - \omega_k - \omega_a - \omega_{ka} + i \epsilon)} 
+ \delta_{k',k}  H(\vec k) \tilde \rho(P_2) \,. 
\label{eq:Fsdef}
\end{align}
Since $E^*$ is fixed, all $N^2$ entries of the matrix 
${\cal K}_{{\rm df},3;k',k}^s$ have the same value.
It therefore has only one non-zero eigenvalue, \(N \Kdfth^{\rm iso}(E^*)\). 
If we work in the basis in which \(\Kdfth^s\) is diagonal,
then, irrespective of the form of $F_3$,
the quantization condition (\ref{eq:mainres}) reduces to the single equation:
\begin{equation}
1 + F_{3}^{\rm iso} \Kdfth^{\rm iso}(E^*)=0  \,.
\label{eq:mainiso}
\end{equation}
Here
\begin{equation}
F_{3}^{\rm iso} \equiv \sum_{\vec k, \vec p} \frac1{2 \omega_k L^3}
\left [ F^s
  \left( - \frac{2}{3} + \frac{1}{1 + [1 + \K^s G^s]^{-1} \K^s F^s}\right)
  \right]_{k,p}
\end{equation}
is (up to a factor of $1/N$)
the projection of $F_3^s$ into the subspace spanned 
by the eigenvector of $\Kdfth^s$ with non-zero eigenvalue.
We stress that the sums over $\vec k$ and $\vec p$ are both truncated
to $N$ contributions by the factors of $H$ contained in $F^s$.\footnote{%
The truncations that enter through the $H$ functions can also be
relaxed in the isotropic limit if desired. Recall that $H(\vec k)$ was
required to vanish for $E^{*2}_{2,k}<0$, see Eq.~(\ref{eq:Jdef}). This
is necessary because otherwise the various starred quantities that
enter $F$ and $G$ become ill-defined. However, as is clear from
Eqs.~(\ref{eq:Gsdef}) and (\ref{eq:Fsdef}), all of these starred
quantities are absent in the isotropic limit. Thus $H(\vec k)$ may
have support for $E^{*2}_{2,k}<0$, as long as $\mathcal K_2^s(E^*_{2,k})$ 
is a well-defined smooth function which is known
over the energy range included. This extension of $H$ is required to
show that our quantization condition reproduces the threshold
expansion of Refs.~\cite{Beane:2007qr,Tan:2007}. This check will be
presented in Ref.~\cite{HSinprep}.}

The result (\ref{eq:mainiso}) is strikingly simple.
If we know \(\K^s\) for two-particle CM energies 
in the range $0 < E_{2,k}^* < E^*-m$, then we can evaluate
\(F_{3,s}\), a real function depending only on \(E\) and \(L\). 
Evaluating this function at a value of \(L_i\) for which \(E_i\) is known to
be in the finite-volume spectrum, then gives, using
Eq.~(\ref{eq:mainiso}), \(\Kdfth^{\rm iso}(E_i^*)=-1/F_3^{\rm iso}(E_i,L_i)\).
This is conceptually very similar to the application of the
two-particle quantization condition, which, in the single-channel
limit, can be written as $1+F \K=0$ 
[see Eq.~(\ref{eq:twoquant}) in the following section].
The difference is that the quantity $F_3$ contains information
about two-particle scattering, while $F$ is simply a kinematic
function. This difference reflects the fact that, in the three-particle
case, particles can interact pairwise as well as all together.

One concern one might have about the isotropic approximation
and the result (\ref{eq:mainiso}) is that one apparently
only obtains a single energy level whereas $N$ free three-particle
levels enter the analysis. It thus seems that some finite-volume
states have been lost. In fact, all but one of the free states
are present once one takes into account that the equality of
all $N^2$ elements of the truncated $\Kdfth$ will not be exact. 
This is shown in a particular example
in Appendix~\ref{app:iso}.\footnote{%
A similar issue arises with the two-particle
quantization condition when one truncates the angular momentum expansion.
The ``lost'' states involving higher angular momenta are recovered
if one reintroduces the higher partial wave amplitudes but with infinitesimal
strength. The quantization condition then has solutions corresponding to
free two-particle states projected onto states in appropriate irreps of
the finite-volume symmetry group.}

%% file: derivation.tex
\section{Derivation}
\label{sec:der}

\indent

In this section we present a derivation of the quantization condition
described in the previous section. Following Ref.~\cite{Kim:2005}, we
obtain the spectrum from the poles in the finite-volume Minkowski-space
correlator\footnote{%
  Minkowski time turns out to be convenient for our
  analysis, even though numerical lattice determinations of the spectrum
  work in Euclidean time.
  The point is that the finite-volume spectrum is the same, however it
  is determined.}
\begin{equation}
C_L(E, \vec P) \equiv \int_{L} d^4 x 
e^{i(E x^0-\vec P \cdot \vec x)} 
\langle 0 \vert \mathrm{T} \sigma(x) \sigma^\dagger(0) \vert 0
\rangle \,.
\end{equation}
Here T indicates time-ordering and \(\sigma(x)\) is an interpolating
field coupling to states with an odd number of particles. 
The Fourier transform, implemented via
an integral over the finite spatial volume, restricts the states to have
total energy $E$ and momentum $\vec P=2 \pi \vec n_P/L$. 

The simplest choice for \(\sigma(x)\) is a one-particle interpolating
field, $\phi(x)$, since in the interacting theory this will couple to states
with any odd number of particles.
In a simulation, however, it is advantageous
to use a choice with larger overlap to the
three-particle states of interest. An example is
\begin{equation}
\label{eq:sigex}
\sigma(x) = \int_L d^4 y d^4 z f(y,z) \phi(x) \phi(x+y) \phi(x+z) \,,
\end{equation}
with $f$ a smooth function with period $L$ in all directions.

At fixed \(\{L, \vec n_P\}\),\footnote{%
  It is more natural to think in
  terms of \(\{L, \vec n_P\}\) rather than \(\{L, \vec P\}\),
  since $\vec n_P$ is quantized whereas $\vec P$ varies with $L$.}
the spectrum of our theory is the set of CM frame
energies \(E^*_j\), $j=1,2,\cdots$ for which
$C_L(E_j,\vec P)$ has a pole, with $E_j= (E_j^{*2} + \vec P^2)^{1/2}$. 
Our goal is thus to include all contributions to \(C_L\) 
which fall at most like a power of $1/L$, and 
determine the pole structure. In the previous section we summarized
the main result of this work, but made no reference to the correlator
in doing so. The connection is given by the following identity, the
demonstration of which is the task of this section:
\begin{equation}
\label{eq:corrresult}
C_{L}(E, \vec P) = C_\infty(E, \vec P) + i A' \frac{1}{1+F_{3} \Kdfth
  {}} F_{3} A \,.
\end{equation}
This result is valid up to terms exponentially suppressed in the
volume, terms which we will discard implicitly throughout this section.
The quantities \(A' \equiv A'_{k',\ell',m'}\) and
\(A \equiv A_{k,\ell,m}\) are, respectively, row and column vectors 
in [finite-volume momentum]\(\times\)[two-particle angular momentum] space.  
Since
\(A\) and \(A'\) do not enter the quantization condition, 
we have not given their definitions above.
Indeed, we think it most useful to
introduce their definitions as they emerge in our all orders
summation. We have also introduced \(C_\infty\), which is an
infinite-volume correlator whose definition we will also
build up over the following subsections.

A key technical issue in the derivation is the need to use
a non-standard pole prescription when defining momentum integrals
in infinite-volume Feynman diagrams. This is at the root of the
complications in defining $A'$, $A$ and $C_\infty$.
Despite these complications, these are infinite-volume
quantities that we do not expect to lead to poles in $C_L$.\footnote{%
We discuss this point, following the derivation,
at the end of this section.}
It follows that, at fixed \(\{L, \vec n_P\}\), 
\(C_{L}\) diverges at all energies for which the matrix
between \(A\) and \(A'\) has a divergent eigenvalue. In addition, as
long as \(\Kdfth\) is nonzero, diverging eigenvalues of \(F_3\) leave
the finite-volume correlator finite. The spectrum is therefore given
by energies for which \([1 + F_{3} \Kdfth ]\) has a
vanishing eigenvalue, which is the quantization condition quoted above.

The demonstration of Eq.~(\ref{eq:corrresult}) proceeds by 
an all-orders analysis of the
Feynman diagrams building up the correlator. As we accommodate any
scalar field theory (assuming only a \(\mathbb{Z}_2\) symmetry), 
Feynman diagrams consist of any number of even-legged vertices, as well
as one each of the interpolating fields \(\sigma\) and
\(\sigma^\dagger\), connected by propagators. The finite-volume
condition enters here only through the prescription of summing (rather
than integrating) the spatial components of all loop momenta, i.e.
\begin{equation}
 \frac{1}{L^3} \sum_{\vec q = 2 \pi \vec n/L} \int \frac{d q^0}{2 \pi}
 \mathrm{\ \ \ over\ all\ \ \ } \vec n \in \mathbb Z^3 \,.
\end{equation}

We now introduce the crucial observation that makes our derivation
possible: Power-law finite-volume effects only enter through on-shell
intermediate states. This motivates a reorganization of the
sum of diagrams into a skeleton expansion that keeps all on-shell
intermediate states explicit, while grouping off-shell states into
Bethe-Salpeter kernels. Heuristically, the importance of on-shell
intermediate states can be understood by noting that on-shell
particles can travel arbitrarily far, and are thus maximally affected
by the periodic boundary conditions. By contrast, off-shell states are
localized so that the effect of finite-volume is smaller (and, indeed,
exponentially suppressed in general).

The technical justification for this description begins by noting
that the difference between a sum and an integral acting on 
a smooth (i.e. infinitely differentiable) function \(f(\vec q \,)\)
falls off faster than any power of \(1/L\)~\cite{Luscher:1991n1}.\footnote{%
This is what we refer to as exponentially suppressed, 
although strictly it is not equivalent.}
As noted above, we treat terms with this
highly suppressed scaling in \(L\) as negligible, and thus set
\begin{equation}
\label{eq:sumintf}
\bigg [\frac{1}{L^3} \sum_{\vec q} - \int_{\vec q} \bigg] f(\vec q\,)
= 0 \,.
\end{equation}
By contrast, if a function \(d(\vec q\,)\) is not continuous but
instead diverges for some real \(\vec q\), or if some derivative
diverges, then the sum-integral difference receives power-law
corrections
\begin{equation}
\label{eq:sumintd}
\bigg [\frac{1}{L^3} \sum_{\vec q} - \int_{\vec q} \bigg] d(\vec q\,)
= \mathcal O(L^{-n}) \,,
\end{equation}
for some positive integer \(n\). We keep all such contributions.

A convenient tool to determine when the summands of Feynman
diagrams are singular is time-ordered perturbation theory (TOPT).\footnote{%
For a clear discussion of this method see Chapter 9 of Ref.~\cite{Stermantext}.}
In this method one first does all $k^0$ integrals, leaving
only the sums over spatial components of loop momenta.
(In a continuum application these would, of course, be replaced
by integrals.)
Each Feynman diagram then becomes
a sum of terms corresponding to the different time-orderings
of the vertices. Within a given time ordering,
each pair of neighboring vertices leads to an energy denominator,
\begin{equation}
\frac{1}{E_{\rm cut}- \sum_{y\in {\rm cut}} \omega_y}
\,.
\end{equation}
Here $E_{\rm cut}$ is total energy flowing through the propagators
in the ``cut'', which is the vertical line between adjacent vertices. The propagators have momenta $\vec p_y$
and on-shell energies $\omega_y$. For our correlator $E_{\rm cut}$
can be $E$, $0$ or $-E$, depending on the time ordering.
All other factors in the summand are non-singular: they arise from
momentum dependence in the vertices or from $1/\omega$ factors.

Given the assumed $Z_2$ symmetry and our choices of $\sigma$ and
$\sigma^\dagger$, the cuts in the diagrams contributing to
$C_L$ can only involve an
odd number of particles. Furthermore, given the restriction
$m <E^* < 5m$, the only energy denominators which can vanish
must involve three particles in the cut, i.e.
\begin{equation}
\label{eq:threeompole}
\frac{1}{E - \omega_k - \omega_a - \omega_{ka}} \,.
\end{equation}
Thus it is only when a three-particle state goes on shell
that replacing the sum over spatial momenta with
an integral can lead to power-law corrections.

The only subtlety in the application of this result to
our analysis is that $m$ 
(which appears in $\omega_k^2=\vec k^2+m^2$ and in the
condition on $E^*$) should be the physical and not the bare mass.
Technically this arises because the usual geometric sum of
irreducible two-point correlation functions shifts
the pole position in the dressed propagator to the physical
mass. This sum should be done before applying the TOPT analysis.\footnote{%
Doing things in this order makes the application of TOPT
more complicated, because the
dressed propagator itself now has multiple-particle poles.
This subtlety does not affect our analysis because all we are
taking from TOPT is the conclusion that divergences occur when
a time integral extends, undamped, over an infinite range.
Thus it is the long-time dependence of the dressed propagator that
matters, and this has the same form as 
that of a free propagator but with the physical mass.
We stress that we use TOPT only to identify diagrams that
can lead to power-law corrections. We do not use TOPT
to do the calculation, but rather use 
standard relativistic Feynman rules.}

\begin{figure}
\begin{center}
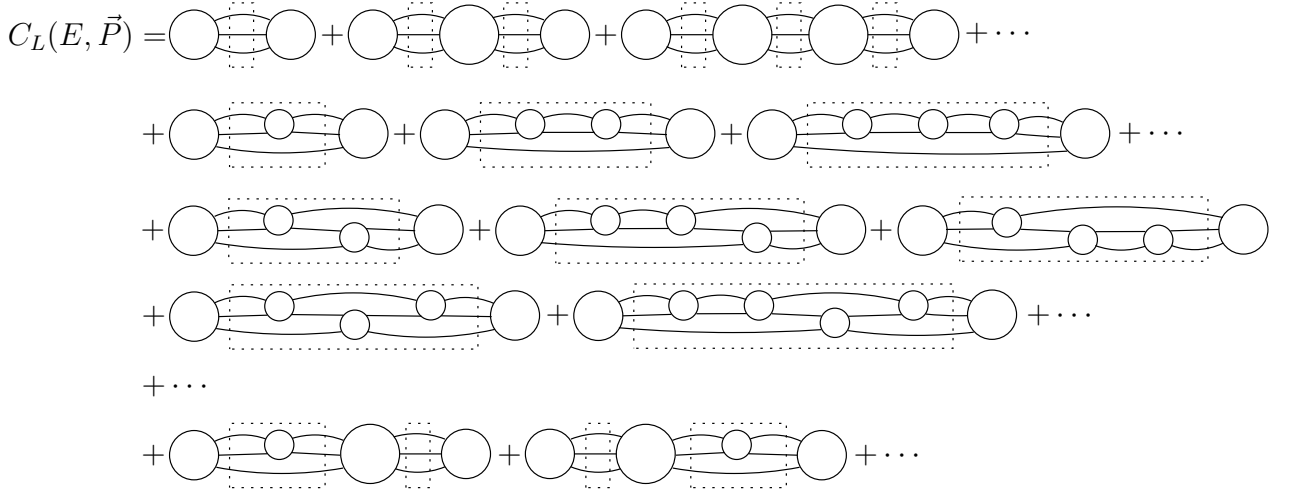
\caption{Skeleton expansion defining the finite-volume correlator. The
  rightmost blob in all diagrams represents a function of momentum
  \(\widetilde \sigma^\dagger\), whose specific form is determined by
  the interpolating fields defining the correlator. The leftmost blob
  represents an analogous function, \(\widetilde \sigma\). Any
  insertion between these with four legs represents a two-to-two
  Bethe-Salpeter kernel \(i B_{2}\). Any insertion with six legs
  represents an analogous three-to-three kernel \(i B_{3}\). All lines
  connecting kernels and \(\widetilde \sigma\)-functions represent
  fully-dressed propagators. The dashed rectangles indicate that all
  loop momenta are summed rather than integrated, due to the
  finite-volume condition. The regions bounded by these rectangles
  also emphasize chains of loops that have common coordinates which
  prevent the diagram from factorizing. This is one of the central
  complications faced in this work. (For example the top line, with
  only three-to-three insertions, does factorize and is therefore a
  straightforward generalization of the two particle case.)}
\label{fig:fullskelexpansion}
\end{center}
\end{figure}

We can now describe the skeleton expansion we use for $C_L$,
which is displayed in Fig.~\ref{fig:fullskelexpansion}.
Since only three-particle intermediate states can go on-shell,
we display them explicitly, and use a notation indicating that
their momenta are summed. Intermediate states with five or more
particles cannot go on shell for our range of $E^*$, and so sums over
the momenta of such intermediate states can be 
replaced by integrals.\footnote{%
Here we are using the language of TOPT although we are calculating using
relativistic propagators in which multiple time-orderings are
contained within a single diagram. If we focus on a particular cut, however, 
then there is only one time ordering in which all particles can go on-shell. 
}
These contributions can be grouped into infinite-volume Bethe-Salpeter
kernels, which are defined below.

Each diagram in the expansion contains ``endcaps''
\(\widetilde \sigma^\dagger\)
and \(\widetilde \sigma\) on the far left and far right,
respectively. These are each functions of the off-shell momenta of
three attached propagators, subject to the constraint that they total
\((E, \vec P)\). Thus they can be written
\(\widetilde \sigma = \widetilde \sigma(q,p)\)
and \(\widetilde \sigma^\dagger = \widetilde\sigma^\dagger(q,p)\). 
For the example of the \( \sigma\) operator given in 
Eq.~(\ref{eq:sigex}),
\begin{equation}
\widetilde \sigma(q,p) = \widetilde f(q,p) + 
\widetilde f(p, P-p-q) + \widetilde f(P-p-q, q) +\widetilde f(p, q) +
\widetilde f( P-p-q, p) + \widetilde f(q,P-p-q)\,,
\end{equation}
where
\begin{equation}
\widetilde f(q,p) = \int_L d^4 x \; d^4 y\; e^{i p x + i q y } f(x,y)
\,.
\end{equation}
Note here that we use the mostly-minus metric, 
\(p x = p^0 x^0 - \vec p \cdot \vec x\). The exact forms of $\widetilde \sigma$ and
$\widetilde \sigma'$ are not important to the final answer. 
We only require that they
are analytic in the complex \(q^0\) and \(p^0\) planes 
and fall off fast enough at infinity to justify the
contour integrals we perform below. 

\begin{figure}
\begin{center}
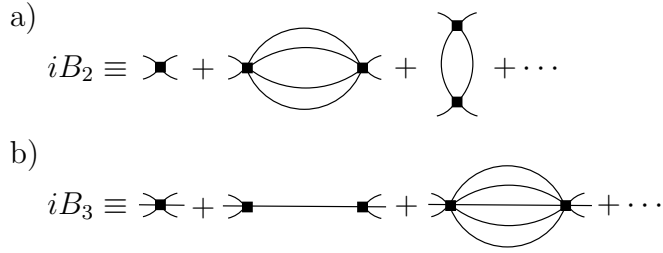
\caption{Examples of Feynman diagrams contributing to (a) 
\(i B_{2}\), the two-to-two Bethe-Salpeter kernel and (b) \(i B_{3}\),
  the analogous three-to-three kernel.}
\label{fig:Kexamps} 
\end{center}
\end{figure}

Between the endcaps, each diagram contains
some number of two-to-two and three-to-three
Bethe-Salpeter kernels. The two-to-two Bethe-Salpeter kernel 
\(i B_{2}\) was introduced in Ref.~\cite{Luscher:1991n1}. It is the sum
of all four-point diagrams
(with external propagators amputated)
that are two-particle irreducible in the
\(s\)-channel, see Fig.~\ref{fig:Kexamps}a. 
Thus this kernel is the sum of all diagrams which
have no on-shell intermediate states when the total CM energy being
fed into the kernel is below \(4m\). Because \(i B_{2}\) contains no
on-shell intermediate states, the summands of all contributing terms
are smooth functions of summed momenta. It follows that finite-volume
corrections are exponentially suppressed and for our purposes
negligible. We thus work from now on with the infinite-volume version
of the kernel.

Similarly, \(i B_{3}\) contains no diagram in which three propagators
carry the total energy and momentum \((E, \vec P)\). Diagrams with one
propagator carrying the total energy and momentum as well as any odd
number greater than three are allowed, see Fig.~\ref{fig:Kexamps}b. 
The technical definition of this quantity is slightly more
complicated because of the possibility of having single-particle
intermediate states. To give the definition, we first introduce three
intermediate quantities 
\(i \widetilde B_{3 \rightarrow 3}\), 
\(i \widetilde B_{1 \rightarrow 3}\), 
\(i \widetilde B_{3 \rightarrow 1}\). 
In each case \(i \widetilde B_{n \rightarrow m}\) is the sum
of all amputated diagrams, with $n$ incoming and $m$ outgoing external lines, which are three-particle irreducible
in the \(s\)-channel. Next we introduce a modified, fully-dressed
propagator \(\widetilde \Delta(q)\). This differs from the standard
propagator, defined in Eq.~(\ref{eq:propdef}) below, only in that its
self-energy graphs are three-particle irreducible (as opposed to the
usual one-particle irreducible). In terms of these ingredients,
our three-to-three kernel is
\begin{equation}
i B_{3} \equiv i \widetilde B_{3 \rightarrow 3} + i \widetilde B_{3
  \rightarrow 1} \widetilde \Delta i \widetilde B_{1 \rightarrow 3}
\,.
\end{equation}
In direct analogy to the two-to-two case, \(i B_{3}\) is the sum of
all diagrams with no on-shell intermediate states when the CM energy
is between \(m\) and \(5m\). Again we drop exponentially suppressed
corrections and work with the infinite-volume version of the kernel.

We stress that the need for two kinds of kernels follows directly from
requiring that both \(i B_2\) and \(i B_3\) contain only connected
diagrams. For example, one might think that only the top line of
Fig.~\ref{fig:fullskelexpansion} is needed, as long as one chooses
an alternative \(i B_3\) which accommodates pairwise scatterings. This
is attractive since the top line closely resembles the two-particle
skeleton expansion of Ref.~\cite{Kim:2005}, in which the correlator is
written as a ladder series of two-particle loops. However, in the
three-particle sector this approach results in \(i B_3\) containing
Dirac delta functions, which is a problem because we rely on smoothness of the kernel in our derivation.
For this reason the three-particle case is
fundamentally different. After much investigation, we found it most
convenient to require that \(i B_3\) only contain connected diagrams
and thus display all pairwise scatterings explicitly.

Finally, in our skeleton expansion all kernels and interpolating
functions are connected by fully-dressed propagators,
\begin{equation}
\label{eq:propdef}
\Delta(q) \equiv \int d^4 x\; e^{i q\cdot x} \langle 0 \vert \mathrm{T}
\phi(x) \phi(0) \vert 0 \rangle\,.
\end{equation}
Here \(\phi(x)\) is a one particle interpolating field defined with
on-shell renormalization such that
\begin{equation}
\lim_{q^0 \rightarrow \omega_q} \Delta (q) [(q^2-m^2)/i] = 1 \,.
\label{eq:propnorm}
\end{equation}
Since we are working with fully dressed
propagators, we do not include self-energy contributions explicitly in
our skeleton expansion. We use infinite-volume fully-dressed
propagators throughout, which is justified because the self-energy
graphs do not contain on-shell intermediate states.


In summary, the skeleton expansion 
of Fig.~\ref{fig:fullskelexpansion} displays explicitly
all the intermediate states that can go on shell and
give rise to power-law corrections. All intermediate states which cannot go on-shell are included in the infinite-volume two-to-two and three-to-three Bethe-Salpeter kernels.

In the remaining subsections, we work through the different
classes of diagrams appearing in this expansion.
First, in Sec.~\ref{sec:noswitch}, we sum
diagrams containing only $iB_2$ kernels on the same pair of propagators 
(second line of Fig.~\ref{fig:fullskelexpansion}). 
Then, in Secs.~\ref{sec:oneswitch} and \ref{sec:twoswitch}, we
sum diagrams with, respectively, one or two changes
in the pair that is being scattered
(third and fourth lines of Fig.~\ref{fig:fullskelexpansion}).
At this stage, we can extend the pattern 
and sum all diagrams built from $iB_2$ kernels
with any number of changes in the scattered pair. 
This is done in Sec.~\ref{sec:remainingdiag}.
Incorporating three-to-three insertions at this point is
relatively easy, and is done in Sec.~\ref{sec:B3},
leading to the final result for \(C_{L}\) given in Eq.~(\ref{eq:corrresult}).

As we proceed we identify the
diagrams contributing to \(\K\) and \(\Kdfth\), as well as
\(A,A'\) and \(C_\infty\). The precise definitions of these
infinite-volume quantities will thus emerge step by step.


%% file: fig25.pdf_t
\begin{picture}(0,0)%
\includegraphics{fig25.pdf}%
\end{picture}%
\setlength{\unitlength}{3947sp}%
\begingroup\makeatletter\ifx\SetFigFont\undefined%
\gdef\SetFigFont#1#2#3#4#5{%
  \reset@font\fontsize{#1}{#2pt}%
  \fontfamily{#3}\fontseries{#4}\fontshape{#5}%
  \selectfont}%
\fi\endgroup%
\begin{picture}(7938,3071)(2347,-3492)
\put(7772,-1912){\makebox(0,0)[lb]{\smash{{\SetFigFont{12}{14.4}{\rmdefault}{\mddefault}{\updefault}{\color[rgb]{0,0,0}$+$}%
}}}}
\put(3202,-1308){\makebox(0,0)[lb]{\smash{{\SetFigFont{12}{14.4}{\rmdefault}{\mddefault}{\updefault}{\color[rgb]{0,0,0}$+$}%
}}}}
\put(5251,-1912){\makebox(0,0)[lb]{\smash{{\SetFigFont{12}{14.4}{\rmdefault}{\mddefault}{\updefault}{\color[rgb]{0,0,0}$+$}%
}}}}
\put(2362,-699){\makebox(0,0)[lb]{\smash{{\SetFigFont{12}{14.4}{\rmdefault}{\mddefault}{\updefault}{\color[rgb]{0,0,0}$C_L(E, \vec P) =$}%
}}}}
\put(6042,-682){\makebox(0,0)[lb]{\smash{{\SetFigFont{12}{14.4}{\rmdefault}{\mddefault}{\updefault}{\color[rgb]{0,0,0}$+$}%
}}}}
\put(4326,-682){\makebox(0,0)[lb]{\smash{{\SetFigFont{12}{14.4}{\rmdefault}{\mddefault}{\updefault}{\color[rgb]{0,0,0}$+$}%
}}}}
\put(4778,-1308){\makebox(0,0)[lb]{\smash{{\SetFigFont{12}{14.4}{\rmdefault}{\mddefault}{\updefault}{\color[rgb]{0,0,0}$+$}%
}}}}
\put(6832,-1308){\makebox(0,0)[lb]{\smash{{\SetFigFont{12}{14.4}{\rmdefault}{\mddefault}{\updefault}{\color[rgb]{0,0,0}$+$}%
}}}}
\put(3197,-1912){\makebox(0,0)[lb]{\smash{{\SetFigFont{12}{14.4}{\rmdefault}{\mddefault}{\updefault}{\color[rgb]{0,0,0}$+$}%
}}}}
\put(3203,-2452){\makebox(0,0)[lb]{\smash{{\SetFigFont{12}{14.4}{\rmdefault}{\mddefault}{\updefault}{\color[rgb]{0,0,0}$+$}%
}}}}
\put(5741,-2447){\makebox(0,0)[lb]{\smash{{\SetFigFont{12}{14.4}{\rmdefault}{\mddefault}{\updefault}{\color[rgb]{0,0,0}$+$}%
}}}}
\put(3203,-2888){\makebox(0,0)[lb]{\smash{{\SetFigFont{12}{14.4}{\rmdefault}{\mddefault}{\updefault}{\color[rgb]{0,0,0}$+ \cdots$}%
}}}}
\put(3202,-3323){\makebox(0,0)[lb]{\smash{{\SetFigFont{12}{14.4}{\rmdefault}{\mddefault}{\updefault}{\color[rgb]{0,0,0}$+$}%
}}}}
\put(5437,-3323){\makebox(0,0)[lb]{\smash{{\SetFigFont{12}{14.4}{\rmdefault}{\mddefault}{\updefault}{\color[rgb]{0,0,0}$+$}%
}}}}
\put(7675,-3323){\makebox(0,0)[lb]{\smash{{\SetFigFont{12}{14.4}{\rmdefault}{\mddefault}{\updefault}{\color[rgb]{0,0,0}$+ \cdots$}%
}}}}
\put(8372,-670){\makebox(0,0)[lb]{\smash{{\SetFigFont{12}{14.4}{\rmdefault}{\mddefault}{\updefault}{\color[rgb]{0,0,0}$+ \cdots$}%
}}}}
\put(8759,-2444){\makebox(0,0)[lb]{\smash{{\SetFigFont{12}{14.4}{\rmdefault}{\mddefault}{\updefault}{\color[rgb]{0,0,0}$+ \cdots$}%
}}}}
\put(9334,-1302){\makebox(0,0)[lb]{\smash{{\SetFigFont{12}{14.4}{\rmdefault}{\mddefault}{\updefault}{\color[rgb]{0,0,0}$+ \cdots$}%
}}}}
\end{picture}%

%% file: fig105.pdf_t
\begin{picture}(0,0)%
\includegraphics{fig105.pdf}%
\end{picture}%
\setlength{\unitlength}{3947sp}%
\begingroup\makeatletter\ifx\SetFigFont\undefined%
\gdef\SetFigFont#1#2#3#4#5{%
  \reset@font\fontsize{#1}{#2pt}%
  \fontfamily{#3}\fontseries{#4}\fontshape{#5}%
  \selectfont}%
\fi\endgroup%
\begin{picture}(3717,1571)(3049,-2496)
\put(3301,-2281){\makebox(0,0)[lb]{\smash{{\SetFigFont{12}{14.4}{\rmdefault}{\mddefault}{\updefault}{\color[rgb]{0,0,0}$i B_{3} \equiv$}%
}}}}
\put(4203,-2290){\makebox(0,0)[lb]{\smash{{\SetFigFont{12}{14.4}{\rmdefault}{\mddefault}{\updefault}{\color[rgb]{0,0,0}$+$}%
}}}}
\put(3064,-1961){\makebox(0,0)[lb]{\smash{{\SetFigFont{12}{14.4}{\rmdefault}{\mddefault}{\updefault}{\color[rgb]{0,0,0}b)}%
}}}}
\put(3064,-1096){\makebox(0,0)[lb]{\smash{{\SetFigFont{12}{14.4}{\rmdefault}{\mddefault}{\updefault}{\color[rgb]{0,0,0}a)}%
}}}}
\put(5479,-2281){\makebox(0,0)[lb]{\smash{{\SetFigFont{12}{14.4}{\rmdefault}{\mddefault}{\updefault}{\color[rgb]{0,0,0}$+$}%
}}}}
\put(6751,-2276){\makebox(0,0)[lb]{\smash{{\SetFigFont{12}{14.4}{\rmdefault}{\mddefault}{\updefault}{\color[rgb]{0,0,0}$+ \cdots$}%
}}}}
\put(4203,-1416){\makebox(0,0)[lb]{\smash{{\SetFigFont{12}{14.4}{\rmdefault}{\mddefault}{\updefault}{\color[rgb]{0,0,0}$+$}%
}}}}
\put(6097,-1411){\makebox(0,0)[lb]{\smash{{\SetFigFont{12}{14.4}{\rmdefault}{\mddefault}{\updefault}{\color[rgb]{0,0,0}$+ \cdots$}%
}}}}
\put(5475,-1411){\makebox(0,0)[lb]{\smash{{\SetFigFont{12}{14.4}{\rmdefault}{\mddefault}{\updefault}{\color[rgb]{0,0,0}$+$}%
}}}}
\put(3301,-1416){\makebox(0,0)[lb]{\smash{{\SetFigFont{12}{14.4}{\rmdefault}{\mddefault}{\updefault}{\color[rgb]{0,0,0}$i B_{2} \equiv$}%
}}}}
\end{picture}%

%% file: noswitches.tex
\subsection{Two-to-two insertions: no switches}

\indent

\label{sec:noswitch}

\begin{figure}
\begin{center}
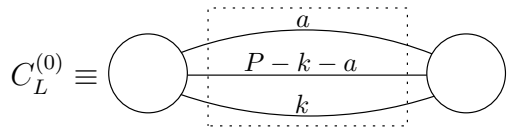
\caption{Finite-volume correlator diagram with no kernel
  insertions.}
\label{fig:noinsertion}
\end{center}
\end{figure}

\begin{figure}
\begin{center}

%
%
%
%
%
%
%
%

\includegraphics[scale=0.25]{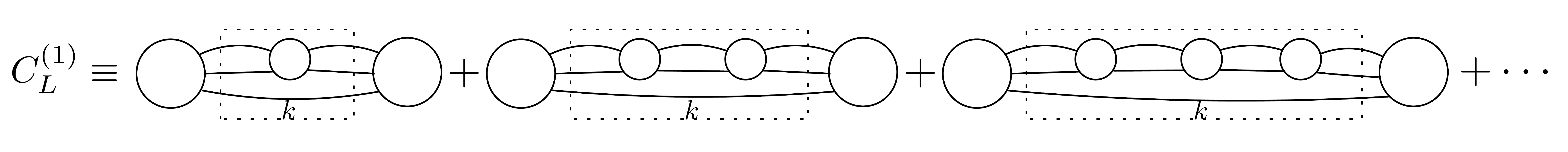}

\caption{Finite-volume correlator diagrams containing only
  two-to-two insertions with no change in the scattered pair.}
\label{fig:noswitches}
\end{center}
\end{figure}

In this section we sum the diagrams of Figs.~\ref{fig:noinsertion}
and \ref{fig:noswitches}. Each diagram contains only \(B_{2}\)
insertions, all of which scatter the same pair of
propagators. We separate the diagram with no \(B_2\) insertions,
labeled \(\CL 0\) [Fig.~\ref{fig:noinsertion}], 
from the sum of diagrams with one or more insertions, denoted \(\CL 1\)
[Fig.~\ref{fig:noswitches}]. 
We refer to these diagrams as having ``no
switches'', meaning that the pair that is scattered does not
change. This designation anticipates subsequent sections in which we sum
diagrams with one or more switches in the scattered pair.

An important check on the calculation of this subsection is obtained
by noting
that the no-switch diagrams are the complete set appearing in
a theory of two different particle types, with one of the types
non-interacting. This is the case provided that the correlator is constructed with fields that
interpolate one free particle and two interacting particles. 
Thus the result for \(\CL 0 + \CL 1\) must be that for the full finite-volume correlator in 
the two-plus-spectator theory. 
This check is discussed below.

We begin our detailed calculation by determining the finite-volume
residue of the no-insertion diagram of Fig.~\ref{fig:noinsertion}. 
This diagram represents the expression\footnote{%
  In the remainder of this article 
  we drop tilde's on the Fourier-transformed interpolating operators, 
  \(\widetilde \sigma(k,a)\) and \(\widetilde \sigma^\dagger(k,a)\),
  since we no longer use the position-space forms.}
\begin{equation}
\label{eq:VLdef}
\CL 0 \equiv \frac{1}{6} \frac{1}{L^6} \sum_{\vec k, \vec a} 
\int_{a^0}\int_{k^0} \sigma(k,a) 
\Delta(k)
\Delta(a) \Delta(P-k-a) \sigma^\dagger(k,a) \,,
\end{equation}
where $\int_{k^0}\equiv \int dk^0/(2\pi)$, etc.,
and the $1/6$ is the symmetry factor.
We stress that the $\Delta$'s are fully dressed propagators,
with the normalization given in Eq.~(\ref{eq:propnorm}).

We first evaluate the \(a^0\) and \(k^0\)
integrals using contour integration, wrapping both contours in the
lower half of the respective complex planes. Each contour encircles a
one-particle pole 
[\(a^0=\omega_a - i \epsilon\) and \(k^0=\omega_k -  i \epsilon \)] 
as well as three-particle (and higher) poles from
excited-state contributions to the propagators. The result of
integration may thus be written
\begin{equation}
\label{eq:VLoneprop}
\CL 0 = \frac{1}{6} \frac{1}{L^6} \sum_{\vec k, \vec a} \bigg [ \frac{
    \sigma([\omega_k, \vec k], [\omega_a, \vec a]) \Delta(P-k-a)
    \sigma^\dagger([\omega_k, \vec k], [\omega_a, \vec a])}{2 \omega_k 2
    \omega_a} + \mathcal R(\vec k, \vec a) \bigg ]\,,
\end{equation}
where \(\mathcal R(\vec k, \vec a)\) is the contribution from
excited-state poles. Here \(k\) and \(a\) appearing in
\(\Delta(P-k-a)\) are now understood as on-shell four-vectors,
a fact that we have made explicit in the arguments of
$\sigma$ and $\sigma^\dagger$. We next
note that \(\Delta(P\!-\!k\!-\!a)\) can be split into its one-particle pole
plus a remainder: 
\begin{equation}
\label{eq:propexp}
\Delta(P\!-k\!-a) = \frac{i}{2 \omega_{ka}(E - \omega_k - \omega_a -
  \omega_{ka} )} + r(\vec k, \vec a)\,.
\end{equation}
Substituting Eq.~(\ref{eq:propexp}) into Eq.~(\ref{eq:VLoneprop})
gives
\begin{equation}
\label{eq:VLthreepartpoleA}
\CL 0 = \frac{1}{6} \frac{1}{L^6} \sum_{\vec k, \vec a} \bigg [
  \frac{i \sigma([\omega_k, \vec k], [\omega_a, \vec a])
    \sigma^\dagger([\omega_k, \vec k], [\omega_a, \vec a])}{2 \omega_k 2
    \omega_a 2 \omega_{ka}(E - \omega_k - \omega_a - \omega_{ka} )} +
  \mathcal R'(\vec k, \vec a) \bigg ]\,,
\end{equation}
where \(\mathcal R'\) is the sum of \(\mathcal R\) and the term
containing \(r\). This grouping is convenient because 
\(\mathcal R'(\vec k, \vec a)\) is a smooth function of 
\(\vec k\) and \(\vec a\) for our range of $E$, since we have explicitly
pulled out the three-particle singularity.
Indeed, we are free to further adjust the
separation between first and second terms, as long as the latter
remains smooth. For the following development we need to include the damping function $H(\vec k)$ in the singular term. 
We recall that $H(\vec k)$,
defined in Eqs.~(\ref{eq:HvalA}) and (\ref{eq:Hdef}), 
is a smooth function which equals unity when the other two particles 
(those with momenta $a$ and $P\!-\!k\!-\!a$) 
are kinematically allowed to be on shell
(for the given values of $E$ and $\vec k$).
In particular, if we multiply the singular term by $1=H(\vec k)+[1-H(\vec k)]$,
then the $1-H(\vec k)$ term cancels the singularity, leading to a
smooth function that can be added to ${\cal R'}$ 
to obtain a new residue ${\cal R''}$:
\begin{equation}
\label{eq:VLthreepartpole}
\CL 0 = \frac{1}{6} \frac{1}{L^6} \sum_{\vec k, \vec a} \bigg [
  \frac{i \sigma([\omega_k, \vec k], [\omega_a, \vec a])
    \sigma^\dagger([\omega_k, \vec k], [\omega_a, \vec a]) 
    H(\vec k)}
   {2\omega_k 2 \omega_a 2 \omega_{ka}(E - \omega_k - \omega_a -
    \omega_{ka} )} + \mathcal R''(\vec k, \vec a) \bigg ]\,.
\end{equation}

At this stage we want to rewrite \(\CL 0\) as an infinite-volume
(\(L\)-independent) quantity plus a remainder. Infinite-volume
quantities differ only in that loop-momenta are integrated rather than
summed. We can thus pull out the infinite-volume object by replacing
each sum with an integral plus a sum-integral difference. We stress
that integrals, unlike sums, require a pole-prescription. We
are free to use any prescription we like, and it turns out to be most
convenient to make a non-standard choice which we call the
``modified principle-value'' or $\widetilde{\rm PV}$ prescription.
This is defined in the present context as follows\footnote{%
In the definition of $\PV$ we are using $\sigma$ and $\sigma^\dagger$ 
which are continuous functions of $\vec a$ and $\vec k$. Since these 
were originally defined only for discrete finite-volume momenta, 
this requires a continuation of the original functions.
 We require only that the continuation is smooth and slowly varying. 
More precisely we demand 
\begin{equation}
\bigg [\frac{1}{L^3} \sum_{\vec a} - \int_{\vec a} \bigg] 
\sigma([\omega_k, \vec k],[\omega_a, \vec a]) = \mathcal O(e^{-mL}) \,.
\end{equation}} 
\begin{multline}
\label{eq:PVtildedef}
\frac12\widetilde{\mathrm{PV}} \int_{\vec a} 
\frac{i\sigma([\omega_k, \vec k], [\omega_a, \vec a]) 
      \sigma^\dagger([\omega_k,\vec k], [\omega_a, \vec a]) H(\vec k)}
     {2 \omega_a 2 \omega_{ka} (E -\omega_k - \omega_a - \omega_{ka} )} 
\equiv \frac12 \int_{\vec  a} 
\frac{i\sigma([\omega_k, \vec k], [\omega_a, \vec a])
      \sigma^\dagger([\omega_k, \vec k], [\omega_a, \vec a]) H(\vec k)}
     {2 \omega_a 2 \omega_{ka} (E - \omega_k - \omega_a - \omega_{ka} + i
      \epsilon)} 
\\ 
- \sigma^*_{\ell',m'}(\vec k)
    i\rho_{\ell',m';\ell,m}(\vec k) 
    \sigma^{\dagger*}_{\ell,m}(\vec k) \,.
\end{multline}
where $\rho$
was introduced in Eq.~(\ref{eq:rhodef}) above.

To complete the definition we need to explain the meanings
of the on-shell quantities \(\sigma^*_{\ell',m'}(\vec k)\) and 
\(\sigma^{\dagger*}_{\ell,m}(\vec k)\). 
Similar quantities will appear many times below so we give here
a detailed description.
First recall that \((\omega_{a}^* , \vec a^*)\) is the four-vector 
obtained by boosting \((\omega_a, \vec a)\) with velocity
$\vec \beta_k = - ({\vec P - \vec k})/({E - \omega_k})$.
This boost is only physical if \(E_{2,k}^*>0\), a constraint which
is guaranteed to be satisfied by the presence of \(H(\vec k)\) in Eq.~(\ref{eq:PVtildedef}).
We now change variables from $\vec a$ to $\vec a^*$ and define 
\begin{equation}
\label{eq:sigstardef}
\sigma^*(\vec k,\vec a^* ) \equiv 
\sigma([\omega_k, \vec k], [\omega_{a}, \vec a]) \,, 
\end{equation}
and similarly for $\sigma^\dagger$.
The left-hand side exemplifies our general notation that,
if the momentum argument is a three-vector, e.g. $\vec k$,
then the momentum is on-shell, e.g. $k^0=\omega_k$.
If the argument is a four-momentum, e.g. $k$, then it is, in general, off shell.
Here we include a superscript $*$ on $\sigma$ to indicate that it is strictly a 
different function from that appearing in say Eq.~(\ref{eq:VLthreepartpole}),
since it depends on different coordinates (in particular on momenta defined 
in different frames). 
Next we decompose $\sigma^*$ and $\sigma^{\dagger*}$
into spherical harmonics in the CM frame
\begin{align}
\sigma^*( \vec k, \vec a^* ) &\equiv
\sqrt{4 \pi} Y_{\ell, m}(\hat a^*) \sigma_{\ell,m}^*(\vec k, a^*)
\\
\sigma^{\dagger*}( \vec k, \vec a^* ) &\equiv
\sqrt{4 \pi} Y^*_{\ell, m}(\hat a^*) \sigma_{\ell,m}^{\dagger*}(\vec k, a^*)
\,,
\label{eq:sigdagYdecom}
\end{align}
where there is an implicit sum over $\ell$ and $m$.
Our convention, used throughout, is that the quantities
to the left of the three particle ``cut'' are decomposed using
$Y_{\ell, m}$'s while those to the right use the complex conjugate
harmonics.
Finally, with the ``starred'' quantities in hand we can define
on-shell restrictions.
As explained in the introduction, $P\!-\!k\!-\!a$ is only on shell if
$a^*=q_k^*$, so we define
\begin{equation}
 \sigma_{\ell,m}^*(\vec k) \equiv 
\sigma_{\ell,m}^*(\vec k, q^*_k)
 \,, \qquad 
\sigma^{\dagger*}_{\ell,m}(\vec k) \equiv 
\sigma^{\dagger*}_{\ell,m}(\vec k, q^*_k) \,.
\label{eq:onshellproj}
\end{equation}
These are the quantities appearing in the
``$\rho$'' term in Eq.~(\ref{eq:PVtildedef}).
If $E_{2,k}^*< 2m$, then the $\vec a,\vec b_{ka}$ pair is
below threshold, and $\sigma^*_{\ell,m}$ and
$\sigma^{\dagger*}_{\ell,m}$ must be obtained by analytic
continuation from above threshold.

The reason for using this rather elaborate pole prescription is that
we want the integral over $\vec a$ to produce a smooth function of $\vec k$.
This allows the sum over $\vec k$ to be replaced by an integral. If we were to instead use the $i \epsilon$ prescription, then the resulting function of $\vec k$ would have a unitary cusp at $E^*_{2,k}=2m$. This observation leads us to consider a principal-value pole prescription instead. We note that $\rho$ is defined so that, for $E^*_{2,k}>2m$, Eq.~(\ref{eq:PVtildedef}) simply gives the standard principal-value prescription. It turns out that this choice gives a smooth function of $\vec k$, provided that one uses analytic continuation to extend from $E^*_{2,k}>2m$ to $E^*_{2,k}<2m$. This is accomplished by our subthreshold definition of $\rho$, which is then
smoothly turned off by the function $H(\vec k)$.
A derivation of the smoothness property is given in Appendix \ref{app:PVtilde}.
We stress that the $\widetilde{\rm PV}$ prescription is always defined
relative to a spectator momentum, here $\vec k$.

A slightly more general form of the $\PV$ prescription
is instructive and will be useful below. For any two-particle
four-momentum $P_2$ for which the only kinematically
allowed cut involves two particles, we can write
\begin{multline}
\PV \int_a A(P_2,a) B(P_2,a) \Delta(a)\Delta(P_2-a)
= \int_a A(P_2,a) B(P_2,a) \Delta(a)\Delta(P_2-a)
\\
- 2 i J(P_2^2/[4m^2])\tilde\rho(P_2)
\left[\int_{\hat a^*} A^*(P_2,\vec a^*) B^*(P_2,\vec a^*)\right] 
\Bigg|_{a^*=\sqrt{P_2^2/4-m^2}}
\,.
\label{eq:PVtildedef2}
\end{multline}
Here $A$ and $B$ are smooth, non-singular functions of their arguments.
The quantities $A^*$ and $B^*$ are defined in a similar
way to $\sigma^*$ above,
e.g. $A^*(P_2,\vec a^*) = A(P_2,[\omega_a,\vec a])$,
where the boost to the two-particle CM has velocity $-\vec P_2/P_2^0$.
The function $J$, defined in Eq.~(\ref{eq:Jdef}), ensures that this boost is
well defined.\footnote{%
Here $J$ is playing the role of $H(\vec k)=J(P_2^2/[4m^2])$ in 
Eq.~(\ref{eq:PVtildedef}).}
Finally, the angular integral is normalized such that $\int_{\hat a^*} 1=1$.
The form (\ref{eq:PVtildedef2}) makes clear that the prescription
can be defined for four-momentum integrals (and not just
three-momentum integrals) and that its dependence on external
momenta enters entirely through $P_2$.
We have also used the angular independence of $\rho$ to rewrite the
subtraction term as an angular average in the CM frame.
The two functions $A$ and $B$ could be combined into one, but are
left separate since in our applications we always have separate
functions to the left and right of the cut.

Returning to the main argument, we now substitute
\begin{equation}
\frac{1}{L^3} \sum_{\vec a} = \widetilde{ \mathrm{PV}} \int_{\vec a} +
\bigg [\frac{1}{L^3} \sum_{\vec a} - \widetilde{ \mathrm{PV}}
  \int_{\vec a} \bigg] \,,
\end{equation}
into Eq.~(\ref{eq:VLthreepartpole}) to reach
\begin{multline}
\label{eq:VLaftersub}
\CL 0 = \frac{1}{6} \frac{1}{L^3} \sum_{\vec k} \widetilde{
  \mathrm{PV}} \int_{\vec a} 
\bigg[ 
\frac{i \sigma([\omega_k, \vec k],  [\omega_a, \vec a]) 
\sigma^\dagger([\omega_k, \vec k], [\omega_a, \vec a]) H(\vec k)}
{2 \omega_k 2 \omega_a 2 \omega_{ka}
(E - \omega_k -     \omega_a - \omega_{ka} )} 
+ \mathcal R''(\vec k, \vec a) 
\bigg]
\\ 
+ \frac{1}{6} \frac{1}{L^3} \sum_{\vec k} 
\bigg[\frac{1}{L^3} \sum_{\vec a} - \widetilde{ \mathrm{PV}} \int_{\vec a}\bigg]
\frac{i \sigma([\omega_k, \vec k], [\omega_a, \vec a])
  \sigma^\dagger([\omega_k, \vec k], [\omega_a, \vec a]) H(\vec k)}
{2\omega_k 2\omega_a 2\omega_{ka}(E - \omega_k - \omega_a -  \omega_{ka} )} 
\,.
\end{multline}
Note that the sum-integral-difference operator annihilates \(\mathcal R''(\vec k, \vec a)\) up to exponentially suppressed terms.
As already noted, we can replace the sum over $\vec k$ with an integral
in the first term, resulting in the infinite-volume quantity
\begin{equation}
\label{eq:Vinfdef}
\CI 0 \equiv \frac{1}{6} \int_{\vec k} \widetilde{ \mathrm{PV}}
\int_{\vec a} \bigg [ 
\frac{i \sigma([\omega_k, \vec k], [\omega_a, \vec a]) 
\sigma^\dagger([\omega_k, \vec k], [\omega_a, \vec a]) H(\vec k)}
{2\omega_k 2\omega_a 2\omega_{ka}(E - \omega_k - \omega_a -\omega_{ka} )} 
+ \mathcal R''(\vec k, \vec a) \bigg ] \,.
\end{equation}
Note that no pole prescription is required for the ${\vec k}$ integral.

The second term in Eq.~(\ref{eq:VLaftersub}) is then the finite-volume
residue. First we note that we can multiply the summand/integrand by
$H(\vec a) H(\vec b_{ka})$, since the remainder cancels the pole and thus
has vanishing sum-integral difference.
Next we use the identity for sum-integral differences presented
in Eq.~(\ref{eq:sumintegral}) of Appendix~\ref{app:sumintegral}. 
This is based on an extension of the work of Ref.~\cite{Kim:2005} 
to include the effects of 
subthreshold momenta and the $\widetilde{\rm PV}$ prescription.
The essence of the identity is that the sum-integral difference picks out the
on-shell residue of the singularity multiplied by a kinematic function.
In more detail the identity makes use of the analytic properties of 
$\sigma_{\ell,m}^*(\vec k, a^*)$ and $\sigma_{\ell,m}^{\dagger*}(\vec k, a^{*})$, 
the functions defined in Eqs.~(\ref{eq:sigstardef}-\ref{eq:sigdagYdecom}) above. 
The result is that
\begin{align}
\label{eq:VLfinalnoind}
\CL 0 & = \CI 0 + \frac{1}{L^3} \sum_{\vec k} \frac{1}{6 \omega_k }
\sigma^*_{ \ell', m'}(\vec k) iF_{ \ell', m'; \ell, m}(\vec k)
\sigma^{\dagger *}_{\ell,m}(\vec k) \,,\\ &= \CI 0 + \sigma^*_{k',
  \ell', m'} \frac{1}{6 \omega_k L^3} iF_{k', \ell', m'; k, \ell, m}
\sigma^{\dagger *}_{k,\ell,m} 
\,, \label{eq:VLfinalwithind}
\end{align}
where the finite-volume kinematical function $F$ is defined
in Eqs.~(\ref{eq:Fdef1})-(\ref{eq:Fdef3}),
%
%
and
\begin{equation}
\sigma^{*}_{k,\ell,m} \equiv \sigma_{\ell,m}^*(\vec k)\,, \ \ 
 \sigma^{\dagger *}_{k,\ell,m} \equiv \sigma_{\ell,m}^{\dagger*}(\vec k) 
\ \ \mathrm{for} \ \  \vec k \in (2 \pi/L) \mathbb Z^3 
\label{eq:sigmadecom2}
\end{equation}
are the restrictions of the on-shell functions to finite-volume momenta. 
All indices in Eq.~(\ref{eq:VLfinalwithind}) are understood to be
summed, including \(k\) and \(k'\) which are summed over the
allowed values of finite-volume momenta. 
This index structure appears repeatedly in our
derivation, and from now on we leave indices implicit.
Indeed, using the matrix notation introduced in Section \ref{sec:res},
we can write the final result compactly as
\begin{equation}
\label{eq:VLfinal}
\CL 0 = \CI 0 + \sigma^* \frac{iF}{6 \omega L^3} \sigma^{\dagger *}
\,.  
\end{equation}
This is the main result of this subsection.

\begin{figure}
\begin{center}
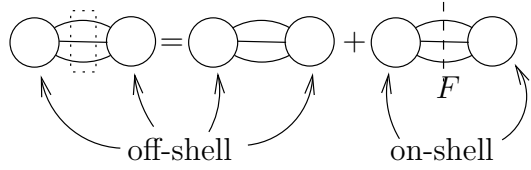
\caption{Diagrammatic representation of Eq.~(\ref{eq:VLfinal}).}
\label{fig:mainident}
\end{center}
\end{figure}

Our treatment of the three-particle cut will be reused repeatedly in
the following, except that $\sigma$ and $\sigma^\dagger$
will be replaced by other smooth functions
of the momenta. Since no properties of
$\sigma$ and $\sigma^\dagger$ other than smoothness were used in the
derivation of Eq.~(\ref{eq:VLfinal}), the result generalizes immediately.
It is useful to have a diagrammatic version, and this
is given in Fig.~\ref{fig:mainident}.
The key feature of the result
is that the finite-volume residue depends only on on-shell restrictions
of the quantities appearing on either side of the cut
(analytically continued below threshold as needed).

\bigskip

Before considering diagrams containing two-to-two insertions,
we take stock of the impact of using the
non-standard $\PV$ pole prescription.
First we relate $\CI 0$ [defined in Eq.~(\ref{eq:Vinfdef})]
to the conventional infinite-volume form which uses
the $i\epsilon$ prescription. The latter is
\begin{align}
\CIeps 0 & \equiv \frac{1}{6} \int_{\vec k, \vec a} \bigg [ \frac{i
    \sigma([\omega_k, \vec k], [\omega_a, \vec a])
    \sigma^\dagger([\omega_k, \vec k], [\omega_a, \vec a]) H(\vec k)}{2
    \omega_k 2 \omega_a 2 \omega_{ka}
  (E - \omega_k - \omega_a - \omega_{ka} + i \epsilon)} 
   + \mathcal R''(\vec k, \vec a) \bigg ]
\,, \\ 
& = \frac{1}{6} \int_{k,a} \sigma(k,a) \Delta(k) \Delta(a)
\Delta(P\!-\!k\!-\!a) \sigma^\dagger(k,a) \,,
\end{align}
where \(\int_k \equiv \int d^4k/(2 \pi)^4\), etc., indicate integrals
over four-momenta. 
To obtain the second line, which is the standard expression for
the Feynman diagram, we have reversed the steps leading from 
Eq.~(\ref{eq:VLdef}) to (\ref{eq:VLthreepartpole}).
It then follows from the definition of the $\widetilde{\rm PV}$ prescription,
Eq.~(\ref{eq:PVtildedef}), that
\begin{equation}
\label{eq:VinftoVeps}
\CI 0 = \CIeps 0 - \int_{\vec k} \sigma^*(\vec k) 
\frac{i \rho(\vec k)}{6 \omega_k} \sigma^{\dagger*}(\vec k) \,.
\end{equation}
This relation is similar in form to Eq.~(\ref{eq:VLfinal}),
with the ``$F$-cut'' being replaced by a ``$\rho$-cut''.
The key point for present purposes is that the $\rho$-cut term
in Eq.~(\ref{eq:VinftoVeps}) does not introduce poles as a
function of $E$. This follows from noting that $\rho$ is a finite function of $(E, \vec P)$ and $\vec k$, which has a finite range of support in the latter.

We can also determine the form of the finite-volume correction if
we use the $i\epsilon$ prescription throughout, 
including in $F$ [see Eq.~(\ref{eq:Fdef3}) above].
This connects our result to earlier work on two-particle quantization
conditions, e.g. Ref.~\cite{Kim:2005}, where $F^{i\epsilon}$ was used.
Defining
\begin{align}
F^{i \epsilon}_{k',\ell',m';k,\ell,m} & 
\equiv \delta_{k',k} F^{i \epsilon}_{\ell',m';\ell,m}(\vec k) \,,
\end{align}
it follows from Eq.~(\ref{eq:Fdef1}) that
\begin{equation}
\label{eq:Fdefshort}
F_{k',\ell',m';k,\ell,m} = 
F^{i \epsilon}_{k',\ell',m';k,\ell,m} + 
\delta_{k',k}\; \rho_{\ell', m'; \ell, m}(\vec k) \,.
\end{equation}
Combining the results above we then find
\begin{equation}
\label{eq:VLtoVieps}
\CL 0 = \CIeps 0 +  
\sigma \frac{i F^{i \epsilon}}{6 \omega L^3} \sigma^\dagger + 
\bigg[ \frac{1}{L^3} \sum_{\vec k} - \int_{\vec k} \bigg ]
\sigma^*(\vec k) \frac{i \rho(\vec k)}{6 \omega_k} \sigma^{\dagger*}(\vec k) \,.
\end{equation}
Thus we see that, were we to use quantities defined using the $i\epsilon$
prescription, we would need to account for the additional finite-volume
correction coming from the last term, which arises due to the cusp at threshold.\footnote{%
This term is absent in the two-particle analysis, 
where there is only a single value of $\vec k$ (the total momentum flowing
through the two-particle system).}
This extra term greatly complicates the all-orders summation of diagrams. 
We have found
it is more convenient to approach the analysis of finite-volume
diagrams in two steps: first relate finite-volume quantities to
\(\PV\)-quantities (for example relating \(\CL 0\) to \(\CI 0\)), and
then relate \(\PV\)-quantities to those defined with the standard 
\(i \epsilon\) prescription (\(\CI 0\) to \(\CIeps 0\)). 
We concentrate on the first step in this article.

\bigskip

\begin{figure}
\begin{center}


\hspace {-420pt} {\large (a)}

\vspace{10pt}

%
%
%
%
%
%

\includegraphics[scale=0.5]{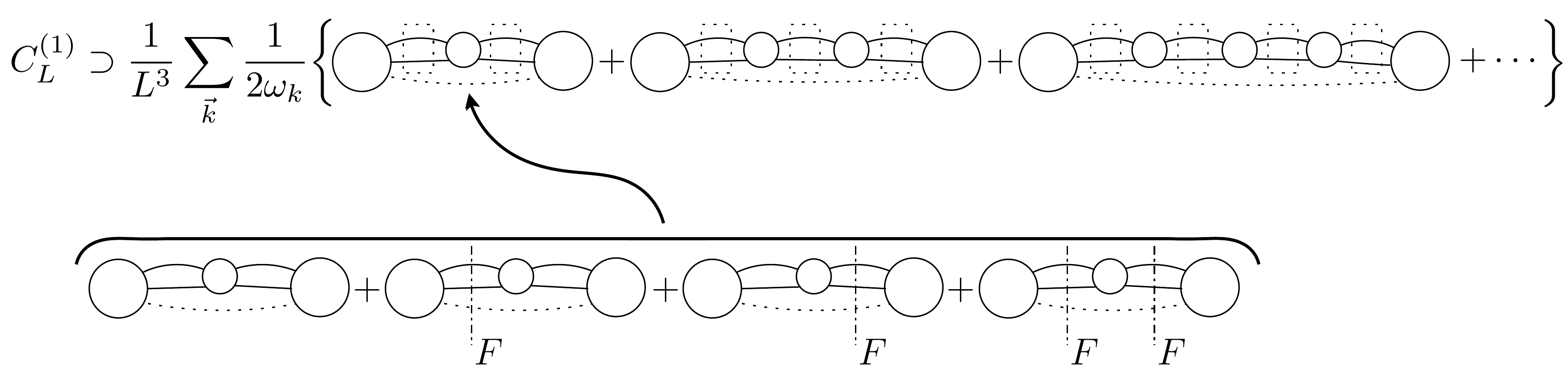}

\vspace{10pt}

\hspace {-420pt} {\large (b)}

\vspace{10pt}

\includegraphics[scale=0.49]{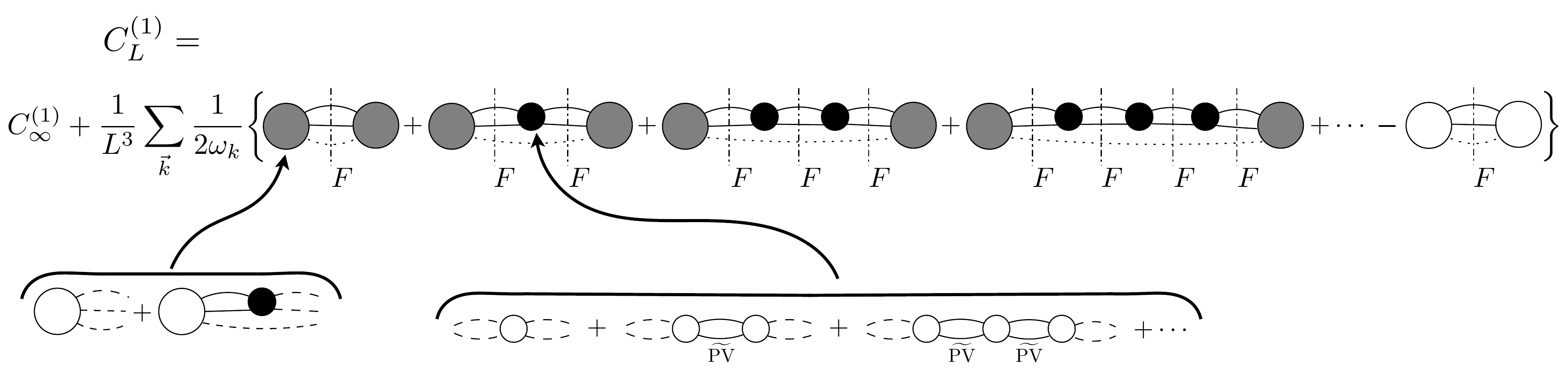}

\vspace{15pt}

%
%
%
%
%
%
%
%

\hspace {-420pt} {\large (c)}


\hspace{-120pt} \resizebox{.170\hsize}{!}{\(A'^{(1,u)}(\vec k, \hat
  a^*) \equiv \)}

\vspace{-2pt}

\hspace{190pt} \resizebox{.012\hsize}{!}{\(\vec k\)} \hspace{5pt}

\vspace{-28pt}

\hspace{194pt} \resizebox{.045\hsize}{!}{\(\}\ \hat a^* \)}

\vspace{-15pt}

\hspace{75pt} \includegraphics[scale=0.5]{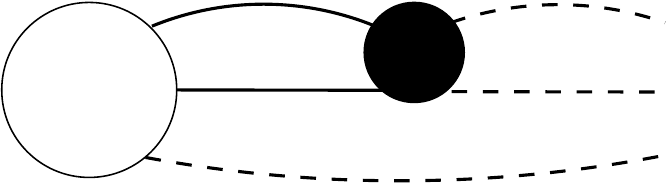}

\vspace{15pt}

\caption{(a) Diagrams contributing power-law 
finite-volume contributions to $\CL 1$. The dashed line for the
bottom propagator indicates that the $k^0$ integration has been done
and only the one-particle pole kept, giving rise to the factor of $1/2\omega_k$.
The inset shows the effect of substituting the identity of
Fig.~\ref{fig:mainident}.
(b) Result for $\CL 1$ after grouping terms according to the number of 
\(F\) insertions. Diagrams with no insertions combine with the 
terms neglected in (a) to give $\CI 1$.
In diagrams with at least one insertion of $F$ the
factors to the left and right are \(\sigma^* + A'^{(1,u)}\)
and \(\sigma^{\dagger*}+A^{(1,u)}\), respectively. 
The factors between \(F\)-insertions (denoted by black circles) 
are two-to-two K-matrices. 
The final term in the curly braces must be subtracted since it is 
included in the first term but is not part of the definition of $C_L^{(1)}$.
(c) Definition of \(A'^{(1,u)}\). The superscript \(u\)
indicates that the unscattered particle is also the particle whose
momentum is singled out by the coordinate system.
Dashed lines for external momenta indicate both that they are on shell
and that they are amputated.}
\label{fig:twopartsub}
\end{center}
\end{figure}

We now turn to diagrams of Fig.~\ref{fig:noswitches}. 
We recall that only three-particle on-shell intermediate states 
lead to power-law finite-volume dependence.
To isolate such terms we first do the $k^0$ integral 
and keep only the pole at \(k^0=\omega_k\).
Other poles will be collected into infinite-volume quantities,
as for $\CL 0$.
This means that we can replace \(\Delta(k)\) with \(1/(2\omega_k)\) 
and set \(k^0 = \omega_k\) in all finite-volume terms. 
Furthermore we can pull out the sum
\begin{equation}
\frac{1}{L^3} \sum_{\vec k} \frac{1}{2 \omega_k}
\end{equation}
and consider the summand at fixed values of $\vec k$. 
The result of these steps is shown in Fig.~\ref{fig:twopartsub}a. 

At each fixed value, we are left precisely with all scattering
diagrams for {\em two particles} with energy-momentum 
\((E - \omega_k, \vec P - \vec k)\). 
We can thus follow the approach of Ref.~\cite{Kim:2005} 
to obtain the answer for this set of diagrams. 
In particular, we can repeatedly use the sum-integral
difference identity of Eq.~(\ref{eq:VLfinal}) and Fig.~\ref{fig:mainident}
to replace sums over the two-particle loop momenta with
integrals plus factors of $F$. As already noted, the identity
holds if either \(\sigma\) or \(\sigma^\dagger\) (or both) are
replaced by \( B_2\). 
This substitution is also indicated in Fig.~\ref{fig:twopartsub}a.

Our next step is to sum all diagrams into a convenient form by
regrouping terms according to the number of \(F\) insertions. This is
depicted in Fig.~\ref{fig:twopartsub}b. We first consider terms
with no \(F\) insertions. These are conveniently combined with the
smooth terms arising when the \(k^0\) contour encircles higher-particle poles,
yielding
\begin{equation}
\label{eq:CI1}
\CI 1 = \int_{\vec k} \PV\!\int_{a} \PV\!\int_{a'} 
\int_{k^0}\sigma(k, a) \Delta(a) \Delta(P\!-\!k\!-\!a)
\Delta(k) i 
\Kin{\mathrm{off}}(a,P\!-\!k\!-\!a,-a') 
\Delta(a') \Delta(P\!-\!k\!-\!a') \sigma^\dagger(k, a')
\,.
\end{equation}
Here we are using the definition of $\PV$ given in
Eq.~(\ref{eq:PVtildedef2}),
while the off-shell K-matrix is
\begin{multline}
i\Kin{\mathrm{off}}(a,b,-a') 
=
i B_2(a,b,-a') + \frac12 
\PV \! \int_{a_1} iB_2(a,b,-a_1) \Delta(a_1) \Delta(b_1) iB_2(a_1,b_1,-a')
\\
+ \left(\frac12\right)^2 \PV \! \int_{a_2} \PV \! \int_{a_1} iB_2(a,b,-a_1) 
\Delta(a_1)\Delta(b_1) iB_2(a_1,b_1,-a_2) \Delta(a_2)\Delta(b_2) 
iB_2(a_2,b_2,-a')
+ \dots \,,
\label{eq:Koffdef}
\end{multline}
or equivalently
\begin{equation}
i\Kin{\mathrm{off}}(a,b,-a') 
=
i B_2(a,b,-a') + \frac12 
\PV \! \int_{a_1} iB_2(a,b,-a_1) \Delta(a_1) \Delta(b_1) i \mathcal K_{2;\mathrm{off}}(a_1,b_1,-a') \,.
\end{equation}
For both $\Kin{\mathrm{off}}$ and $B_2$ 
we display only three of the (inflowing) momentum arguments,
the fourth being given by momentum conservation:
$a+b=a_1+b_1=a_2+b_2$.
If all external momenta are on-shell, $\Kin{\mathrm{off}}$ becomes the
usual physical two-particle K-matrix ${\cal K}_2$, which is real and smooth
(in our kinematic range) because the $\PV$ prescription is identical
to the PV prescription in this regime.
Within $\CI 1$, the K-matrix is needed also below threshold, and our
use of the $\PV$ prescription ensures that $\Kin{\mathrm{off}}$ is smooth
(cusp-free) in this regime as well. 
These results allow the overall sum over \(\vec k\) to be
replaced with an integral (for which no pole prescription is needed).

We stress that in Eq.~(\ref{eq:CI1}) the integral over $k^0$ must be done
before the other loop integrals.
This either puts the lower line on shell (leading to the
cuts which are dealt with by the $\PV$ prescription) or leads 
to intermediate states without a singularity 
(for which no pole prescription is needed).
The need to keep track of the ordering of integrals is an
unpleasant feature of the $\PV$ prescription.

We next sum all terms with exactly one \(F\) insertion, obtaining
\begin{align}
\label{eq:CLF11}
\CLF 1 1 & = \sigma^* \frac{i F}{2 \omega L^3} A^{(1,u)} + A'^{(1,u)}
\frac{i F}{2 \omega L^3} \sigma^{\dagger *} + A'^{(1,u)} \frac{i F}{2
  \omega L^3} A^{(1,u)} \,, \\ & = (\sigma^* + A'^{(1,u)}) \frac{i
  F}{2 \omega L^3} (\sigma^{\dagger *} + A^{(1,u)}) - \sigma^* \frac{i
  F}{2 \omega L^3} \sigma^{\dagger *} \,.
\end{align}
Here \(\sigma^*\) and \(A'^{(1,u)}\) 
(\(\sigma^{\dagger *}\) and \(A^{(1,u)}\)) are understood as 
row (column) vectors in the $k,\ell,m$ space introduced above.
The vectors \(\sigma^*\) and \(\sigma^{\dagger *}\) have been
defined in Eq.~(\ref{eq:sigmadecom2}), while
\(A'^{(1,u)}\) and \(A^{(1,u)}\) are new.
To define these, we begin with the functions 
\begin{align}
\label{eq:Ap1}
 A'^{(1,u)}(\vec k, a) & \equiv \frac{1}{2} \widetilde{\mathrm{PV}}
 \int_{a'} \sigma(a',k) \Delta(a') \Delta(P\!-\!k\!-\!a') 
i \mathcal  K_{2;\mathrm{off}}(a',P\!-\!k\!-\!a',-a) \,, 
\\[5pt] 
A^{(1,u)}( \vec k, a) & \equiv \frac{1}{2}
 \widetilde{\mathrm{PV}} \int_{a'} 
i \mathcal K_{2;\mathrm{off}}(a,P\!-\!k\!-\!a,-a')
 \Delta(a') \Delta(P\!-\!k\!-\!a') \sigma^{\dagger }(a',k) \,,
\label{eq:A1}
\end{align}
in which $k=[\omega_k,\vec k]$ is on shell while $a$ is not.
The superscripts \(u\) 
indicate that the first momentum argument (here \(\vec k\)) is
also the momentum of the particle that is {\em unscattered} by the
two-to-two K-matrix. 
We next set the momentum $a$ on shell, convert to CM coordinates
for the scattered particles, and decompose in spherical harmonics:\footnote{%
Note that here we do not add a superscript ${}^*$ to $A$ and $A'$
when one of the momenta is in the CM frame. This would make the
notation too heavy. The presence of the harmonic subscripts
$\ell, m$ serves as an alternative indicator that we are using a
CM momentum.}
\begin{align}
\label{eq:Ap1b}
A'^{(1,u)}_{\ell',m'}(\vec k, a^*) \sqrt{4 \pi} Y_{\ell',m'}(\hat a^*)
\equiv 
A'^{(1,u)}(\vec k, [\omega_a, \vec a]) \,, 
\\ 
\sqrt{4 \pi} Y^*_{\ell,m}(\hat a^*)A^{(1,u)}_{\ell,m}(\vec k, a^*)
\equiv 
A^{(1,u)}(\vec k, [\omega_a,\vec a]) \,.
\end{align}
Finally we project on-shell and restrict to finite-volume momenta
\begin{equation}
A'^{(1,u)}_{k,\ell',m'} \equiv A'^{(1,u)}_{\ell',m'}(\vec k, q^*_k)
\ \ \mathrm{and} \ \
A^{(1,u)}_{k,\ell,m} \equiv A^{(1,u)}_{\ell,m}(\vec k, q^*_k)\,, 
\ \ {\rm with}\ \ 
\vec k,\vec k' \in (2\pi/L) \mathbb Z^3\,.
\end{equation}
This gives the vectors appearing in Eq.~(\ref{eq:CLF11}).
The definition of $A'^{(1,u)}$ is shown diagrammatically in
Fig.~\ref{fig:twopartsub}c.

To see that Eq.~(\ref{eq:CLF11}) is valid, first observe that terms
with a single \(F\) insertion fall into three classes: (1) those with
no \(B_2\) kernels to the left of the \(F\) insertion but one or more
to the right; (2) those with no kernels to the right but one or more
to the left; (3) those with one or more \(B_2\) kernels on both sides
of the single \(F\) insertion. 
These give rise, respectively, to the three terms in Eq.~(\ref{eq:CLF11}),
after performing the sums over insertions of $B_2$ to obtain
the factors of $\Kin{\mathrm{off}}$ contained in $A'^{(1,u)}$ 
and $A^{(1,u)}$. Finally, observe that
coordinates that are common with the single $F$-insertion
are projected onto the on-shell, finite-volume phase space,
leading to the now-familiar matrix structure.

At this stage we can easily generalize to terms with \(n>1\) \(F\)
insertions between \(B_2\) kernels. We find
\begin{equation}
\CLF n 1 = (\sigma^* + A'^{(1,u)}) \frac{i F}{2 \omega L^3} 
[i \K i F]^{n-1} (\sigma^{\dagger *} + A^{(1,u)}) \,.
\end{equation}
Here we are using the matrix definition of $\K$ given in
Eq.~(\ref{eq:K2deflong}).
In words, this says that, between insertions of $F$, one can
have any number of $B_2$'s connected by $\PV$ integrals,
and these sum to give $\K$.
Summing over \(n\), including the $n=0$ result $\CI 1$, we obtain
\begin{equation}
\label{eq:C1Lint}
\CL 1 = \CI 1 + (\sigma^* + A'^{(1,u)}) \big [ \mathcal A \big ]
(\sigma^{\dagger *} + A^{(1,u)}) - \sigma^* \frac{i F}{2 \omega L^3}
\sigma^{\dagger *} \,, 
\end{equation}
where
\begin{equation}
\label{eq:Adef}
\mathcal A \equiv \frac{i F }{2 \omega L^3} \frac{1}{1 + \K F } 
= \frac{1}{1+F \K} \frac{iF}{2\omega L^3}\,.
\end{equation} 
Combining with our earlier expression (\ref{eq:VLfinal}) for \(\CL 0\) 
gives the main result of this subsection
\begin{equation}
\label{eq:CL0plusCL1}
\CL 0 + \CL 1 = \CI 0 + \CI 1 + (\sigma^* + A'^{(1,u)}) 
\big [ \mathcal A \big ] (\sigma^{\dagger *} + A^{(1,u)}) - 
(2/3) \sigma^* \frac{i F}{2 \omega L^3} \sigma^{\dagger *} \,.
\end{equation} 
We have succeeded in separating the correlator into factors
of $F$, which depend on the volume, and infinite-volume quantities.

The calculation just described follows very closely
the derivation of the two-particle quantization condition
in a moving frame given in Ref.~\cite{Kim:2005}.
This is because, for the diagrams of Fig.~\ref{fig:noswitches},
the third particle is a spectator whose main impact is to
take momentum away from the other two particles.
One difference in the present calculation, however, is that the symmetry
factor of $1/6$ for the no-insertion diagram, Fig.~\ref{fig:noinsertion},
does not match with those in the geometric sum
leading to the factor of $\big[{\cal A}\big]$ in the second term in
Eq.~(\ref{eq:CL0plusCL1}).
This is the reason for the appearance of the last term in
our result.

We can make the connection to the result of Ref.~\cite{Kim:2005}
more precise by considering instead the theory in which the spectator
is of a different type from the other two particles and does not
interact. For such a theory the symmetry factor for
Fig.~\ref{fig:noinsertion} is $1/2$, and the last term in
Eq.~(\ref{eq:CL0plusCL1}) is absent.
Indeed, for this theory we have already calculated {\em all} possible
diagrams, with the final result
\begin{equation}
C_{L}^{2+\mathrm{spec}} - C_{\infty}^{2+\mathrm{spec}} =
(\sigma^*+A'^{(1,u)}) \big [ \mathcal A \big] 
(\sigma^{\dagger   *}+A^{(1,u)}) \,.
\end{equation}
The spectrum is given by the poles of $C_L$. Since
infinite-volume quantities do not lead to poles, $C_L$
diverges if and only if \( \big[ \mathcal A \big ]\) 
has a divergent eigenvalue. This gives the quantization condition
\begin{equation}
\label{eq:specquant1}
\det \big [ \K^{-1} + F \big ] = 0 \,,
\end{equation}
where the determinant is over our
[finite-volume momentum]\(\times\)[angular momentum] space. 
Because both \(i \mathcal K_{2;k', \ell', m'; k, \ell, m}\) and 
\(i F_{k',\ell', m'; k, \ell, m}\) are
diagonal in \(k, k'\) space, this condition may be rewritten as
\begin{equation}
\label{eq:specquant2}
\prod_{\vec k} D(\vec k)= 0 \,,
\end{equation}
where
\begin{equation}
\label{eq:twoquant}
D(\vec k) \equiv \det_{\mathrm{ang\ mom}} 
\big [\K(\vec k)^{-1} +  F(\vec k) \big ] \,.
\end{equation}
The quantities appearing in this equation are defined in
Eqs.~(\ref{eq:K2def}) and (\ref{eq:sigmadecom2}), and
have only angular-momentum indices, since $\vec k$ is fixed.

This result is exactly what we expect given
given the two-particle quantization condition of Ref.~\cite{Kim:2005}. 
To see this, we note that,
using Eqs.~(\ref{eq:PVtildedef}) and (\ref{eq:rhodef})
to convert the $\PV$ into the $i\epsilon$ prescription,
${\mathcal M}_{2;\ell',m';\ell,m}(\vec k)$
is related to ${\mathcal K}_{2;\ell',m';\ell,m}(\vec k)$ by
\begin{equation}
\label{eq:Krhoseries}
i\M = i\K + i\K (i\rho) i\K + \cdots
= i\K \frac1{1+\rho\K}
\,.
\end{equation}
Here all arguments and indices are implicit.
It follows that
\begin{equation}
\label{eq:MKdif}
\M^{-1}(\vec k)-\K^{-1}(\vec k) = \rho(\vec k) 
= F(\vec k) - F^{i\epsilon}(\vec k)
\,,
\end{equation}
where the last equality follows from Eq.~(\ref{eq:Fdef2}).
Thus we can rewrite the quantity appearing in
the ``2+spec'' quantization condition as
\begin{equation}
\label{eq:DdefwithM2}
D(\vec k) \equiv \det_{\mathrm{ang\ mom}} 
\big [\M(\vec k)^{-1} +   F^{i \epsilon}(\vec k) \big ] \,.
\end{equation}
If this vanishes for one of the finite-volume choices of $\vec k$,
then there is a finite-volume state in the ``2+spec'' theory.

The connection to the result of Ref.~\cite{Kim:2005} can now be made.
If the spectator, which is necessarily on-shell
since it is non-interacting, has momentum $[\omega_k,\vec k]$,
then the total momentum of the other two particles is
$P_2=[E\!-\!\omega_k,\vec P-\vec k]$. For the full ``2+spec'' theory
to have a finite-volume state, the two interacting particles with
momentum $P_2$ must have a finite-volume state. The condition for
this, as given in Ref.~\cite{Kim:2005}, is exactly $D(\vec k)=0$.
This agreement  provides a useful check on our formalism.\footnote{%
Note that $F^{i\epsilon}$ (and not $F$) is the kinematic factor derived
in Ref.~\cite{Kim:2005}--see Eq.~(\ref{eq:comparetoKim}) for the exact relation.
We note that there is a potential confusion
regarding the definitions in earlier papers of $F^{i \epsilon}$ below
two-particle threshold. In particular, in 
in Eqs.~(24) and (25) of Ref.~\cite{Hansen:2012tf},
the {\em above-threshold}
definition of $F^{i \epsilon}$ is split into real and imaginary parts,
with the principal-value pole prescription used to define the
latter. In contrast to the $\PV$ prescription of the present article,
the principal-value in \cite{Hansen:2012tf} is replaced with a simple
prescription-free integral below threshold. In addition, the imaginary
part of $F^{i \epsilon}$, the term that we call $\rho$ here, is set to
zero below threshold in Ref.~\cite{Hansen:2012tf}. The upshot is that
the difference between $\PV$ used here and principal-value in
\cite{Hansen:2012tf} exactly cancels the difference between $\rho$
defined here and the analog in \cite{Hansen:2012tf}, so that the
definition of $F^{i \epsilon}$ is consistent in the two papers.}
%

%% file: fig104.pdf_t
\begin{picture}(0,0)%
\includegraphics{fig104.pdf}%
\end{picture}%
\setlength{\unitlength}{3947sp}%
\begingroup\makeatletter\ifx\SetFigFont\undefined%
\gdef\SetFigFont#1#2#3#4#5{%
  \reset@font\fontsize{#1}{#2pt}%
  \fontfamily{#3}\fontseries{#4}\fontshape{#5}%
  \selectfont}%
\fi\endgroup%
\begin{picture}(3032,765)(2899,-966)
\put(4607,-335){\makebox(0,0)[lb]{\smash{{\SetFigFont{10}{12.0}{\rmdefault}{\mddefault}{\updefault}{\color[rgb]{0,0,0}$a$}%
}}}}
\put(2814,-699){\makebox(0,0)[lb]{\smash{{\SetFigFont{12}{14.4}{\rmdefault}{\mddefault}{\updefault}{\color[rgb]{0,0,0}$ C_L^{(0)} \equiv $}%
}}}}
\put(4279,-612){\makebox(0,0)[lb]{\smash{{\SetFigFont{10}{12.0}{\rmdefault}{\mddefault}{\updefault}{\color[rgb]{0,0,0}$P-k-a$}%
}}}}
\put(4599,-871){\makebox(0,0)[lb]{\smash{{\SetFigFont{10}{12.0}{\rmdefault}{\mddefault}{\updefault}{\color[rgb]{0,0,0}$k$}%
}}}}
\end{picture}%

%% file: fig103v5copy.pdf_t
\begin{picture}(0,0)%
\includegraphics{fig103v5copy.pdf}%
\end{picture}%
\setlength{\unitlength}{3947sp}%
\begingroup\makeatletter\ifx\SetFigFont\undefined%
\gdef\SetFigFont#1#2#3#4#5{%
  \reset@font\fontsize{#1}{#2pt}%
  \fontfamily{#3}\fontseries{#4}\fontshape{#5}%
  \selectfont}%
\fi\endgroup%
\begin{picture}(3276,1020)(2250,-2014)
\put(2997,-1999){\makebox(0,0)[lb]{\smash{{\SetFigFont{12}{14.4}{\rmdefault}{\mddefault}{\updefault}{\color[rgb]{0,0,0}off-shell}%
}}}}
\put(3200,-1308){\makebox(0,0)[lb]{\smash{{\SetFigFont{12}{14.4}{\rmdefault}{\mddefault}{\updefault}{\color[rgb]{0,0,0}$=$}%
}}}}
\put(4347,-1308){\makebox(0,0)[lb]{\smash{{\SetFigFont{12}{14.4}{\rmdefault}{\mddefault}{\updefault}{\color[rgb]{0,0,0}$+$}%
}}}}
\put(4933,-1610){\makebox(0,0)[lb]{\smash{{\SetFigFont{12}{14.4}{\rmdefault}{\mddefault}{\updefault}{\color[rgb]{0,0,0}$F$}%
}}}}
\put(4647,-1999){\makebox(0,0)[lb]{\smash{{\SetFigFont{12}{14.4}{\rmdefault}{\mddefault}{\updefault}{\color[rgb]{0,0,0}on-shell}%
}}}}
\end{picture}%

%% file: oneswitch.tex
\subsection{Two-to-two insertions: one switch}

\indent

\label{sec:oneswitch}

\begin{figure}
\begin{center}

 \includegraphics[scale=0.26]{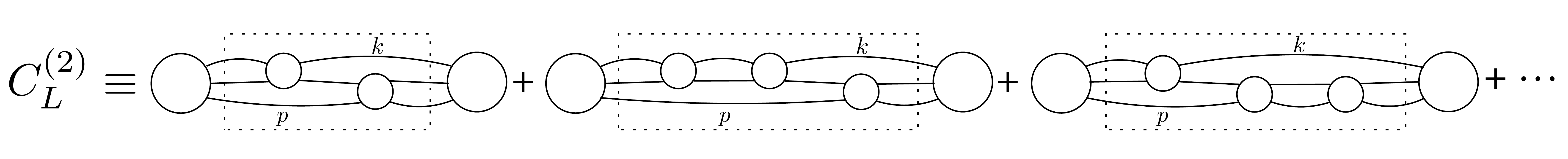}

\vspace{-10pt}

\caption{Finite-volume correlator diagrams containing only two-to-two insertions, with one switch in the scattered pair.}
\label{fig:oneswitch}
\end{center}
\end{figure}

In this section we sum the diagrams of Fig.~\ref{fig:oneswitch}. 
Each diagram has at least one \(B_{2}\)
insertion on exactly two different pairs of particles. 
In other words, the
diagrams have one switch in the pair that is scattered. We denote the
sum of all such diagrams by \(C_L^{(2)}\). Throughout this section we
call the momentum of the incoming spectator particle \(k\) and that of
the outgoing spectator \(p\), as shown in the figure. 
We refer to the three propagators which appear at the 
location where the scattered pair changes as the ``switch state''. 
The presence of a switch leads to the first appearance of a three-particle scattering quantity in our analysis.

To determine the volume-dependent contribution of these diagrams we first
evaluate the \(p^0\) and \(k^0\) integrals. Since we know from earlier
considerations that intermediate states with three on-shell particles
are needed to obtain power-law volume dependence, at least 
one of the two poles at \(p^0 = \omega_p\) and \(k^0 = \omega_k\) 
must be encircled. For concreteness we enumerate the
four types of terms: (a) each contour encircles its one-particle
pole; (b) the \(p^0\) contour encircles its pole but the \(k^0\)
contour encircles all other contributions; (c) as in (b) but with \(k^0\)
and \(p^0\) exchanged; and (d) both contours encircle everything
but the one-particle poles. We now consider the loop sums/integrals
that remain when holding $\vec p$ and $\vec k$ fixed; these are all
two-particle loops involving either the upper two particles
(to the left of the switch state) or the lower two (to the right).
For type (d) terms the summands have no singularities
and thus all sums can be replaced with integrals. Similarly, in type (b) and (c) terms,
the two-particle loops on one side of the switch state cannot go on-shell and may thus be replaced by integrals. For all remaining two-particle loops in terms of types (a), (b) and (c), 
the summand is singular.
Here we substitute the identity of Eq.~(\ref{eq:VLfinal}), thereby separating 
each loop into an infinite-volume contribution and an \(F\)-factor residue. 

There are thus two disjoint regions where insertions of $F$ appear:
to the left of the switch state and to the right. 
It is useful to break our analysis into four classes, 
defined by whether or not each side of the switch state has
at least one insertion. We label these as
\begin{equation}
(1)\ F, F,\ \ (2)\ -, F,\ \ (3)\ F, -,\ \ (4)\ -, -\,,
\end{equation}
so that class (1) contains all terms with at least
one \(F\) insertion both to the left and right of the switch state,
class (2) contains terms with no such insertions to the left but at
least one to the right, etc. Observe that type (a) terms appear in all
four classes, while types (b) and (c) only appear in classes (2+4) and 
(3+4) respectively.

\begin{figure}
\begin{center}

\includegraphics[scale=0.35]{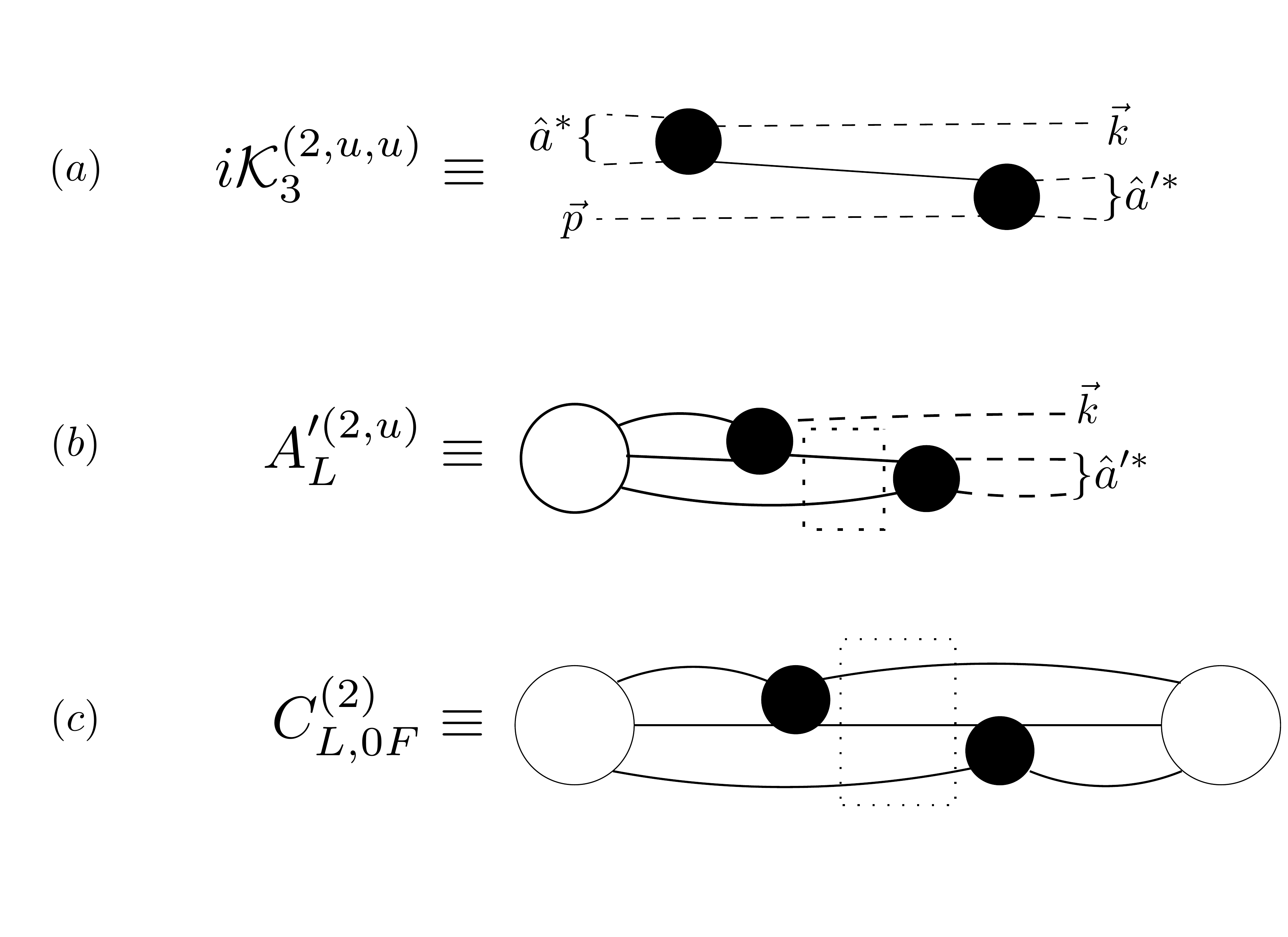}

\vspace{-20pt}

\caption{Diagrammatic definitions of (a) 
\( i \mathcal K_{3 \rightarrow 3}^{(2,u,u)}\),
(b) $A'^{(2,u)}_L$ and (c) $C_{L,0F}^{(2)}$. 
In (b) and (c) the dotted box encloses momenta that are summed
rather than integrated. The solid circle represents the two-particle
K-matrix. Other notation as above.
}
\label{fig:M2def}
\end{center}
\end{figure}

We now analyze the four classes in turn, starting with (1).
Because all terms in this class have both \(k^0\) and \(p^0\)
one-particle poles, the chains of $F$'s, $B_2$'s and $\PV$-integrated
loops to the left and right of the switch state can each be 
independently summed exactly as in the previous subsection.
This leads to 
\begin{equation}
\label{eq:case1}
(\sigma^* + A'^{(1,u)}) \big [ \mathcal A \big ] i \Kth^{(2,u,u)} \big
      [ \mathcal A \big ] (\sigma^{\dagger *} + A^{(1,u)}) \,.
\end{equation}
The new feature here is the quantity
\(\Kth^{(2,u,u)} \equiv \Kthin {p,\ell,m;k,\ell',m'}^{(2,u,u)}\)
which arises from the switch state, and
is a contribution to the three-to-three scattering amplitude.
It is shown diagrammatically in Fig.~\ref{fig:M2def}a,
to which we refer for the momentum labels.
To define it we proceed in the by-now familiar steps,
beginning with the partially off-shell quantity
\begin{equation}
\label{eq:K32uudef} 
i\Kth^{(2,u,u)}(\vec p, \vec a, \vec k, \vec a')
  \equiv 
i \mathcal K_{2;\mathrm{off}}(a,P\!-\!p\!-\!a,-k)
\Delta(P\!-\!p\!-\!k)
i \mathcal K_{2;\mathrm{off}}(P\!-\!p\!-\!k,p, -a')\,.
\end{equation}
At this stage $p$, $k$, $a$ and $a'$ are on shell,
while $P\!-\!p\!-\!k$, $P\!-\!p\!-\!a$ and $P\!-\!k\!-\!a'$ are not.
We have parametrized \(i \mathcal K_3^{(2,u,u)}\)
with incoming and outgoing spectator momenta, $\vec k$ and $\vec p$,
as well as incoming and outgoing momenta of one of the scatterers, 
$\vec a'$ and $\vec a$. In the second step we boost $\vec a'$ and $\vec a$
to the appropriate CM frames and then decompose in spherical harmonics:
\begin{equation}
\label{eq:M332decom}
\Kth^{(2,u,u)}(\vec p, \vec a, \vec k, \vec a') 
\equiv
4 \pi Y^*_{\ell,m}(\hat a^*) 
\Kthin { \ell, m;  \ell', m'}^{(2,u,u)}(\vec p, a^*, \vec k, a'^*) 
Y_{\ell,m}(\hat a'^*)  
\,,
\end{equation}
where $\vec a^*$ is defined by boosting 
$(\omega_a, \vec a) \rightarrow (\omega_{a}^*, \vec a^*)$ 
with velocity $\vec \beta_p$, and $\vec a'^*$ is defined 
by boosting the corresponding primed vector with $\vec \beta_k$. 
Next we recall that all incoming and outgoing particles
are on-shell if and only if \(a'^*=q^*_k\) and \(a^* = q^*_p\). 
Thus we define the on-shell version of \(\mathcal K_3^{(2,u,u)}\) as
\begin{equation}
\label{eq:K3onshell}
\Kthin { \ell, m;  \ell', m'}^{(2,u,u)}(\vec p, \vec k)  
\equiv 
\Kthin { \ell, m;  \ell', m'}^{(2,u,u)}(\vec p, q^*_p, \vec k, q^*_k) \,.
\end{equation}
The final step is to restrict to finite-volume momenta
\begin{equation}
\Kthin{p, \ell, m; k, \ell', m'}^{(2,u,u)} 
\equiv  
\Kthin{\ell, m; \ell', m'}^{(2,u,u)}(\vec p, \vec k\,) 
\ \ \ \mathrm{for} \ \ \ \vec k, \vec p \in (2 \pi/L) \mathbb Z^3 \,.
\end{equation}
This gives the matrix contained in the result Eq.~(\ref{eq:case1}).

Several further explanations are in order.
First, $\Kth^{(2,u,u)}$ in Eq.~(\ref{eq:case1}) is
on-shell on both ``sides'' because it is
sandwiched between factors of $F$.
This is because $[{\cal A}]$, defined in Eq.~(\ref{eq:Adef}),
has an $F$ on both ends.
Second, the boosts to CM momenta $\vec a^*$ and $\vec a'^*$ are always
well defined because $F$ contains factors of $H(\vec p)$
(on the left) and $H(\vec k)$ (on the right).
Third, sub-threshold momenta occur in both left and right CM frames
as $\vec p$ and $\vec k$ are varied,
requiring analytic continuation of the $\Kth^{(2,u,u)}$.
Fourth, all factors from external propagators are contained in
the $[{\cal A}]$'s, so $\Kth^{(2,u,u)}$ is a contribution to the
{\em amputated} three-to-three scattering amplitude.
Fifth, the superscript ``$(2,u,u)$'' indicates that this
contribution involves two factors of 
$\mathcal K_{2;\mathrm{off}}$, and that, on both sides,
the particles singled out by the label ($\vec p$ on the left,
$\vec k$ on the right) are unscattered.
And, finally, although the result (\ref{eq:case1})
has a symmetric form, it is important to note that
$\Kth^{(2,u,u)}$ switches the spectator momentum index from 
$p$ to $k$.

We now turn our attention to class (2) contributions,
i.e. those with no \(F\) insertions to the left of the switch state
but at least one such insertion on the right. As noted above,
these contributions come from terms of types (a) and (b).
In the former, the
\(p^0\) and \(k^0\) integrals both encircle one-particle poles,
but all two-particle loops with $p$ as the spectator are integrated
using the $\PV$ prescription.
In the latter, only the $k^0$ integral encircles the one-particle pole,
so all two-particle loop sums to the left of the switch state 
can be replaced by integrals.
Combining these contributions, we find
\begin{equation}
\label{eq:C2sum}
\ApnL 2 {} \big [ \mathcal A \big ]  (\sigma^{\dagger *} +A^{(1,u)}) \,,
\end{equation}
where the new quantity $\ApnL 2 {}$ is shown diagrammatically
in Fig.~\ref{fig:M2def}b. It is a contribution to the left endcap
involving one switch. It is given by
\begin{equation}
\label{eq:ApL2a}
A'^{(2,u)}_{L;k,\ell',m'}   \equiv A'^{(2,u)}_{L;\ell',m'}(\vec k, q^*_k) 
\ \ [\textrm{with}\ k\in (2\pi/L)\mathbb Z^3 ]\,,
\end{equation}
where
\begin{equation}
\label{eq:ApL2b}
A'^{(2,u)}_{L;\ell',m'}(\vec k, a'^*) 
\sqrt{4 \pi} Y_{\ell',m'}(\hat a'^*)  
\equiv 
A_L'^{(2,u)}(\vec k,  [\omega_{a'},\vec a'])
\end{equation}
and
\begin{multline}
\label{eq:ApL2c}
A_L'^{(2,u)}(\vec k,  a')
\equiv 
\frac{1}{2} \frac{1}{L^3} \sum_{\vec p} \PV \int_{a} \int_{p^0}
\sigma(p,a) \Delta(a) \Delta(P\!-\!p\!-\!a) \\ 
\times i \mathcal K_{2;\mathrm{off}}(a, P\!-\!p\!-\!a, -k)  
\Delta(p) \Delta(P\!-\!p\!-\!k)  
i \mathcal K_{2;\mathrm{off}}(p, P\!-\!p\!-\!k, -a')
\end{multline}
is the endcap amplitude with $k$ on shell but $a'$ not.
The subscript $L$ is a reminder that this quantity 
contains important finite-volume effects. These arise
from the sum over $\vec p$ with a singular summand 
(from the switch state).
The superscript $(2,u)$ refers to the presence of two factors
of $\mathcal K_{2;\mathrm{off}}$ and the fact that the 
particle carrying the momentum that is
singled out by the coordinate system (here $\vec k$) is unscattered.

Class (3) contributions mirror those from class (2), with the roles
of the parts of the diagrams to the left and right of the switch state 
interchanged. The total result is
\begin{equation}
(\sigma^* +A'^{(1,u)}) \big [ \mathcal A \big ] \AnL 2 {} \,,
\end{equation}
where  
\begin{equation}
A^{(2,u)}_{L;p,\ell,m}   \equiv A^{(2,u)}_{L;\ell,m}(\vec p, q^*_p) 
\ \ [\textrm{with}\ p\in (2\pi/L)\mathbb Z^3 ]\,,
\end{equation}
with
\begin{equation}
\AnL 2 {; \ell, m}(\vec p, a^*) \sqrt{4 \pi} Y_{\ell',m'}(\hat a^*) 
\equiv 
\AnL 2 {} (\vec p, [\omega_a,\vec a]) 
\end{equation}
and
\begin{multline}
\AnL 2 {} (\vec p, a)
\equiv 
\frac{1}{2} \frac{1}{L^3} \sum_{\vec k} \PV \int_{a'} \int_{k^0}
i \mathcal K_{2;\mathrm{off}}(a,P\!-\!p\!-\!a,-k) 
\Delta(k) \Delta(P\!-\!p\!-\!k) 
\\ 
\times i\mathcal K_{2;\mathrm{off}}(p,P\!-\!p\!-\!k,-a') 
\Delta(a') \Delta(P\!-\!k\!-\!a') \sigma^\dagger (k,a') \,.
\end{multline}

Finally, we turn to class (4) contributions,
which have no \(F\) insertions on either side of the switch state. 
Combining contributions from types (a-d), we find
[see Fig.~\ref{fig:M2def}c]
\begin{multline}
\label{eq:CL0F2}
\CLF 0 2 = \frac{1}{4} \frac{1}{L^6} \sum_{\vec p, \vec k}  \PV \int_{a, a'}   
\int_{p^0} \int_{k^0}
 \sigma(p,a) \Delta(a) \Delta(P\!-\!p\!-\!a) \Delta(p) 
\\
\times i \mathcal K_{2;\mathrm{off}}(a,P\!-\!p\!-\!a,-k) \Delta(P\!-\!p\!-\!k)
 i \mathcal K_{2;\mathrm{off}}(P\!-\!p\!-\!k,p,-a')
 \Delta(a') \Delta(P\!-\!k\!-\!a') \Delta(k) \sigma^\dagger(k,a') \,.
\end{multline}
Adding this to the results from the other classes, we obtain
\begin{equation}
\label{eq:C2Lint}
 C^{(2)}_{L}  = (\sigma^* + A'^{(1,u)}) \big [ \mathcal A \big ] 
i \Kth^{(2,u,u)}  \big [ \mathcal A \big ]  (\sigma^{\dagger *} + A^{(1,u)})  
+ \ApnL 2 {}  \big [ \mathcal A \big ]  (\sigma^{\dagger *} +A^{(1,u)}) 
+(\sigma^* +A'^{(1,u)}) \big [ \mathcal A \big ] \AnL 2 {} 
+  \CLF 0 2 \,.
\end{equation}

At this stage we have achieved only a partial separation of
finite-volume effects, because
\(\ApnL 2 {}\), \(\AnL 2 {}\) and \( \CLF 0 2\) still contain
momentum sums that cannot be replaced by integrals.
In addition, \(\Kth^{(2,u,u)}\) suffers from the problem,
discussed in the introduction, of diverging for certain physical
momenta. In the remainder of this section we derive identities
for these four quantities that allow a complete separation of
finite-volume effects and avoid divergences in the $3\to3$ scattering
amplitude.

We begin with \(\Kth^{(2,u,u)}\), and separate it into two terms,
one which is singular but only depends on the on-shell $\mathcal K_2$,
and another which is regular. We do this separation in a way that allows generalization to diagrams
with more switches.
In particular, we will analyze the partially off-shell quantity
$\Kth^{(2,u,u)}(\vec p, \vec a, \vec k, \vec a')$,
defined in Eq.~(\ref{eq:K32uudef}),
although for this subsection we only need the on-shell version of
this quantity [as in Eq.~(\ref{eq:K3onshell})].
In fact, we keep the four momentum arguments completely general
so that the boosts to the CM-frames for $a$ and $a'$ need not
be defined.

Our first step is to write the intermediate propagator as
\begin{equation}
\Delta(P\!-\!p\!-\!k) = \frac{iH(\vec p) H(\vec k)}
{2\omega_{pk}(E-\omega_p-\omega_k-\omega_{kp})}
+ {\cal R}^a(\vec p, \vec k)
\,.
\label{eq:Deltadecomp}
\end{equation}
The first term contains the on-shell singularity, while the
second is smooth. 
We focus for now on the singular term in (\ref{eq:Deltadecomp})
and substitute this into $\Kth^{(2,u,u)}$, Eq.~(\ref{eq:K32uudef}).
The presence of the $H$ factors means that we can boost to the CM
frames for the $\{k,P\!-\!p\!-\!k\}$ 
and the $\{P\!-\!p\!-\!k,p\}$ pairs, and decompose the dependence
on $\vec k^*$ and $\vec p^*$ into spherical harmonics.
The result is that the singular contribution becomes
\begin{equation}
\label{eq:K32uustep1}
i\Kth^{(2,u,u)}(\vec p, \vec a, \vec k, \vec a') \supset
i \mathcal K_{2 \dl \mathrm{off} \dl  {\rm off}\,\ell m}(\vec p;\vec a \dl k^*)
i G^a_{\ell,m;\ell',m'}(\vec p,\vec k)
i \mathcal K_{2 \dl \mathrm{off}\, \ell' m' \dl {\rm off}}(\vec k;p^* \dl \vec a')\,,
\end{equation}
where 
\begin{equation}
i G^a_{\ell,m;\ell',m'}(\vec p,\vec k)
\equiv
\frac{i 4 \pi Y_{\ell,m}(\hat k^*) 
H(\vec p)H(\vec k)Y_{\ell',m'}^*(\hat p^*)}
{2 \omega_{pk}(E -  \omega_p - \omega_k - \omega_{pk} )} 
\,,
\end{equation}
and
\begin{align}
\sqrt{4\pi}
\mathcal K_{2 \dl \mathrm{off} \dl {\rm off} \, \ell m}(\vec p;\vec a \dl k^*) 
Y_{\ell,m}(\hat k^*)
&\equiv
\mathcal K_{2;\mathrm{off}}(a,P\!-\!p\!-\!a,-k)
\\
\sqrt{4\pi} Y^*_{\ell',m'}(\hat p^*)
\mathcal K_{2 \dl \mathrm{off} \, \ell' m' \dl {\rm off}}(\vec k; p^* \dl \vec a') 
&\equiv
\mathcal K_{2;\mathrm{off}}(P\!-\!p\!-\!k,p, -a')\,.
\end{align}
In the latter two definitions, the two subscripts ``off'' are a reminder
that both incoming and outgoing scattering pairs
have one particle off-shell. 
If the ``off'' is followed by angular momentum indices,
this indicates that the scattered pair has been boosted to its CM
frame and the angular dependence decomposed into spherical harmonics.
The arguments of ${\cal K}_{2\dl{\rm off}\dl{\rm off}}$
list respectively the spectator momentum,
the momentum of one of the incoming scattered pair, 
and the momentum of one of the outgoing pair.
If a CM-frame boost has been done, the argument is 
the magnitude of the CM-frame momentum.
This hybrid notation is needed to maintain generality.

The next step is to write the singular part of
(\ref{eq:K32uustep1}) in terms of on-shell K-matrices.
This is straightforward as we can expand the boosted
momenta $k^*$ and $p^*$ about their on-shell values,
$q_p^*$ and $q_k^*$ respectively.
At the same time, we want the remaining non-singular term
to be a smooth function of $\vec p$ and $\vec k$, since
this is required below. The spherical harmonics
$Y_{\ell,m}(\hat k^*)$ and $Y^*_{\ell',m'}(\hat p^*)$
are not, however, smooth at $\vec k^*=0$ (for $\ell>0$)
and $\vec p^*=0$ (for $\ell'>0$), respectively.
To resolve this problem, and pull out an appropriate singular term,
we introduce the finite difference operator $\delta$.
This can act to the right or left on $\K$, with its action being
\begin{align}
\delta\; \mathcal K_{2 \dl \mathrm{on} \, \ell' m' \dl {\rm off}}(\vec k ; \vec a') 
& \equiv 
\mathcal K_{2 \dl \mathrm{off} \, \ell' m' \dl {\rm off}}(\vec k; p^* \dl \vec a') -
({p^*}/{q_k^*})^{\ell'}
\mathcal K_{2 \dl \mathrm{on} \,\ell' m' \dl {\rm off}}(\vec k;\vec a')  \,,
\label{eq:deltadefleft}\\
\mathcal K_{2 \dl \mathrm{off} \dl {\rm on} \, \ell m}(\vec p; \vec a) \; \delta 
& \equiv 
\mathcal K_{2 \dl \mathrm{off} \dl {\rm off} \, \ell m}(\vec p; \vec a \dl k^*) -
\mathcal K_{2 \dl \mathrm{off} \dl {\rm on} \,\ell m}(\vec p; \vec a)
({k^*}/{q_p^*})^{\ell}
 \,.
\label{eq:deltadefright}
\end{align}
Here we have defined the ``on-off'' and ``off-on'' K-matrices as
\begin{equation}
\mathcal K_{2 \dl \mathrm{on}\,\ell' m' \dl {\rm off}}(\vec k; \vec a')  
\equiv
\mathcal K_{2 \dl \mathrm{off} \, \ell' m' \dl {\rm off}}(\vec k; q_k^* \dl \vec a')
\ \ {\rm and}\ \ 
\mathcal K_{2 \dl \mathrm{off} \dl {\rm on} \, \ell m}(\vec p; \vec a)  
\equiv
\mathcal K_{2 \dl \mathrm{off} \dl {\rm off}\, \ell m}(\vec p; \vec a \dl q_p^*)
\,.
\end{equation}
Note that if a scattering
pair is on shell then it does not have a corresponding momentum 
argument (since the latter is fixed by kinematics).

Inserting Eqs.~(\ref{eq:deltadefleft}) and (\ref{eq:deltadefright})
into Eq.~(\ref{eq:K32uustep1}) we obtain
\begin{equation}
\label{eq:K32uustep2}
i\Kth^{(2,u,u)}(\vec p, \vec a, \vec k, \vec a') 
\supset
i \mathcal K_{2 \dl \mathrm{off} \dl {\rm on}\, \ell m}(\vec p; \vec a)
\left[
i G^b_{\ell,m;\ell',m'}(\vec p,\vec k)
+ {\cal R}^b_{\ell,m;\ell',m'}(\vec p,\vec k)
\right]
i \mathcal K_{2 \dl \mathrm{on} \, \ell' m' \dl {\rm off}}(\vec k; \vec a')\,,
\end{equation}
with
\begin{equation}
\label{eq:Gbdef}
i G^b_{\ell,m;\ell',m'}(\vec p,\vec k)
\equiv
\left(\frac{k^*}{q_p^*}\right)^\ell
\frac{i 4 \pi Y_{\ell,m}(\hat k^*) 
H(\vec p)H(\vec k)Y_{\ell',m'}^*(\hat p^*)}
{2 \omega_{pk}(E -  \omega_p - \omega_k - \omega_{pk} )} 
\left(\frac{p^*}{q_k^*}\right)^{\ell'}
\,,
\end{equation}
and
\begin{equation}
{\cal R}^b_{\ell,m;\ell',m'}(\vec p,\vec k)
\equiv
\delta\; i G^a_{\ell,m;\ell',m'}(\vec p,\vec k) ({p^*}/{q_k^*})^{\ell'}
+ ({k^*}/{q_p^*})^{\ell} i G^a_{\ell,m;\ell',m'}(\vec p,\vec k)\; \delta
+ \delta\; i G^a_{\ell,m;\ell',m'}(\vec p,\vec k) \;\delta 
\,.
\end{equation}
The result (\ref{eq:K32uustep2}) has achieved our goals.
Only the $G^b$ term is singular, because the factors of $\delta$
in ${\cal R}^b$ [which act on the K-matrices appearing in 
Eq.~(\ref{eq:K32uustep2})] give differences which vanish 
when $P\!-\!p\!-\!k$ goes on shell
and thus cancel the singularity in $G^a$. More precisely the 
analyticity of $\mathcal K_2$ near the on-shell point is required to 
demonstrate the cancellation. For example, the difference defined in 
Eq.~(\ref{eq:deltadefleft}) scales as $p^*-q^*_k$, 
the same scaling as the denominator of $G^b$, so that the product is a 
finite smooth function. This is discussed in detail in Appendix~\ref{app:sumintegral}.
Furthermore, the extra powers of $k^*$ and $p^*$ ensure that $G^b$ is smooth
when $\vec k^*$ or $\vec p^*$ vanish. Finally, the
$G^b$ term multiplies K-matrices in which $k^*$ (to the left)
and $p^*$ (to the right) are on shell.

\begin{figure}
\begin{center}
\includegraphics[scale=0.8]{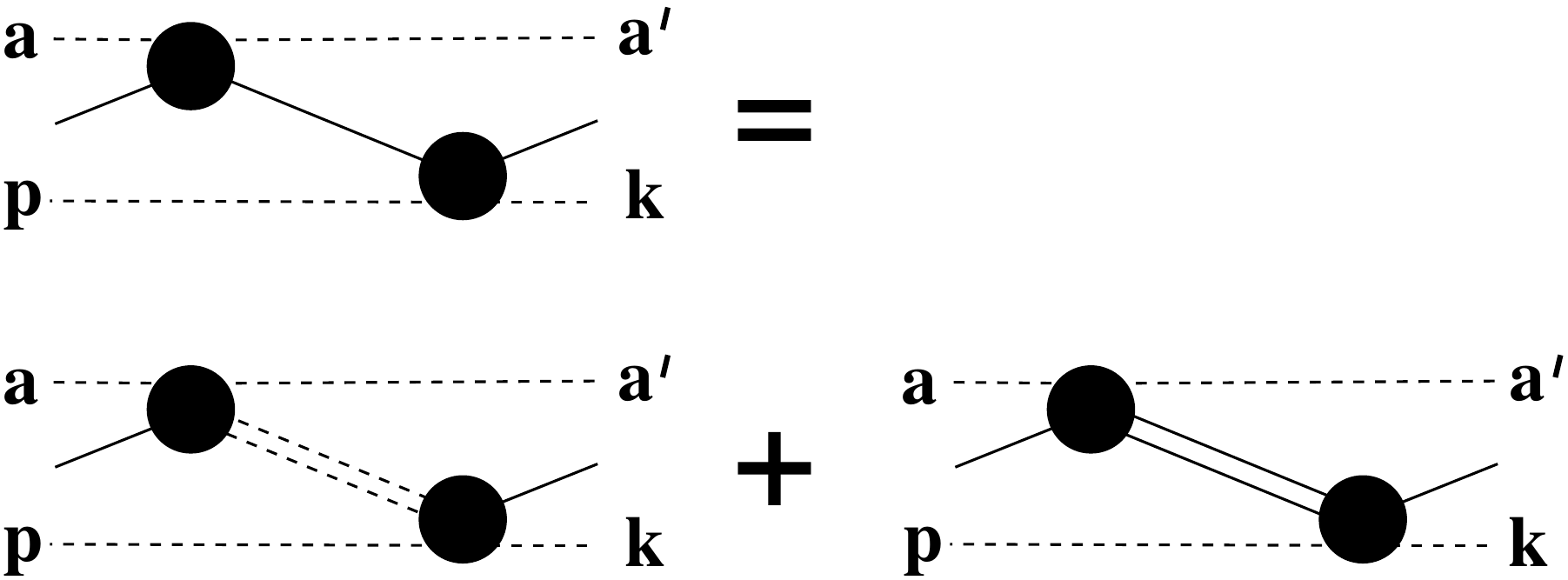}
\caption{Diagrammatic version of the decomposition
of $\Kth^{(2,u,u)}(\vec p, \vec a, \vec k, \vec a')$ given in
Eq.~(\ref{eq:decompfinala}). External dashed lines indicate
on-shell momenta, whereas momenta flowing along the solid 
external lines are, in general, off shell.
All external propagators are amputated. 
The first term on the right-hand side is the singular term
${\cal D}^{(2,u,u)}$, with the double dashed lines representing $G^b$. 
The two $\K$ are evaluated on-shell for all momenta that flow along dashed propagators. 
The second term represents the divergence-free amplitude ${\cal K}_{{\rm df},3}^{(2,u,u)}$.
}
\label{fig:decompK3}
\end{center}
\end{figure}

The end result of this analysis is that
\begin{equation}
\label{eq:decompfinala}
\Kth^{(2,u,u)}(\vec p, \vec a, \vec k, \vec a') 
=
{\cal D}^{(2,u,u)}(\vec p, \vec a, \vec k, \vec a') 
+
{\cal K}_{{\rm df},3}^{(2,u,u)}(\vec p, \vec a, \vec k, \vec a') 
\end{equation}
where the singular part is
\begin{equation}
i{\cal D}^{(2,u,u)}(\vec p, \vec a, \vec k, \vec a') 
=
i \mathcal K_{2 \dl \mathrm{off} \dl {\rm on} \, \ell m}(\vec p; \vec a)
i G^b_{\ell,m;\ell',m'}(\vec p,\vec k)
i \mathcal K_{2 \dl \mathrm{on} \, \ell' m' \dl {\rm off}}(\vec k,\vec a')
\end{equation}
and the divergence-free part of the amplitude is
\begin{multline}
i{\cal K}_{{\rm df},3}^{(2,u,u)}(\vec p, \vec a, \vec k, \vec a') 
\equiv
i \mathcal K_{2;\mathrm{off}}(a,P\!-\!p\!-\!a,-k)
{\cal R}^a(\vec p,\vec k)
i \mathcal K_{2;\mathrm{off}}(P\!-\!p\!-\!k,p, -a')
\\
+
i \mathcal K_{2 \dl \mathrm{off} \dl {\rm on} \, \ell m}(\vec p,\vec a)
{\cal R}^b_{\ell,m;\ell',m'}(\vec p,\vec k)
i \mathcal K_{2 \dl \mathrm{on} \, \ell' m' \dl {\rm off}}(\vec k,\vec a')\,.
\label{eq:Kdf2uudef}
\end{multline}
The relation (\ref{eq:decompfinala}) is shown diagrammatically
in Fig.~\ref{fig:decompK3}.
The key property of ${\cal K}_{{\rm df},3}^{(2,u,u)}$
is that it is a smooth, non-singular function of its arguments.
It is smooth when $\vec k^*$ or $\vec p^*$ vanish
because $\Kth^{(2,u,u)}$ is smooth at these values and, as just discussed,
this is also true of the $G^b$ term.

The quantity $G^b$ is closely related to the matrix $G$ introduced
in Eq.~(\ref{eq:Gdef}). In particular,
\begin{equation}
G^b_{\ell,m;\ell',m'}(\vec p, \vec k)
=
G_{p,\ell,m;k_1,\ell_1,m_1} [2\omega L^3]_{k_1,\ell_1,m_1;k,\ell',m'}
\ \ 
{\rm for}\ p,k\in (2\pi/L)\mathbb Z^3\,,
\end{equation}
where
\begin{equation}
[2 \omega L^3]_{k_1,\ell_1,m_1;k,\ell',m'} \equiv
\delta_{k_1,k}\delta_{\ell_1, \ell'}\delta_{m_1,m'} 2 \omega_k L^3
\,.
\end{equation}

Finally, with this groundwork laid, we can return to the quantity
relevant for the one-switch analysis, namely $\Kth^{(2,u,u)}$
with external momenta on shell and taking finite-volume values. 
In this case we can decompose
the external CM-frame momenta in spherical harmonics, and connect
back to our matrix notation:
\begin{align}
\mathcal K_{2 \dl \mathrm{off} \dl {\rm on} \, \ell m}%
(\vec p;\vec a)\Big|_{a^*=q_p^*,\vec p \, \in (2\pi/L)\mathbb Z^3}
&=
\sqrt{4\pi} Y^*_{\ell',m'}(\hat a^*)\mathcal K_{2;p,\ell',m';p,\ell,m}
\\[5pt]
\mathcal K_{2 \dl {\rm on} \, \ell m \dl \mathrm{off}}%
(\vec k; \vec a')\Big|_{a'^*=q_k^*,\vec k \, \in (2\pi/L)\mathbb Z^3}
&=
\mathcal K_{2;k,\ell,m;k,\ell',m'}
\sqrt{4\pi} Y_{\ell',m'}(\hat a'^*)
\,.
\end{align}
This allows us to write the decomposition into singular and smooth
parts in matrix form
\begin{equation}
\label{eq:Kdf2decom}
i \Kthin{ p, \ell, m;   k, \ell', m'}^{(2,u,u)} 
\equiv 
i \mathcal K_2 i G [2\omega L^3] i \mathcal  K_2
+
i \Kdfthin{ p, \ell, m; k, \ell', m'}^{(2,u,u)} 
\,,
\end{equation}
where, as usual,
\begin{equation}
\label{eq:Kdf2uudecomp}
4\pi Y^*_{\ell,m}(\hat a^*) \Kdfthin{ p, \ell, m; k, \ell', m'}^{(2,u,u)} 
Y_{\ell',m'}(\hat a'^*)
=
{\cal K}_{{\rm df},3}^{(2,u,u)}(\vec p, \vec a, \vec k, \vec a')\Big|_{%
a^*=q_p^*,a'^*=q_k^*,\{\vec p,\vec k\} \in (2\pi/L)\mathbb Z^3}
\,.
\end{equation}

We next derive an identity for \(\ApnL 2 {}\), 
which is defined in Eqs.~(\ref{eq:ApL2a}-\ref{eq:ApL2c}). 
The basic approach is our standard move of replacing the sum over $\vec p$
with a $\PV$ integral and a sum-integral difference, the latter giving
rise to a factor of $F$.
However, the presence of the switch state introduces new features
compared to previous applications, so we work through
the steps in some detail.

We first introduce the fully-integrated counterpart to
$\ApnL 2 {}$, which we call $A'^{(2,u)}$. It is defined exactly
as for $\ApnL 2 {}$ [Eqs.~(\ref{eq:ApL2a}-\ref{eq:ApL2c})
and Fig.~\ref{fig:M2def}b]
except that the sum over $\vec p$ is replaced by a $\PV$-integral. 
This is the first example of an infinite-volume quantity with multiple $\PV$-integrals. 
As we have already mentioned, a consequence of our nonstandard regulator is that the 
order of integration is important. In the definition of $\ApnL 2 {}$, the integral 
over $\vec k$ is done last.
The difference between the two quantities can be written as
\begin{equation}
\ApnL 2 {}(\vec k,a') - A'^{(2,u)}(\vec k,a') =
\bigg [ \frac{1}{L^3} \sum_{\vec p} - \PV \int_{\vec p} \bigg ]
A'^{(1,u)}(\vec p, k)
\frac{H(\vec k)\Delta(P\!-\!p\!-\!k)}{2\omega_p}
i \mathcal K_{2; \mathrm{off}}(k, P\!-\!p\!-\!k,a')
\,,
\label{eq:ApnLdiff1}
\end{equation}
where $k=[\omega_k,\vec k]$.
To obtain this form we have used the fact that 
$A'^{(1,u)}(\vec p,k)$ [defined in Eq.~(\ref{eq:Ap1})]
is a smooth function of $\vec p$, 
so that the only singularity in $p$ comes from the switch state.
Also, we have done the $p^0$ integral and kept only the particle pole,
since other poles give non-singular contributions which
have vanishing sum-integral differences.
Finally, we have added in the cut-off function $H(\vec k)$, which
is allowed since it does not change the singularity.

\begin{figure}
\begin{center}
\vspace{20pt}
\hspace {-350pt} {\large (a)}

\vspace{-15pt}

\hspace{-160pt}
\resizebox{.170\hsize}{!}{\(A'^{(1,s)}(\vec k, p^*) \equiv \)}

\vspace{-25pt}

\hspace{75pt} \includegraphics[scale=0.6]{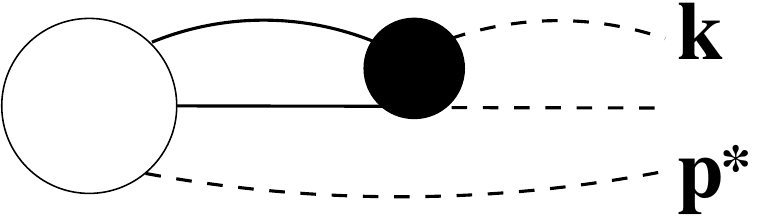}

\vspace{45pt}

\hspace {-350pt} {\large (b)}

\vspace{-15pt}

\hspace{-160pt}
\resizebox{.16\hsize}{!}{\(A'^{(1)}(\vec k, p^*) \equiv \)}

\vspace{-25pt}

\hspace{0pt} \includegraphics[scale=0.6]{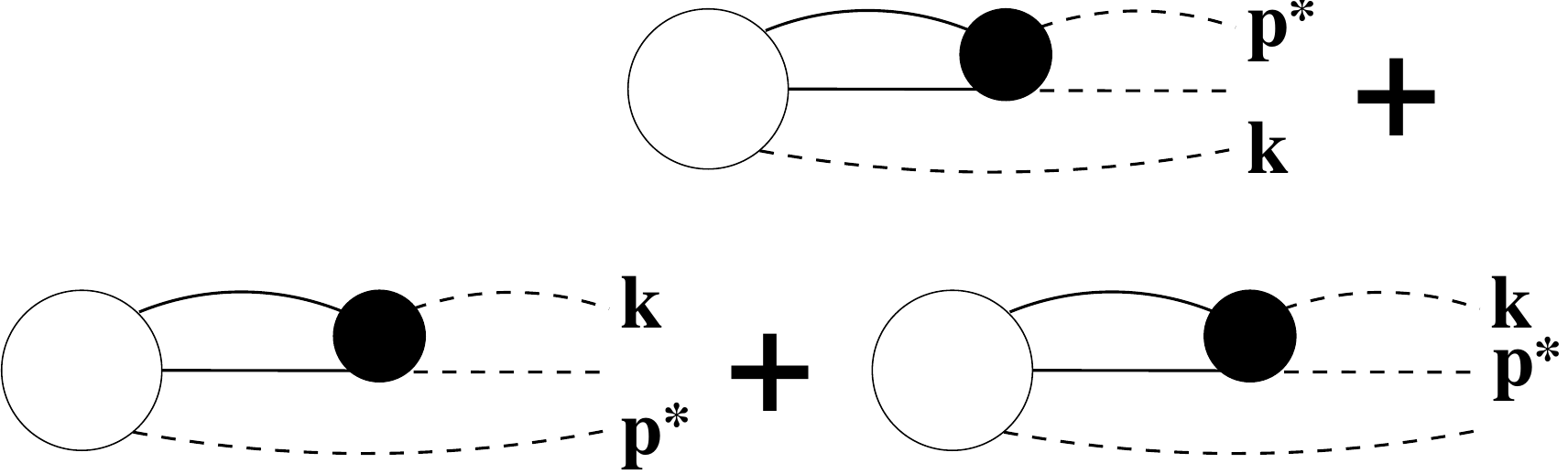}

\vspace{5pt}

\caption{(a) Definition of \(A'^{(1,s)}\), in which the
  momentum singled out by the coordinate system (here $\vec k$)
  is that of a particle that scatters. 
(b) Definition of the symmetrized quantity \(A'^{(1)}\).}
\label{fig:A1sdef}
\end{center}
\end{figure}

To use the sum-integral identity, we need to expand
$A'^{(1,u)}(\vec p, k)$ in spherical harmonics with respect to
$\vec p^*$, i.e. treat $k$ as the spectator and boost to the
CM frame of the other two particles
[with boost velocity $-(\vec P-\vec k)/(E-\omega_k)$].
This is different
from the expansion used earlier, in Eq.~(\ref{eq:Ap1b}),
where $p$ was treated as the spectator.
Thus we define [see Fig.~\ref{fig:A1sdef}a]
\begin{equation}
\label{eq:Acoordchange}
A'^{(1, s)}_{\ell,m}(\vec k, p^*) \sqrt{4 \pi} Y_{\ell,m}(\hat p^*)   
\equiv 
A'^{(1,u)}(\vec p,[\omega_k,\vec k])
\,.
\end{equation}
where the superscript ``$s$'' indicates that the
particle carrying the momentum singled out by the coordinate
system, here $\vec k$, is one of those {\em scattered} by
the $\K$ inside $A'^{(1,u)}$.
We stress that $A'^{(1,s)}_{\ell,m}$ and
$A'^{(1,u)}_{\ell,m}$ are different expansions of the same
underlying function---we are just using different coordinate systems.
We also note that the boost defining $\vec p^*$ is
well defined because of the presence of $H(\vec k)$.

As a final step, we must also decompose 
the off-shell two-particle K-matrix into spherical harmonics:
\begin{equation}
4 \pi Y^*_{\ell,m}(\hat p^*) 
 \mathcal K_{2;\mathrm{off};\ell,m;\ell',m'}(\vec k, p^*, a'^*) 
Y_{\ell',m'}(\hat a'^*) 
\equiv 
 \mathcal K_{2;\mathrm{off}}
([\omega_p,\vec p \,],P\!-\!p\!-\!k, -[\omega_{a'},\vec a']) \,,
\end{equation}
This allows us to write
\begin{multline}
\ApnL 2 {;k, \ell', m'} - A'^{(2,u)}_{k, \ell', m'} =
\\ 
\bigg [ \frac{1}{L^3} \sum_{\vec p} - \PV \int_{\vec p} \bigg ]
A'^{(1,s)}_{\ell_1,m_1}(\vec k, p^*) 
\frac{i 4 \pi Y_{\ell_1,m_1}(\hat p^*) H(\vec k)
Y^*_{\ell_2,m_2}(\hat p^*)}
{2 \omega_{p}   2 \omega_{P-p-k}(E-\omega_{k}-\omega_{p}-\omega_{pk})} 
i \mathcal K_{2; \mathrm{off};\ell_2,m_2;\ell',m'}(\vec k, p^*,q^*_k)
\,,
\label{eq:ApnLdiff3}
\end{multline}
where we have explicitly pulled out the particle
pole in the $P\!-\!p\!-\!k$ propagator, since the remainder
is non-singular.
This has the form for which we can apply the sum integral identity,
from which we deduce
\begin{equation}
\ApnL 2 {} = A'^{(2,u)} + 2 A'^{(1,s)} i F i \K \,,
\label{eq:ApnLfinal}
\end{equation}
where the on-shell matrix form of $A'^{(1,s)}$ is
\begin{equation}
A'^{(1,s)}_{p,\ell, m} = A'^{(1,s)}_{\ell,m}(\vec p,q^*_p)
\ \ {\rm with}\ \ \vec p \in (2\pi/L)\mathbb Z^3\,.
\ \end{equation}
The factor of two  in (\ref{eq:ApnLfinal}) appears because $F$ contains a
symmetry factor of $1/2$ which is absent in the switch-state contribution. 

The new quantity $2A'^{(1,s)}$ will later be combined with $A'^{(1,u)}$
in order to form an object which, aside from one subtlety, is 
symmetric under particle interchange. To understand this point, first observe that there are three independent ways that the external momenta can be
assigned to the diagram:
(i) $\vec p$ is the spectator
with $\vec k$ one of the scattered pair (giving $A'^{(1,u)}$),
(ii) $\vec p$ is one of the scattered pair with $\vec k$ 
the spectator (giving $A'^{(1,s)}$), and
(iii) $\vec p$ {\em and} $\vec k$ form the scattered pair.
This is illustrated in Fig.~\ref{fig:A1sdef}b.
For the symmetry to hold we must sum these three with 
equal weights: (i)+(ii)+(iii).\footnote{%
We stress that here we are discussing on-shell quantities;
the symmetry cannot hold if one of the particles is off shell.}
This differs from the combination that arises naturally in our derivation,
(i)+2(ii). It turns out, however, that we can replace 2 (ii) with (ii)+(iii),
and thus obtain a truly symmetric combination. 
We do this repeatedly below, and thus explain here the justification for
this change.

Momentum assignment (iii) leads to a quantity
we call $A'^{(1,\tilde s)}$ that is related to $A'^{(1,s)}$ as follows:
\begin{equation}
\label{eq:stildedef}
A'^{(1,\tilde s)}_{p,\ell,m} = (-1)^\ell A'^{(1, s)}_{p,\ell,m}
\,.
\end{equation}
This is because the assignments (iii) and (ii) differ
simply by the interchange of the two particles 
that have been boosted to their CM frame.
(These are the particles with momenta $k$ and $P-p-k$.)
This interchange is the same as a parity transform in the CM frame,
leading to the result (\ref{eq:stildedef}).
We also note that $A'^{(1,u)}_{p,\ell,m}$ is only
non-vanishing for even $\ell$ given the symmetry of $\K$.
Thus the desired combination
\begin{equation}
A'^{(1)}_{p,\ell,m}
\equiv
A'^{(1, u)}_{p,\ell,m}+
A'^{(1, s)}_{p,\ell,m}+
A'^{(1,\tilde s)}_{p,\ell,m} 
\end{equation}
satisfies
\begin{equation}
A'^{(1)}_{p,\ell,m} = 
\begin{cases}
A'^{(1, u)}_{p,\ell,m}+
2 A'^{(1, s)}_{p,\ell,m} & \ell\ {\rm even}\,,
\\
0 & \ell\ {\rm odd} \,.
\end{cases}
\end{equation}
This means that we can make the replacements
\begin{equation}
\label{eq:stostilde}
2 A'^{(1, s)}_{p,\ell,m}
\longrightarrow
A'^{(1, s)}_{p,\ell,m}+ A'^{(1,\tilde s)}_{p,\ell,m} 
\ \ {\rm and}\ \
A'^{(1, u)}_{p,\ell,m}+ 2 A'^{(1, s)}_{p,\ell,m}
\longrightarrow
A'^{(1)}_{p,\ell,m}
\end{equation}
as long as only even values of $\ell$ contribute.

To see that only even values of $\ell$ contribute, first recall from Eq.~(\ref{eq:ApnLfinal}) that $A'^{(1,s)}$ is connected
by an $F$ to a factor of $\K$. Next, note that the symmetry of $\K$ implies that only even
angular momenta appear in its expansion.
Finally, use the result Eq.~(\ref{eq:Foddl})
in Appendix~\ref{app:sumintegral} 
that $F_{p,\ell,m;k\ell',m'}$ vanishes if $\ell+\ell'$ is odd.
Together these imply that, since $\ell'$ is even, $\ell$ is also. 

It turns out that, throughout the derivation, ``$(s)$'' quantities
always appear opposite those with a ``$(u)$'' superscript.
The latter always have the requisite symmetry so that only
even angular momenta contribute. Consequently, by the argument just
given, we can always replace $2(s)$ with $(s)+(\tilde s)$.
For the sake of brevity, we do not do this explicitly,
but keep in mind that this replacement is allowed.
At the end of the derivation, once we have summed 
contributions from any number of switches, we make the replacement
explicit so as to allow symmetrization.

The identity for \(\AnL 2 {}\) is derived in exactly the same way
as that for $\ApnL 2 {}$ and we simply state the result:
\begin{equation}
\label{eq:A2Lfinres}
\AnL 2 {}= A^{(2,u)} + i \K  i F  2 A^{(1,s)} \,.
\end{equation}
Here $A^{(2,u)}_{p,\ell',m'}$ and $A^{(1,s)}$ are
the left-right ``reflections'' of the corresponding $A'$ quantities.

Finally we consider \(\CLF 0 2\),
defined in Eq.~(\ref{eq:CL0F2}) and Fig.~\ref{fig:M2def}c.
Our aim is to determine the finite-volume residue
that results when we convert the two momentum sums into integrals. 
The two-particle loops are both rendered non-singular by the
$\PV$ integrals over $a$ and $a'$, so the only
singularity is that in $\Delta(P\!-\!p\!-\!k)$.
To isolate this, both $p^0$ and $k^0$ integrals must circle
the particle poles. (If other poles are encircled in either integral,
the remaining summand is non-singular and both sums can be immediately
changed to integrals.)
We then have to choose which sum to evaluate first.
Our convention, here and below, is to work from left to right. 
Thus we first convert the sum over $\vec p$ into an integral
plus an $F$ term.
The detailed steps are exactly as for $A'^{(2,u)}_L$, except
that here we have $A^{(1,u)}$ on the right rather than $\K$.
For the contribution in which $\vec p$ is integrated, there are no
more singularities, so the sum over $\vec k$ can be converted
directly into an integral. For the $F$-term, however, the sum over $\vec k$
must remain. The result of this analysis is that
\begin{equation}
\CLF 0 2 = C_\infty^{(2)} + 2 A'^{(1,s)} \frac{iF}{2 \omega L^3} A^{(1,u)} 
\,,
\label{eq:CL0F2final}
\end{equation}
where $C_\infty^{(2)}$ is the infinite-volume version of the single-switch
correlator:
\begin{multline}
\label{eq:CI02}
\CI 2 = \frac{1}{4} \PV \int_{\vec k} \PV \int_{\vec p} \PV \int_{a, a'}   
\int_{p^0} \int_{k^0}
 \sigma(p,a) \Delta(a) \Delta(P\!-\!p\!-\!a) \Delta(p) 
\\
\times i \mathcal K_{2;\mathrm{off}}(a,P\!-\!p\!-\!a,-k) \Delta(P\!-\!p\!-\!k)
 i \mathcal K_{2;\mathrm{off}}(P\!-\!p\!-\!k,p,-a')
 \Delta(a') \Delta(P\!-\!k\!-\!a') \Delta(k) \sigma^\dagger(k,a') \,.
\end{multline}
The factor of $1/(2\omega L^3)$ in the last term in
Eq.~(\ref{eq:CL0F2final}) arises because $F$ is defined to contain the
contributions from only two of the three propagators in the switch state.
The overall factor of $2$ in this term arises because
$F$ contains a symmetry factor of $1/2$ that is absent in the switch state. 
Concerning $C_\infty^{(2)}$, we stress again that the order of $\PV$-integration 
matters in the definition of this infinite-volume quantity.

Our ``left to right convention'' has given an asymmetric result,
with $A'^{(1,s)}$ to the left of $A^{(1,u)}$ and no ``$u F s$'' term.
This lack of symmetry can,
however, be corrected {\em a posteriori}, as will be explained when
we consider the result from any number of switches.

Inserting the identities (\ref{eq:Kdf2decom}), (\ref{eq:ApnLfinal}), 
(\ref{eq:A2Lfinres}) and (\ref{eq:CL0F2final})
into Eq.~(\ref{eq:C2Lint}) we find the final result of this section
\begin{align}
\CL 2  - \CI 2  &= (\sigma^* + A'^{(1,u)}) \big [ \mathcal A \big ] 
\left[i \K i G 2 \omega L^3 i \K + i \Kdfth {}^{(2,u,u)} \right]
\big [ \mathcal A \big ]  (\sigma^{\dagger *} + A^{(1,u)}) 
\nonumber\\
&
 + A'^{(2,u)} \big [ \mathcal A \big ] (\sigma^{\dagger *}+A^{(1,u)})
 + (\sigma^*+A'^{(1,u)})  \big [ \mathcal A \big ] A^{(2,u)} 
\nonumber\\ 
&
+ [2 A'^{(1,s)}] i F i\K \big[\mathcal A \big] (\sigma^{\dagger *}+A^{(1,u)})  
+ (\sigma^*+A'^{(1,u)}) \big [ \mathcal A \big ]  i \K i F [2 A^{(1, s)}] 
\nonumber\\
&+ 2 A'^{(1,s)} \frac{iF}{2 \omega L^3} A^{(1,u)}\,.
\label{eq:C2def4}
\end{align}
This is the main result of this subsection. 
The right-hand side is the finite-volume residue of all one-switch diagrams.

%% file: twoswitches.tex
\subsection{Two-to-two insertions: two switches}

\indent

\label{sec:twoswitch}
  
\begin{figure}
\begin{center}
%

\includegraphics[scale=0.24]{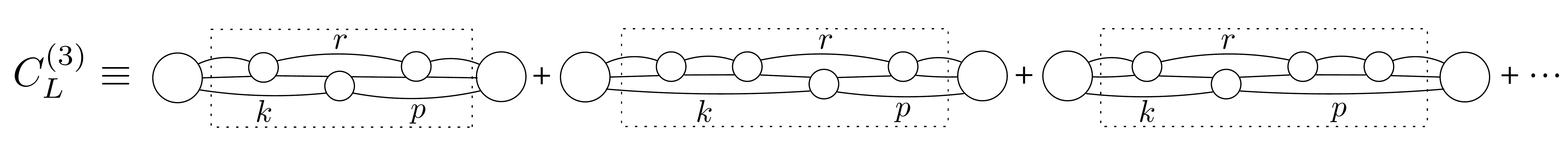}

\caption{Finite-volume correlator diagrams containing only
  two-to-two insertions and with two switches in the scattered pair.}
\label{fig:twoswitch}
\end{center}
\end{figure}

In this section we sum the diagrams of Fig.~\ref{fig:twoswitch}. 
These are all diagrams that have two switches in
the pair that is scattered. We denote the sum of all such diagrams by
\(C_L^{(3)}\). Throughout this section we refer to leftmost
(rightmost) triplet of propagators, at the point where the scattering
pair changes, as the left (right) switch state. 
We label the three different spectator momenta \(p\), \(r\) and \(k\),
as shown in the figure.

We provide a detailed analysis of two-switch diagrams before analyzing
diagrams with any number of switches for two reasons. First, a new
type of intermediate quantity with finite-volume dependence
arises at this order. This is \(\KthnL 3\), a contribution to
three-to-three scattering. 
Second,
a number of new complications enter at this stage with the derivation
of identities relating the intermediate quantities (with \(L\)
subscripts) to infinite-volume quantities. We think it clearer to
analyze these in isolation before generalizing to all orders.

As in the previous sections we evaluate \(p^0\), \(r^0\) and \(k^0\)
integrals and then substitute the identity of Eq.~(\ref{eq:VLfinal})
for all two-particle loops for which the spectator is on shell.
There are three locations where insertions of $F$ can appear:
to the left of the left switch state, between switch states, and to
the right of the right switch state.
We define eight different classes of terms, based on whether
or not at least one \(F\) insertion appears in each of the three locations:
\begin{gather}
(1)\ F,F,F\ \ \ (2)\ -,F,F,  \ \ (3)\ -,-,F,\ \ (4)\ F,F,-,\ \ 
\nonumber\\
(5)\ F,-,-\ \ (6)\ -,F,-,\ \ (7)\ F,-,F \ \ (8)\ -,-,-
  \,.
\end{gather}
For example in class (2) there is at least one insertion to the right
of the right switch state and between the switches, but no insertion to the
left of the left switch state.

\begin{figure}
\begin{center}

\includegraphics[scale=0.5]{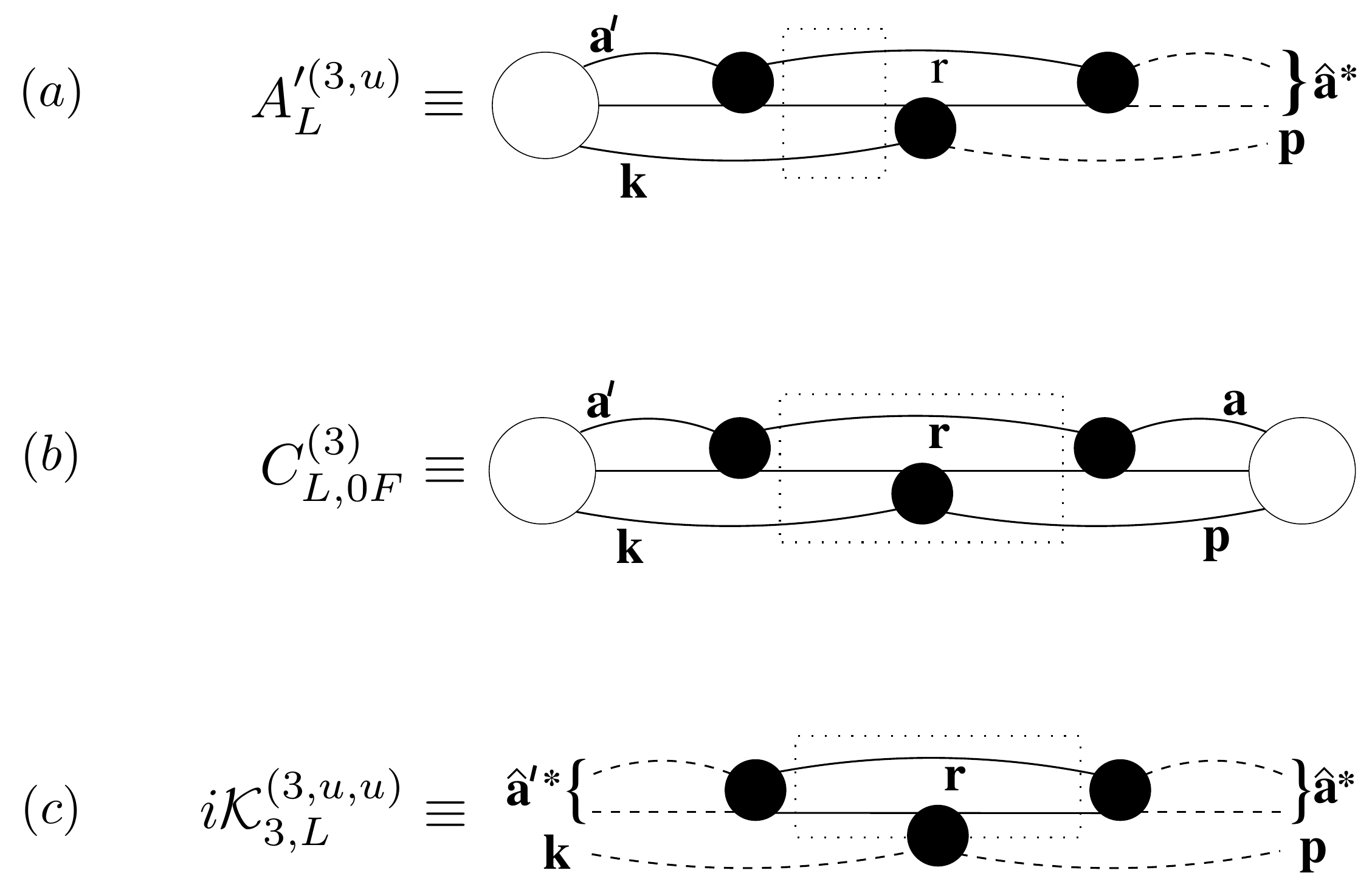}

\caption{
(a) Definition of \(\ApnL 3 {}\).
  The dotted rectangle contains momenta which are summed;
  thus only the leftmost two-particle loop is integrated. 
  (b) Definition of $C_{L,0F}^{(3)}$, which has two integrated and three summed
  loop momenta.
  (c) Definition of ${\cal K}^{(3,u,u)}_{3,L}$, which has a single summed momentum.}
\label{fig:L3defs}
\end{center}
\end{figure}

Using the methods of the previous subsections, it is straightforward
to obtain the results from these classes:
\begin{equation}
\label{eq:twoswitchC}
\begin{split}
\CL 3 & =  (\sigma^* + A'^{(1,u)}) \big [ \mathcal A \big ] 
i \mathcal K_{3 }^{(2,u,u)}   \big [ \mathcal A \big ]  
i \mathcal K_{3 }^{(2,u,u)}  \big [ \mathcal A \big ] 
(\sigma^{\dagger *} + A^{(1,u)}) 
\\
& + \ApnL 2 {} \big [ \mathcal A \big ]  i \mathcal K_{3 }^{(2,u,u)} 
\big [ \mathcal A \big ] (\sigma^{\dagger *} +  A^{(1,u)}) + 
\ApnL 3 {}\big [ \mathcal A \big ] (\sigma^{\dagger *} + A^{(1,u)}) 
\\
& +  (\sigma^* + A'^{(1,u)})  \big [ \mathcal A \big ]  
i \mathcal K_{3 }^{(2,u,u)}  \big [ \mathcal A \big ] \AnL 2 {} 
+  (\sigma^* + A'^{(1,u)}) \big [ \mathcal A \big ] \AnL 3 {}
\\
& + \ApnL 2 {} \big [ \mathcal A \big ] \AnL 2 {} + 
(\sigma^* + A'^{(1,u)})\big [ \mathcal A \big ] i \KthnL 3  
\big [ \mathcal A \big ] (\sigma^{\dagger *} + A^{(1,u)}) 
+\CLF 0 3 
\,.
\end{split}
\end{equation}
Here the eight terms are the results, in turn,
from the eight classes of contribution identified above. 
The four new quantities appearing in
Eq.~(\ref{eq:twoswitchC}) are 
\(\ApnL 3 {}\), \(\AnL 3 {}\), \(\CLF 0 3\) and \( \KthnL 3 \). 
These are defined as (see also Fig.~\ref{fig:L3defs})
\begin{equation}
\label{eq:ApNL3a}
\ApnL 3 {;p, \ell', m'} \equiv \ApnL 3 {; \ell', m'}(\vec p, q^*_p) \ \mathrm{with}\ \vec p \in (2 \pi/L) \mathbb Z^3
\,,
\end{equation}
where
\begin{multline}
\label{eq:ApNL3b}
\ApnL 3 {; \ell', m'}(\vec p, a^*) \sqrt{4 \pi} Y_{\ell',m'}(\hat a^*)
\equiv \frac{1}{2} \frac1{L^6}\sum_{\vec k,\vec p} \PV \int_{a'} 
\int_{k_0} \int_{r^0} \sigma(k,a') \Delta(a') \Delta(P\!-\!k\!-\!a') 
i \mathcal K_{2;\mathrm{off}}(a',P\!-\!k\!-\!a',-r)
\\ 
\times  
\Delta(k) \Delta(P\!-\!r\!-\!k) 
i \mathcal K_{2;\mathrm{off}}(k,P\!-\!k\!-\!r,-p)
\Delta(r) \Delta(P\!-\!p\!-\!r) 
i \mathcal K_{2;\mathrm{off}}(r,P\!-\!p\!-\!r,-a)
\,,
\end{multline}
with $\AnL 3 {;k,\ell, m}$ defined analogously, 
\begin{multline}
\CLF 0 3 \equiv 
\frac{1}{4} \frac{1}{L^9} \sum_{\vec k, \vec p, \vec  r} 
\PV\int_{a'} \PV \int_a 
\int_{k^0}\int_{p^0}\int_{r^0}
\sigma(k, a') \Delta(a')\Delta(P\!-\!k\!-\!a') 
i \mathcal K_{2;\mathrm{off}}(a',P\!-\!k\!-\!a',-r)
\Delta(k) \Delta(P\!-\!k\!-\!r) 
\\ \times 
i \mathcal K_{2;\mathrm{off}}(k,P\!-\!r\!-\!k,-p)
\Delta(r) \Delta(P\!-\!p\!-\!r)
i \mathcal K_{2;\mathrm{off}}(k,P\!-\!p\!-\!r,-a)
\Delta(p) \Delta(P\!-\!p\!-\!a) \Delta(a) \sigma^\dagger(p,a)
\,,
\label{eq:CLF3def}
\end{multline}
and finally
\begin{equation}
\KthinnL 3{k, \ell', m'; p, \ell, m} \equiv 
\KthinnL 3{\ell', m';  \ell, m}(\vec k, q^{*}_k, \vec p, q^*_p)   \ \ \mathrm{with}\ \ \vec p, \vec k \in (2 \pi/L) \mathbb Z^3 \,,
\label{eq:K3uudef}
\end{equation}
with
\begin{multline}
4 \pi Y^*_{\ell', m'}(\hat a'^*) 
i \KthinnL 3 {\ell', m'; \ell, m}(\vec k, a'^{*}, \vec p, a^*) 
Y_{\ell, m}(\hat a^*) 
\equiv \frac{1}{L^3} \sum_{\vec r} \int_{r^0}
i \mathcal K_{2;\mathrm{off}}(a',P\!-\!k\!-\!a',-r)
\\ \times 
\Delta(P\!-\!k\!-\!r) \Delta(r)
i \mathcal K_{2;\mathrm{off}}(k,P\!-\!k\!-\!r,-p)
\Delta(P\!-\!p\!-\!r) 
i \mathcal K_{2;\mathrm{off}}(r,P\!-\!p\!-\!r,-a)
\,.
\label{eq:MML3}
\end{multline}
To obtain these results we have summed $B_2$ kernels into
two-particle K-matrices, and used the fact that
\(\big [ \mathcal  A]\) amputates and puts on-shell 
both factors adjacent to it. 

We now derive identities relating the quantities 
\(\ApnL 3 {}\), \(\AnL 3 {}\), \(\CLF 0 3\) and \( \KthnL 3 \)
to infinite-volume observables. 
We begin with $\ApnL 3 {}$ (see Fig.~\ref{fig:L3defs}a), 
and work from left to right  converting sums into integrals.
The $\vec k$ sum leaves a finite-volume residue because of
the singular propagator $\Delta(P\!-\!k\!-\!r)$,
while the $\vec r$ sum leaves a residue because of
$\Delta(P\!-\!p\!-\!r)$. The infinite-volume quantity that results, which we
call $A'^{(3,u)}$, is thus given by the same expressions as
Eqs.~(\ref{eq:ApNL3a}) and (\ref{eq:ApNL3b})
except that $\sum_{\vec k,\vec p}$ is replaced by
$\PV\int_{\vec k} \PV\int_{\vec p}$ in Eq.~(\ref{eq:ApNL3b}).
The finite-volume residues can be obtained using
the same argumentation as for $\ApnL 2{}$ in the previous subsection.
The result is
\begin{equation}
\label{eq:ApL3intb}
\ApnL 3 {} = A'^{(3,u)} +
2 A'^{(1,s)} \frac{i F}{2\omega L^3} i \Kth^{(2,u,u)} +
2 A'^{(2,s)} \frac{i F}{2\omega L^3} i \K
\,.
\end{equation}
Note the superscripts ``$s$'' on the $A_L$'s and the factors
of 2 due to the missing symmetry factor at the switch states.
The new quantity $A'^{(2,s)}$
is simply $A'^{(2,u)}$ expressed in the alternative coordinate
system, just as in the definition of $A'^{(1,s)}$, Eq.~(\ref{eq:ApLndef}).
We can now use the result from the previous subsection
for ${\cal K}_3^{(2,u,u)}$, Eq.~(\ref{eq:Kdf2decom}),
to obtain the desired identity
\begin{equation}
\label{eq:ApL3res}
\ApnL 3 {}  = A'^{(3,u)} + 2A'^{(2,s)} iF  i  \K  +  
2 A'^{(1,s)} \frac{i F }{2 \omega L^3}  i \K i G 2 \omega L^3 i \K +
2 A'^{(1,s)} \frac{iF}{2 \omega L^3}  i \Kdfth^{(2,u,u)}\,.
\end{equation}
This derivation naturally lends itself to a recursive
extension to higher order, as we explain in the
next subsection.

The result for \(\AnL 3 {}\) is given simply by reversing 
the order of factors in each term:
\begin{equation}
\label{eq:AL3res}
 \AnL 3 {}  = A^{(3,u)} +   i  \K i F 2A^{(2,s)}  
+  i \K  i G  i \K  i F 2 A^{(1,s)} 
+  i \Kdfth^{(2,u,u)} \frac{iF}{2 \omega L^3}  2A^{(1,s)} \,.
\end{equation}

We next consider \(\CLF 0 3\) (see Fig.~\ref{fig:L3defs}b). 
Working from left to right we obtain
the infinite-volume quantity plus one finite-volume residue from the
$\vec k$ sum and another from the $\vec r$ sum. 
Following our by now standard manipulations, this leads to
\begin{equation}
\label{eq:CL3resint}
\CLF 0 3 = \CI 3 + 2 A'^{(1,s)}  \frac{iF }{2 \omega L^3} A^{(2,u)}_L
+ 2 A'^{(2,s)}  \frac{i F}{2 \omega L^3} A^{(1,u)} \,.
\end{equation}
Here $\CI 3$ is defined as in Eq.~(\ref{eq:CLF3def}) except that
the momentum sums are replaced by the ordered integrals
$\PV\int_{\vec p}\PV\int_{\vec r}\PV\int_{\vec k}$.
Note that $A^{(2,u)}_L$ still contains a momentum sum, but we can
obtain a complete decomposition using Eq.~(\ref{eq:A2Lfinres}) 
from the previous subsection. This leads to
\begin{equation}
\label{eq:CLF03res}
\CLF 0 3 - \CI 3 = 
2 A'^{(1,s)}  \frac{iF }{2 \omega L^3}  i \K i F 2 A^{(1,s)} 
+ 2 A'^{(1,s)}  \frac{i F}{2 \omega L^3}  A^{(2,u)} 
+ 2 A'^{(2,s)}  \frac{i F}{2 \omega L^3}  A^{(1,u)} \,.
\end{equation}

\begin{figure}
\begin{center}
\vspace{20pt}
\hspace {-480pt} {\Large (a)}
\vspace{-30pt}

\hspace{20pt}\includegraphics[scale=0.45]{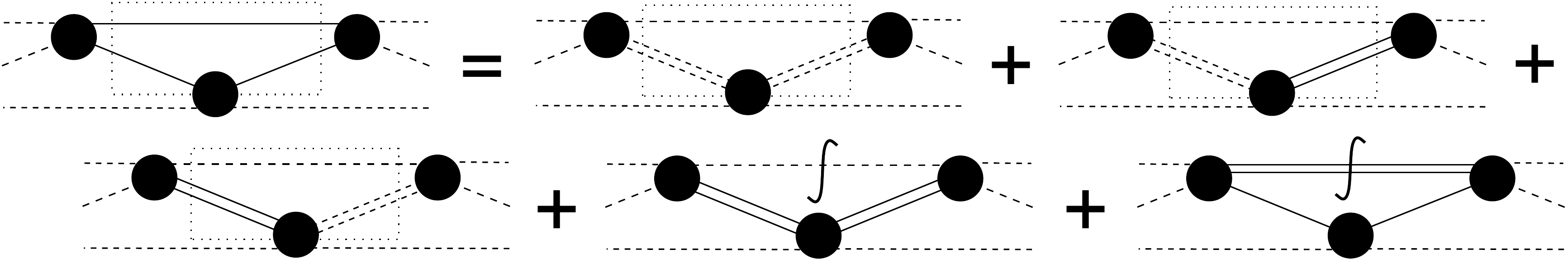}

\vspace{10pt}

\vspace{20pt}
\hspace {-480pt} {\Large (b)}

\vspace{30pt}

\hspace {-480pt} {\Large (c)}

\vspace{-70pt}

\includegraphics[scale=0.45]{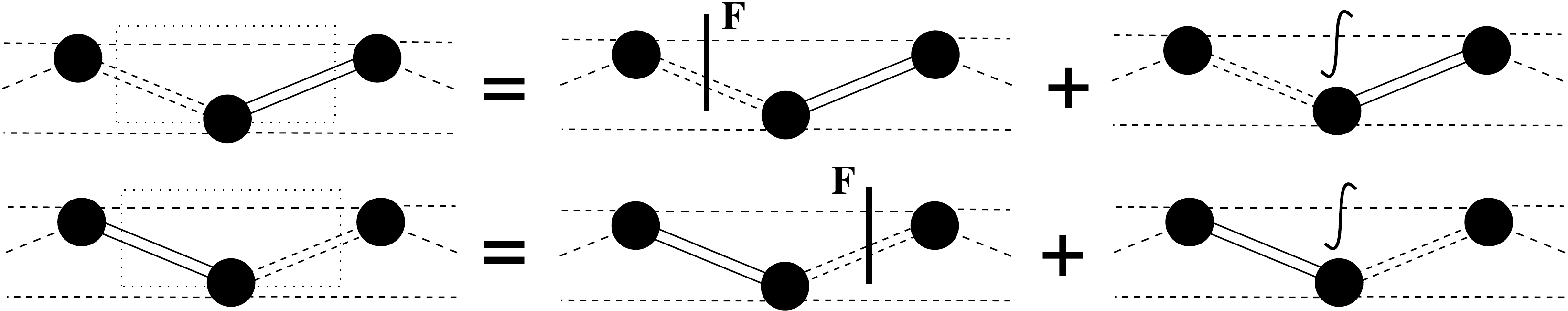}

\caption{Decomposition of \(\KthnL 3 \). See Fig.~\ref{fig:L3defs}(c)
for momentum labels. Double solid lines indicate non-singular
terms. On the top line these come from the $r^0$ contour circling
poles other than the single-particle pole, while for the diagonal lines
in the switch state the notation is as in Fig.~\ref{fig:decompK3}.
Double dashed lines represent the singular quantity $G^b$ sandwiched
between on-shell amplitudes,
as in Fig.~\ref{fig:decompK3}. The single dashed line within the loop
(top propagator) indicates the on-shell propagator-factor $1/(2\omega_r)$.
(a) Initial decomposition. Loop momenta inside dotted boxes are
summed, while those not in a box are integrated.
(b) and (c): Use of the sum-minus-integral identity, as indicated by
the vertical bars and factors of $F$, leaving a remainder which is
integrated. The vertical bar crosses the two propagators whose momenta
are  projected on-shell by $F$, 
so that the uncrossed propagator is the spectator.
$\Kdfth^{(3,u,u)}$ is given by the sum of the four terms containing
loop integrals (two in (a) and one each in (b) and (c).}

\label{fig:K33anal}
\end{center}
\end{figure}

We are thus left with \(\KthnL 3 \) 
[Eqs.~(\ref{eq:K3uudef},\ref{eq:MML3}) and Fig.~\ref{fig:L3defs}c].
As always, our method is to replace sums with integrals while 
keeping track of finite-volume remainders.
The analysis is shown diagrammatically in Fig.~\ref{fig:K33anal}.
The first step is to do the $r^0$ integral. Singular terms occur
only if the contour circles the $r^0=\omega_r$ pole; for the remainder
we can replace the sum over $\vec r$ with an integral (giving the
last term on the right-hand side in Fig.~\ref{fig:K33anal}a).
Thus to study the singular terms we can replace $\Delta(r)$ with 
$1/(2\omega_r)$ and set $r=[\omega_r,\vec r\,]$
(indicated by the dashed top line in the figure).
The sum over $\vec r$ runs over two potential singularities,
one in $\Delta(P\!-\!k\!-\!r)$ and the other in $\Delta(P\!-\!p\!-\!r)$.
To use the sum-minus-integral identity, 
we must pull out the double singularity 
(the first term on the right-hand side in Fig.~\ref{fig:K33anal}a),
leaving a remainder with at most single singularities.
To do so we follow the analysis of the previous subsection
[see Eqs.~(\ref{eq:Deltadecomp}-\ref{eq:Kdf2decom})], 
applied separately to the two propagators,
both of which are sandwiched between factors of $\K$.
This analysis can be applied {\em independently} 
to the contributions associated with each propagator, 
with each separated into
into an on-shell singular part and a divergence-free quantity.
This leads to the decomposition
\begin{equation}
 i \KthnL 3 {}   =  i \K i G i \K  iG [2 \omega L^3] i \K 
+ i \K i G i {\cal K}_{{\rm df},3}^{(2,u,u)}
+ i {\cal K}_{{\rm df},3}^{(2,u,u)} [1/(2\omega L^3)] iG [2\omega L^3]
+ {\cal R}
\,,
\label{eq:M3Lsub1}
\end{equation}
where the first three terms correspond to the first three terms
on the right-hand side of Fig.~\ref{fig:K33anal}a,\footnote{%
The appearance of $[2\omega L^3]$ and its inverse in the third 
but not the second term is due to the facts that $[1/(2\omega L^3)]$
appears on the right in the definition of $G$, Eq.~(\ref{eq:Gdef}),
and that $[2\omega L^3]$ does not commute with $G$.}
while ${\cal R}$ is the sum of the last two diagrams in the figure.
The only properties of ${\cal R}$ that we will need are that it is
an infinite-volume quantity (since the sum over $\vec r$ 
which it contains can be replaced by an integral)
and that it is a smooth function of its arguments.
An explicit form for ${\cal R}$ is not needed---Eq.~(\ref{eq:M3Lsub1}) 
serves as sufficient definition since all terms other than ${\cal R}$
are known.

As noted above, we can use the sum-minus-integral identity on
the two terms in Eq.~(\ref{eq:M3Lsub1}) containing a single factor of $G$.
Unpacking our abbreviated notation, the first such term can be written
\begin{multline}
\left[i \K i G i {\cal K}_{{\rm df},3}^{(2,u,u)}\right]_{p,\ell,m;k,\ell',m'}
 = 
\PV \int_{\vec r} \frac1{2\omega_r} 
i {\cal K}_{2;\ell,m;\ell_1,m_1}(\vec p)
i G^b_{\ell_1,m_1;\ell_2,m_2}(\vec p,\vec r)
i {\cal K}_{{\rm df},3;\ell_2,m_2;\ell'm'}^{(2,u,u)}(\vec r, q_r^*,\vec k,q_k^*)
\\
+
\left[\frac{1}{L^3} \sum_{\vec r} -\PV\int_{\vec r}\right]
i {\cal K}_{2;\ell,m;\ell_1,m_1}(\vec p)
\left(\frac{r^*}{q_p^*}\right)^{\ell_1}
\frac{i 4 \pi Y_{\ell_1,m_1}(\hat r^*) 
H(\vec p)H(\vec r)Y^*_{\ell_2,m_2}(\hat p^*)}
{2\omega_r 2 \omega_{pr}(E -  \omega_p - \omega_r - \omega_{pr} )} 
\left(\frac{p^*}{q_r^*}\right)^{\ell_2}
i {\cal K}_{{\rm df},3;r,\ell_2,m_2;k,\ell'm'}^{(2,u,u)}
\,.
\label{eq:singlesing}
\end{multline}
The integral (the last term in Fig.~\ref{fig:K33anal}b)
is combined with the corresponding integral
from the third term in Eq.~(\ref{eq:M3Lsub1})
(the last term in Fig.~\ref{fig:K33anal}c), and with ${\cal R}$,
to define ${\cal K}_{df,3}^{(3,u,u)}$.
This is the two-switch contribution to the continuum divergence-free
amplitude.
The sum-integral difference requires some adjustments to
allow the use of our identity. First we make the substitution
\begin{multline}
\sqrt{4\pi} Y^*_{\ell_2,m_2}(\hat p^*) (p^*/q^*_r)^{\ell_2} 
\mathcal K^{(2,u,u)}_{\mathrm{df}, 3; r,\ell_2,m_2; k,\ell',m'}
=
\sqrt{4\pi} Y^*_{\ell_2,m_2}(\hat p^*) (p^*/q^*_r)^{\ell_2} 
\mathcal K^{(2,u,u)}_{\mathrm{df}, 3; \ell_2,m_2; \ell',m'}
(\vec r, q_r^*,\vec k, q_k^*)
\\
\longrightarrow 
\sqrt{4\pi} Y^*_{\ell_2,m_2}(\hat p^*) 
\mathcal K^{(2,u,u)}_{\mathrm{df}, 3; \ell_2,m_2; \ell',m'}
(\vec r, p^*,\vec k, q_k^*)
\equiv
\mathcal K^{(2,u,u)}_{\mathrm{df}, 3; \ ; \ell',m'}
(\vec r, \vec p,\vec k, q_k^*)
\,.
\end{multline}
Here we are changing $q_r^*\to p^*$,
which is allowed because the difference between the old and new forms
is proportional to $p^{*2}-q_r^{*2}$, which cancels the singularity. 
We explain in Appendix~\ref{app:sumintegral} why the difference has 
this particular scaling. After this change, the sum over $\ell_2$ and $m_2$ can be done,
leading to a version of $\mathcal K^{(2,u,u)}_{\mathrm{df},3}$ which is
off shell on the left.\footnote{%
Our notation for this quantity,
$\mathcal K^{(2,u,u)}_{\mathrm{df}, 3; \ ; \ell',m'}
(\vec r, \vec p,\vec k, q_k^*)$
indicates through the absence of a subscript between the two
semi-colons that no angular decomposition of the outgoing 
coordinates is being done.}
At this stage we can drop the $H(\vec r)$ factor from the summand
of the sum-minus-integral term in Eq.~(\ref{eq:singlesing}),
since it is not needed to define the boosts, and the difference
$1-H(\vec r)$ cancels the singularity.
These manipulations bring the sum-minus-integral into a form where
we can apply our standard identity. 
In this way we find that the sum-minus-integral in Eq.~(\ref{eq:singlesing})
can be written
\begin{equation}
i \K i F 2 i \Kdfth^{(2,s,u)} \,,
\label{eq:firstGterm}
\end{equation}
where $K^{(2,s,u)}_{\mathrm{df}, 3}$ is defined by re-expanding
$K^{(2,u,u)}_{\mathrm{df}, 3}$ in spherical harmonics in $\vec r$.
This is shown by the first diagram on the right-hand side
in Fig.~\ref{fig:K33anal}b.
Specifically, the off-shell form 
of $K^{(2,u,u)}_{\mathrm{df}, 3}$ is expanded in harmonics
\begin{equation}
\sqrt{4\pi} Y^*_{\ell_1,m_1}(\hat r^*)
\mathcal K^{(2,s,u)}_{\mathrm{df}, 3;\ell_1,m_1;\ell',m'}
(\vec p, r^*,\vec k, q_k^*)
\equiv
\mathcal K^{(2,u,u)}_{\mathrm{df}, 3; \ ; \ell',m'}
(\vec r, \vec p,\vec k, q_k^*)
\,,
\end{equation}
and then put on shell and restricted to finite-volume momenta,
\begin{equation}
\mathcal K^{(2,s,u)}_{\mathrm{df}, 3;p,\ell_1,m_1;k,\ell',m'}
\equiv
\mathcal K^{(2,s,u)}_{\mathrm{df}, 3;\ell_1,m_1;\ell',m'}
(\vec p, q_p^*,\vec k, q_k^*)
\ \ 
[\vec p,\vec k \in (2\pi/L)\mathbb Z^3]\,.
\end{equation}
The superscript ``s'' once again indicates that the momentum
singled-out by the coordinate system on the left, here $\vec p$,
is one of the scattered outgoing particles.

We stress that the validity of Eq.~(\ref{eq:firstGterm}) 
requires two properties of ${\cal K}_{{\rm df},3}^{(2,u,u)}$.
First, it must be a smooth function of its arguments, for otherwise
there would be additional contributions to the sum-integral difference.
As discussed in the previous subsection, smoothness requires that
$G$ be defined including the factors of $(p^*)^{\ell_1+\ell_2}$.
Second, it must be divergence-free, and thus local in position-space,
so that the expansion in spherical harmonics of $\hat r^*$ is convergent.
This is one of the ways that our analysis forces us to use
divergence-free quantities, as announced in the introduction.

The other term in Eq.~(\ref{eq:M3Lsub1}) containing a single factor of $G$
can be analyzed in a similar fashion, leading to an integral plus
the finite-volume residue $2 i \Kdfth^{(2,u,s)} i F  i \K$.
This is shown in Fig.~\ref{fig:K33anal}c.
Here $\Kdfth^{(2,u,s)}$ is defined in an analogous way to
$\Kdfth^{(2,s,u)}$, but with the re-expansion in new coordinates occurring
for the incoming momenta.
As already noted, the integral is part of the non-singular remainder
which builds up $\Kdfth^{(3,u,u)}$.
Combining all elements, we finally reach
\begin{equation}
\label{eq:M33Lid}
i \KthnL 3  = i  \K i G i  \K  iG [2 \omega L^3]i  \K +  
i \K i F  2 i \Kdfth^{(2,s,u)}  + 2 i \Kdfth^{(2,u,s)} i F  i \K 
+ i \Kdfth^{(3,u,u)}\,.
\end{equation}
This has the desired form in which each
term is a product of on-shell,
infinite-volume quantities and kinematic factors. 

The result (\ref{eq:M33Lid}) and the similar 
decompositions in Eqs.~(\ref{eq:ApL3res}), (\ref{eq:AL3res})
and (\ref{eq:CLF03res}) can now be substituted
in Eq.~(\ref{eq:twoswitchC}) to obtain our final result for
the two-switch correlator $C_L^{(3)}$. The result is lengthy
and, at this stage, unilluminating. We hold off on making such
substitutions until we are working to all orders, in the next
subsection, for then the result simplifies.

%% file: remdiagrams.tex
\subsection{Two-to-two insertions: any number of switches}

\indent

\label{sec:remainingdiag}

In this section we sum all remaining contributions to the
finite-volume correlator containing only \(B_2\) kernels, 
allowing any number of switches in scattered pair.

\begin{figure}
\begin{center}

\includegraphics[scale=0.6]{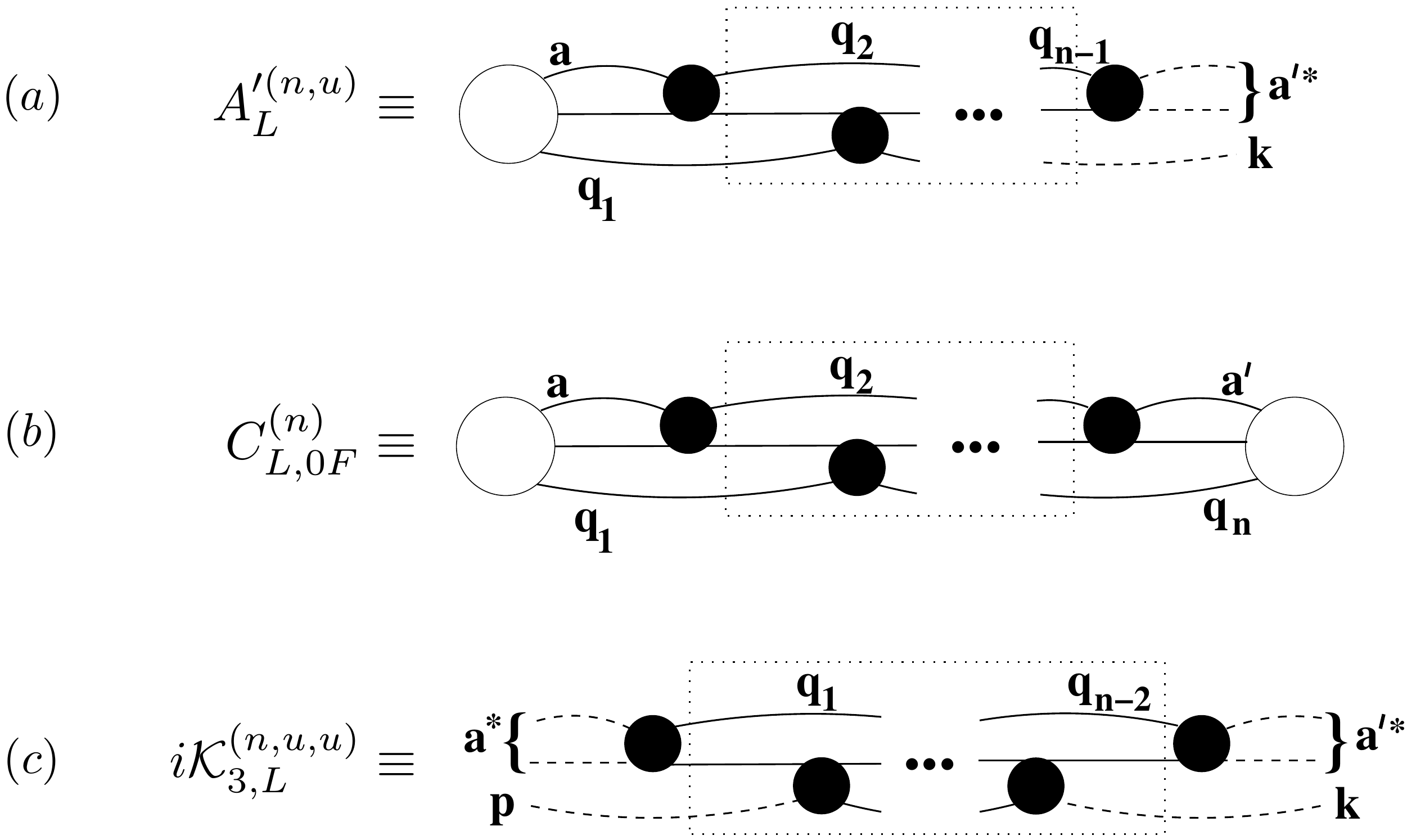}

\caption{Diagrammatic definitions of 
(a) \(\ApnL n {}\), 
(b) \(\CLF 0 n\), and (c) \(i\KthnL n {}\).}
\label{fig:Ldefs}
\end{center}
\end{figure}

The first step, as before, is to replace sums on two-particle
loops with integrals plus factors of $F$. This leads to the
appearance of 
\(\CLF 0 n\), \(\ApnL n {}\), \(\AnL n {}\) and \(\KthnL n {}\),
with $n\ge 3$, 
which are generalizations of the quantities found earlier.
Their definitions, sketched in Fig.~\ref{fig:Ldefs},
are
\begin{multline}
\label{eq:CL0Fndef}
 \CLF 0 n \equiv \frac{1}{4} 
\left[\prod_{m=1}^{n} \frac{1}{L^3} \sum_{\vec q_m}\right]
 \PV \int_{a} \PV \int_{a'}
\left[\prod_{m=1}^{n} \int_{q_m^0}\right]
 \sigma(q_1, a) \Delta(a) \Delta(P\!-\!q_1\!-\!a) 
i \mathcal K_{2,\mathrm{off}}(a, P\!-\!q_1\!-\!a, -q_{2}) 
\\ 
\times
\Delta(q_1) \Delta(P\!-\!q_{1}-q_2) 
i \mathcal K_{2, \mathrm{off}}(q_1, P\!-\!q_{1}\!-\!q_2, -q_{3}) 
\times \cdots \times
 \Delta(q_{n-1})\Delta(P\!-\!q_n\!-\!q_{n-1}) 
\\ 
\times i \mathcal K_{2,\mathrm{off}}(q_{n-1}, P\!-\!q_n\!-\!q_{n-1},-a') 
\Delta(q_n) \Delta(P\!-\!q_n\!-\!a') \Delta(a') \sigma^\dagger(q_n,a')\,,
\end{multline} 
\begin{multline}
 \ApnL n {;\ell',m'}(\vec k, a'^*) \sqrt{4 \pi} Y_{\ell',m'}(\hat a'^*)
\equiv \frac{1}{2} 
\left[\prod_{m=1}^{n-1} \frac{1}{L^3}  \sum_{\vec q_m} \right] 
\PV \int_{a}
\left[\prod_{m=1}^{n-1} \int_{q_m^0} \right]
\sigma(q_{1},a) \Delta(a) \Delta(P\!-\!q_{1}\!-\!a) 
\\ 
\times i \mathcal K_{2, \mathrm{off}}(a,P\!-\!q_{1}\!-\!a, -q_{2}) 
\Delta(q_{1}) \Delta(P\!-\!q_{1}\!-\!q_{2}) 
i \mathcal K_{2, \mathrm{off}}(q_{1}, P\!-\!q_{1}\!-\!q_{2}, -q_{3})
\Delta(q_{2}) \Delta(P\!-\!q_{3}\!-\!q_{2}) \times \cdots
 \\ 
\cdots
\times i \mathcal K_{2, \mathrm{off}}(q_{n-2}, P\!-\!q_{n-2}\!-\!q_{n-1}, -p)
\Delta(q_{n-1}) \Delta(P\!-\!p\!-\!q_{n-1})  
i \mathcal K_{2,\mathrm{off}}(q_{n-1}, P\!-\!p\!-\!q_{n-1}, -a') \,,
\label{eq:ApLndef}
\end{multline}
with $\AnL n{}$ defined analogously by reflection, and
\begin{multline}
4 \pi Y^*_{\ell', m'}(\hat a^*) 
i \mathcal K_{3,L; \ell', m'; \ell,  m}^{(n,u,u)}(\vec p, a^*,\vec k, a'^*) 
Y_{\ell, m}(\hat a'^*) 
\equiv 
\left[ \prod_{m=1}^{n-2} \frac{1}{L^3} \sum_{\vec q_m} \right] 
\left[\prod_{m=1}^{n-2} \int_{q_m^0}\right]
i \mathcal K_{2, \mathrm{off}}(a, P\!-\!p\!-\!a,-q_{1}) 
\Delta(P\!-\!p\!-\!q_{1}) 
\\
\times i \mathcal K_{2, \mathrm{off}}(p,P\!-\!p\!-\!q_{1},-q_{2}) 
\Delta(q_{2}) \Delta(P\!-\!q_{1}\!-\!q_{2}) 
\times \cdots 
\times \Delta(q_{n-2}) \Delta(P\!-\!k\!-\!q_{n-2}) 
i \mathcal K_{2,\mathrm{off}}(q_{n-2}, P\!-\!p\!-\!q_{n-2},-a')
\,,
\label{eq:Kndef}
\end{multline}
The above definitions give partially off shell versions of $\ApnL n {}$
and \(\KthnL n {}\). The on-shell versions are defined as usual by
\begin{equation}
\ApnL n {;k,\ell',m'} \equiv \ApnL n {;\ell',m'}(\vec k, q_k^*)
\ {\rm and}\ 
\mathcal K_{3, L;p, \ell', m';k,\ell,m}^{(n,u,u)}
\equiv
\mathcal K_{3, L; \ell', m'; \ell, m }^{(n,u,u)}
(\vec p, q_p^*,  \vec k, q_k^*)
\ {\rm with}\ 
\vec p,\vec k\in (2\pi/L)\mathbb Z^3
\,,
\end{equation}
with an analogous definition for $\AnL n {}$.

It is simpler to write down the all-orders form of
\begin{equation}
C_L^{[B_2]}= \sum_{n=0}^\infty C^{(n)}_{L} \,,
\end{equation}
than it is to write down $C^{(n)}_L$ itself.
The superscript ``$[B_2]$'' here is a reminder that no
$B_3$ kernels have yet been included.
We find
\begin{equation}
C^{[B_2]}_{L}  = 
\sum_{n=0}^\infty \CLF 0 n 
- (2/3) \sigma^* \frac{i F}{2 \omega  L^3} \sigma^{\dagger *} 
+ 
\left[ \sum_{i=0}^\infty  \ApnL i {} \right ] 
\big [ \mathcal A \big ] 
\left\{\sum_{n=0}^\infty \left( i \mathcal K_{3,L}^{(u,u)}  
\big[\mathcal A \big]\right)^n\right\}
\left[ \sum_{k=0}^\infty \AnL k {} \right] \,.
\label{eq:Cno3MLres}
\end{equation}
Here we have made the definitions
\begin{gather}
\AnL 0 {} \equiv \sigma^{*\dagger},\ \ \AnL 1 {} \equiv
A^{(1,u)},\ \ \ApnL 0 {} \equiv \sigma^*,\ \ \ApnL 1 {} \equiv
A'^{(1,u)},\\ i \KthnL 2 {} \equiv i \mathcal K_{3 }^{(2,u,u)},
\ \ \CLF 0 0 \equiv  C_\infty^{(0)}, \ \ \CLF 0 1 \equiv
C_\infty^{(1)}\,,
\end{gather}
in which infinite-volume quantities are relabeled as though they 
have volume dependence,
in order to simplify the form of the result.
We have also introduced
\begin{equation}
\mathcal K_{3,L}^{(u,u)}  \equiv 
\sum_{n=2}^\infty \mathcal K_{3 ,  L}^{(n,u,u)} \,.
\end{equation}
The way in which (\ref{eq:Cno3MLres}) arises should be clear
by generalizing the discussion leading
to the results for $C_L^{(0)}+C_L^{(1)}$ [Eq.~(\ref{eq:CL0plusCL1})], 
$C_L^{(2)}$ [Eq.~(\ref{eq:C2Lint})] 
and $C_L^{(3)}$ [Eq.~(\ref{eq:twoswitchC})] above.
In words, one has endcaps, involving any number of switches,
connected to any number of
finite-volume three-particle scattering amplitudes with 
intermediate factors of $\big[{\cal A}\big]$.
Recall that $\big[{\cal A}\big]$, defined in Eq.~(\ref{eq:Adef}),
is closely related to the two-particle finite-volume propagator.

To bring Eq.~(\ref{eq:Cno3MLres}) into a useful form we need
identities relating all quantities with \(L\) subscripts to
infinite-volume quantities and finite-volume remainders.

We first consider \(\ApnL n {}\), and for now seek only to rewrite
this in terms of \(\KthnL j {}\) as well as infinite-volume quantities. 
Our basic strategy is to move from left to right, replacing sums with
integrals plus sum-integral differences.
Then each sum then has a summand with only one singular factor, which we know how to
handle. All double
singularities are removed, because each sum is adjacent to an integral which removes the singularities in one of the two switch-states containing the summed coordinate.

We describe in some detail how the process works for $q_1$ and then state
the final result. The sum over $\vec q_1$ has a potentially
singular summand in the propagator $\Delta(P\!-\!q_1\!-\!q_2)$
(the singularity in $\Delta(P\!-\!q_1\!-\!a)$ being removed by the the $\PV$ integral).
For this singularity to be present, both $q_1$ and $q_2$ must be
on shell, so we must first do the $q_1^0$ and $q_2^0$ integrals
and pick out the particle poles. In this singular term we can replace the
sum over $\vec q_1$ with an integral plus the sum-integral difference.
Generalizing the analysis given earlier, we find 
that the sum-integral difference gives
\begin{equation}
 2 A'^{(1,s)} \frac{i F}{2 \omega L^3} i \KthnL {n-1} {} \,.
\end{equation}
What remains are terms involving the $\PV$ integral over $\vec q_1$,
and these can be repackaged into a quantity with
exactly the form of $\ApnL n {}$ [Eq.~(\ref{eq:ApLndef})] except that
the sum-integral $1/L^3\sum_{\vec q_1} \PV \int_{q_1}$ is replaced by
the four-momentum integral $\PV \int_{q_1}$.
In this way the finite-volume residue from the sum over $\vec q_1$ has been determined.

We now repeat this analysis for $q_2$, finding that the $F$ term is
\begin{equation}
2 A'^{(2,s)} \frac{i F}{2 \omega L^3} i \KthnL {n-2} {} \,,
\end{equation}
while the remainder has the form of $\ApnL n{}$
with sum-integrals over both $q_1$ and $q_2$ replaced by integrals.
Continuing in this way, we deduce 
\begin{equation}
\label{eq:ApnLinML}
\ApnL n {} = \sum_{i=1}^{n-2}  2 A'^{(i,s)} \frac{i F}{2 \omega  L^3} 
i \KthnL {n-i} {} + 2 A'^{(n-1,s)} i F i \K + A'^{(n,u)} \,.
\end{equation}
This result holds for $n>2$, and agrees with Eq.(\ref{eq:ApL3intb}) for $n=3$.
The infinite-volume quantity $A'^{(n,u)}$ is given by the
same expression as $\ApnL n {}$ [Eq.~(\ref{eq:ApLndef})] except that all
sums are replaced with $\PV$ integrals, with the order being
\begin{equation}
\PV \int_{q_n} \cdots \PV \int_{q_1} \PV \int_{a} \,.
\end{equation}
The quantities $A'^{(n,s)}$ are defined in terms of $A'^{(n,u)}$
by changing variables exactly as for $A'^{(1,s)}$ 
[see Eq.~(\ref{eq:Acoordchange})].

The analysis for $\AnL n {}$ is the mirror image of that for $\ApnL n {}$,
so that the sums are now dealt with moving from right to left.
The result is
\begin{equation}
\label{eq:AnLinML}
\AnL n {} = \sum_{i=1}^{n-2}  
i \KthnL {n-i} {}\frac{i F}{2 \omega  L^3} 2 A^{(i,s)} 
 + i \K iF 2 A^{(n-1,s)}  + A^{(n,u)} \,.
\end{equation}

We treat \(\CLF 0 n\) in a similar fashion.
Here we can choose to work from left to right or {\em vice versa}---both
choices lead to single singularities for each loop sum.
As above, our convention is to work from left to right.
Since the analysis follows that for $A'^{(n,u)}_L$ very closely,
we simply quote the result,
\begin{align}
\CLF 0 n &= 
\sum_{i=1}^{n-2} 2 A'^{(i,s)} \frac{i F}{2 \omega L^3} \AnL {n-i} {} 
+ 2 A'^{(n-1,s)} \frac{i F}{2 \omega L^3} A^{(1,u)} 
+ \CI n \,,
\\
&=
\sum_{i=1}^{n-1} 2 A'^{(i,s)} \frac{i F}{2 \omega L^3} \AnL {n-i} {} 
+ \CI n \,.
\label{eq:CnLinML}
\end{align}
To obtain the second form we have used 
$\AnL 1 {}\equiv A^{(1,u)}$.
The quantity $\CI n$ takes the same form as $\CLF 0 n$
[Eq.~(\ref{eq:CL0Fndef})] except that all sums are replaced by integrals,
ordered as
\begin{equation}
\PV\int_{q_n} \cdots \PV\int_{q_1}\PV \int_{a} \PV \int_{a'} \,.
\end{equation}
The result (\ref{eq:CnLinML}) is valid for \(n > 1\),
and agrees with Eqs.~(\ref{eq:CL0F2final}) and (\ref{eq:CL3resint}) for $n=2$
and $3$, respectively.

\bigskip
The next step towards simplifying the result for $C^{[B_2]}_{L}$,
Eq.~(\ref{eq:Cno3MLres}), is to perform the sums over the number of switches.
In particular, from Eq.~(\ref{eq:ApnLinML}) we find
\begin{equation}
\label{eq:ApLures}
A'^{(u)}_L \equiv \sum_{n=1}^\infty \ApnL n {}
= A'^{(u)} + 2 A'^{(s)} 
\left[ iF i\K + \frac{iF}{2\omega L^3} i {\cal K}_{3,L}^{(u,u)}
\right]\,,
\end{equation}
where
\begin{equation}
A'^{(u)}\equiv \sum_{n=1}^\infty A'^{(n,u)}
\ {\rm and}\
A'^{(s)}\equiv \sum_{n=1}^\infty A'^{(n,s)}\,.
\end{equation}
Similar definitions will be used for $A^{(u)}$ and $A^{(s)}$,
and also for the amplitudes corresponding to the third
choice of momentum assignments, $A'^{(\tilde s)}$
and $A^{(\tilde s)}$. The latter were introduced in
Eq.~(\ref{eq:stildedef}).

An analogous result to Eq.~(\ref{eq:ApLures}) holds for the other endcap
\begin{equation}
\label{eq:ALures}
A^{(u)}_L  
\equiv \sum_{n=1}^\infty \AnL n {}
= A^{(u)} + 
\left[ i\K iF +  i {\cal K}_{3,L}^{(u,u)}\frac{iF}{2\omega L^3}
2 A^{(s)} 
\right]\,,
\end{equation}
while for the correlator sum we obtain
\begin{equation}
\label{eq:CL0Fsum}
\sum_{n=0}^\infty C_{L,0F}^{(n)}
= C^{[B_2]}_\infty + 2 A'^{(s)} \frac{iF}{2\omega L^3} A^{(u)}_L
\,,
\end{equation}
where
\begin{equation}
C^{[B_2]}_\infty = \sum_{n=0}^\infty C_\infty^{(n)}
\,,
\end{equation}

We can now express $C^{[B_2]}_{L}$ in terms of infinite-volume
quantities together with ${\cal K}_{3,L}^{(u,u)}$. This requires substituting
Eqs.~(\ref{eq:ApLures}), (\ref{eq:ALures}) and (\ref{eq:CL0Fsum}) 
into Eq.~(\ref{eq:Cno3MLres}) and using the following identities
\begin{align}
\left( iF i\K + \frac{iF}{2\omega L^3} i{\cal K}_{3,L}^{(u,u)}\right)
\big[\mathcal A \big] \sum_{n=0}^\infty 
\left(i\mathcal K_{3,L}^{(u,u)} \big[\mathcal A \big]\right)^n
&=
\big[\mathcal A \big] \sum_{n=0}^\infty 
\left(i \mathcal K_{3,L}^{(u,u)} \big[\mathcal A \big]\right)^n
- \frac{iF}{2\omega L^3}
\\
&=
\big[\mathcal A \big] \sum_{n=0}^\infty 
\left(i\mathcal K_{3,L}^{(u,u)} \big[\mathcal A \big]\right)^n
\left( i\K iF + i{\cal K}_{3,L}^{(u,u)}\frac{iF}{2\omega L^3}\right)
\,.
\end{align}
After some algebra, we obtain a relatively simple form
\begin{equation}
\label{eq:CLB2res}
C_L^{[B_2]} = C^{[B_2]}_\infty + \delta C^{[B_2]}_\infty +
A'^{[B_2]} \left[
-\frac23\frac{iF}{2\omega L^3}
+\big[\mathcal A \big] \sum_{n=0}^\infty 
\left(i \mathcal K_{3,L}^{(u,u)} \big[\mathcal A \big]\right)^n
\right] A^{[B_2]}
\,,
\end{equation}
where
\begin{equation}
A'^{[B_2]} = \sigma^* + A'^{(u)}+A'^{(s)}+A'^{(\tilde s)}\,,
\quad
A'^{[B_2]} = \sigma^{\dagger *} + A^{(u)}+A^{(s)}+A^{(\tilde s)}
\,,
\end{equation}
and
\begin{equation}
\label{eq:deltaCdef}
\delta C^{[B_2]}_\infty =
\frac23 A'^{[B_2]} \frac{iF}{2\omega L^3} (A^{(u)}-A^{(s)})
+
\frac23 (A'^{(u)}-A'^{(s)}) \frac{iF}{2\omega L^3} 
\sigma^{\dagger *}
\,.
\end{equation}

Several comments are in order.
First, we observe that
summing over all switches has led to a dramatic simplification
in the expression for the correlator. This can be seen, for example,
by comparing even the one-switch expression (\ref{eq:C2def4}) to
Eq.~(\ref{eq:CLB2res}).
Second, to obtain Eq.~(\ref{eq:CLB2res}) we have made use of
the fact, explained after Eq.~(\ref{eq:stostilde}), that,
within our derivation thus far, 
superscripts $(s)$ and $(\tilde s)$ are interchangeable.
This allows us to write the result in terms of endcaps,
$A'^{[B_2]}$ and $A^{[B_2]}$, which are symmetric under particle
interchange.
We stress that this symmetrization occurs only when working to
all orders in the number of switches, since it requires combining
terms with different numbers of switches.
Our third comment also concerns symmetrization, or rather the its absence in Eq.~(\ref{eq:deltaCdef}).
Recall that particle-interchange symmetry was violated 
when we chose to analyze the loops in $C_{L,0F}^{(n)}$ moving from
left to right, since this led to $(s)$ quantities always being to
the left of those with superscripts $(u)$.
Forcing the endcaps into symmetric form
leads to the remainder $\delta C^{[B_2]}_\infty$.
Note that in the terms involving a $(u)-(s)$ difference, we can
freely interchange $(s)$ and $(\tilde s)$, and we have used this
freedom to choose both terms to involve $(s)$. 
Although $\delta C_\infty$ appears to be a finite-volume term
(since it contains factors of $F$),
in fact, as we show below, it can be rewritten as an infinite-volume quantity.
This means that $\delta C_\infty^{[B_2]}$ can be absorbed into an alternative 
infinite-volume quantity, used in place of $C_\infty^{[B_2]}$. Since other contributions of this type arise in the analysis that follows, we delay our definition of the replacement until Eq.~(\ref{eq:CB2rho}) below. 
We note that our job is not done, because the result
(\ref{eq:CLB2res}) still contains the asymmetric three-particle
finite-volume scattering amplitude
$\mathcal K_{3,L}^{(u,u)}$. We return shortly to the task
of rewriting this in terms of infinite-volume quantities.

First, however, we rewrite $\delta C_\infty^{[B_2]}$ in a manifestly
infinite-volume form. We show how this works for
the first term in (\ref{eq:deltaCdef}) from which the generalization to
the second term is immediate.
The steps are as follows:
\begin{align}
\frac23 A'^{[B_2]}& \frac{iF}{2\omega L^3} (A^{(u)}\!-\!A^{(s)}) 
\nonumber\\
&=\frac13\frac{1}{L^3}\! \sum_{\vec k}
\left[\frac{1}{L^3} \!\sum_{\vec a} - \PV \!\int_{\vec a} \right] 
A'^{[B_2]}(\vec k, \vec a)
\frac{iH(\vec k) H(\vec a)H(\vec b_{ka})}
{2 \omega_k 2 \omega_a 2 \omega_{ka}
(E\!-\!\omega_k\!-\!\omega_a\!-\!\omega_{ka})}
[A^{(u)}(\vec k,\vec a)\!-\!A^{(u)}(\vec a,\vec k)]
\\
&=-\frac13\int_{\vec k} \PV \!\int_{\vec a}
A'^{[B_2]}(\vec k, \vec a)
\frac{iH(\vec k) H(\vec a)H(\vec b_{ka})}
{2 \omega_k 2 \omega_a 2 \omega_{ka}
(E\!-\!\omega_k\!-\!\omega_a\!-\!\omega_{ka})}
[A^{(u)}(\vec k,\vec a)\!-\!A^{(u)}(\vec a,\vec k)]
\label{eq:rhosteptwo}
\\
&= 
\int_{\vec k} A'^{[B_2]}(\vec k)  \frac{i\rho(\vec k) }{3\omega_k }
[A^{(u)}(\vec k) \!-\!A^{(s)}(\vec k) ]
\equiv 
A'^{[B_2]}  \frac{i\rho }{3\omega }
[A^{(u)} \!-\!A^{(s)} ]
\,,
\label{eq:deltaCinftyanal}
\end{align}
where in an abuse of notation, in the last line we have introduced the shorthand that integration over $\vec k$ is implicit for a product involving $\rho$.

In the first step, we use the sum-minus-integral
identity in reverse, as well as the definition of
$A^{(s)}$ [see Eq.~(\ref{eq:Acoordchange})].
The momenta $\vec k$ and $\vec a$ are on shell, but,
in general, the third four-momentum, $b_{ka}=P-k-a$, is now off shell, 
Thus amplitudes are not invariant under the full particle-interchange symmetry.
Nevertheless, $A'^{[B_2]}(\vec k,\vec a)$ remains symmetric under
the interchange $\vec k\leftrightarrow \vec a$,
while the $A^{(u)}-A^{(s)}$ becomes a term which is
manifestly antisymmetric under this interchange.
Since the remaining terms are symmetric, 
the entire summand/integrand is antisymmetric.
This observation allows us to drop the double sum, 
since a symmetric sum over an antisymmetric summand clearly vanishes.
The sum over $\vec k$ can now be replaced by an integral, since the
$\PV$ integral over $\vec a$
leads to a smooth function of $\vec k$. 
At this stage we obtain the second form of the right-hand side,
Eq.~(\ref{eq:rhosteptwo}).
The final step is to notice that, if an $i\epsilon$ pole prescription 
were used, then the double integration would also vanish by symmetry.
Thus it is only the $\rho$ term in the definition of $\PV$ integration,
Eq.~(\ref{eq:PVtildedef}), that survives.

Applying a similar analysis to the second term in Eq.~(\ref{eq:deltaCdef}), 
we find, in total,
\begin{equation}
\delta C^{[B_2]}_\infty 
=
A'^{[B_2]} \frac{i\rho }{3\omega }
[A^{(u)} \!-\!A^{(s)} ]
+
[A'^{(u)} \!-\!A'^{(s)} ]
\frac{i\rho }{3\omega }
\sigma^{\dagger *} 
\,,
\end{equation}
where, as above, integration over $\vec k$ is implicit.
This is a manifestly infinite-volume quantity depending only
on on-shell (but not symmetric) amplitudes.

\begin{figure}
\begin{center}
\includegraphics[scale=0.4]{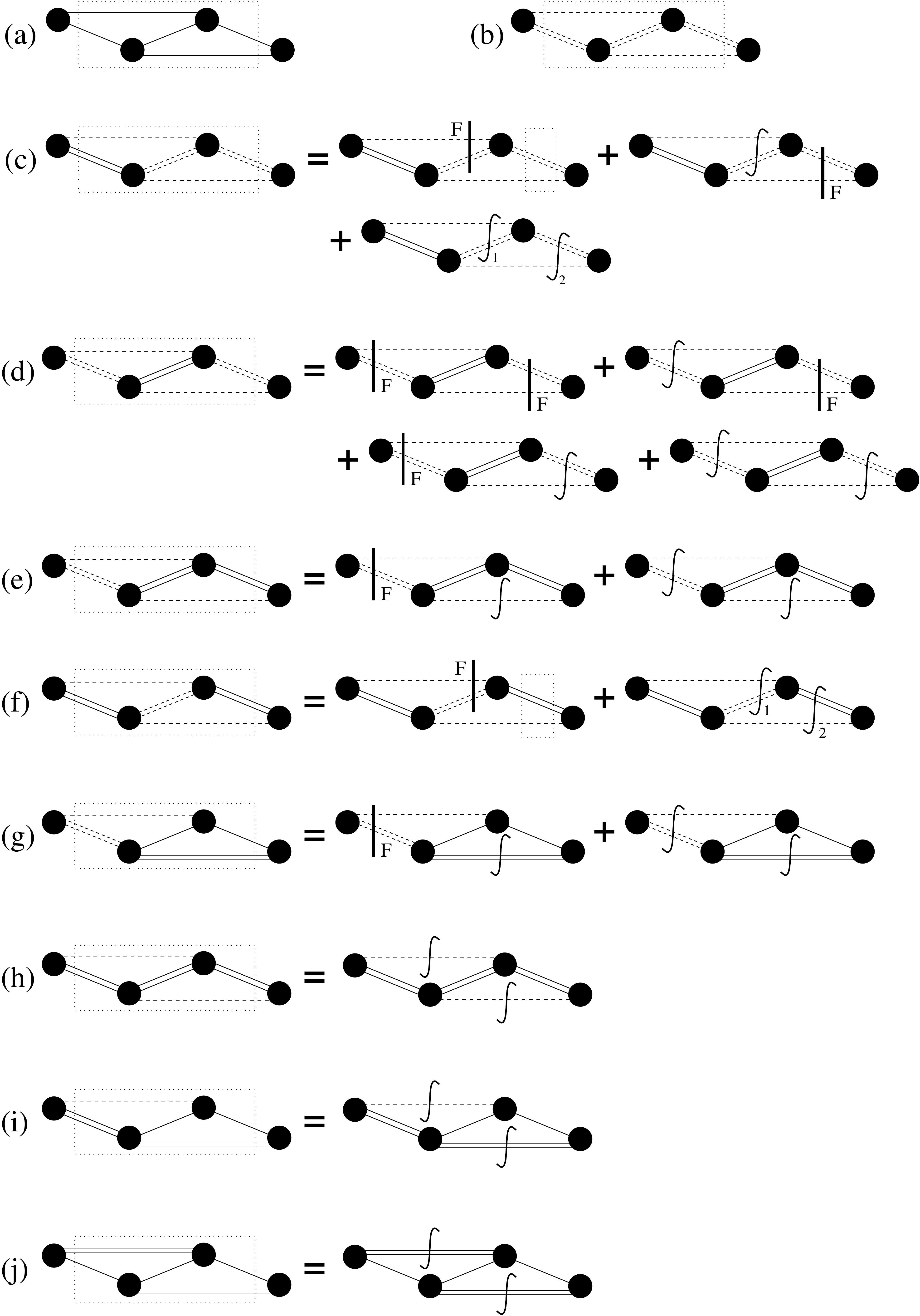}

\caption{Decomposition of  \(\mathcal K_{3 , L}^{(4,u,u)} \).
All external propagators are dropped, and the notation of 
Figs.~\ref{fig:decompK3} and \ref{fig:K33anal} is used.
(a) \(\mathcal K_{3 , L}^{(4,u,u)} \) itself [see Eq.~(\ref{eq:Kndef})];
(b) the most singular term (with three singular propagators);
(c) and (d): terms with two singular propagators and their decompositions;
(e), (f) and (g): 
terms with one singular propagator and their decompositions;
(h), (i) and (j): non-singular terms.
Terms in the decompositions are always ordered from most to least singular.
The treatment of loop momenta is indicated explicitly: they
are either summed (dashed box), integrated (integral sign) or
the sum-minus-integral identity is used (factor of $F$).
Where the order of integrals matters it is shown explicitly.}
\label{fig:K34anal}
\end{center}
\end{figure}

\bigskip
The final identity we require is that
for \(\mathcal K_{3 , L}^{(n,u,u)} \).
The identities for \(n=2\) and $3$ are given, respectively,
by Eqs.~(\ref{eq:Kdf2decom}) and (\ref{eq:M33Lid}).
To understand how the pattern generalizes to arbitrary $n$ it is useful
to first work out explicitly the result for $n=4$, 
since new effects occur at this order.
The decomposition of $\mathcal K_{3 , L}^{(4,u,u)}$ is shown
diagrammatically in Fig.~\ref{fig:K34anal}. Here we are using a
stripped-down diagrammatic notation in which external lines and
momentum labels are implicit. The basic method, however, is exactly
as used earlier for $n=2$ and $3$: (i) do the time-component integrals
over the loop momenta, and separate the result into on-shell particle
contributions and the remainders; (ii) separate each of the remaining
``diagonal'' propagators and their attached factors of $\K$ into
a singular part (containing $G^b$) and the non-singular remainder;
(iii) pull out the most singular term; 
(iv) analyze the remainder by converting sums into integrals where
possible, which in some cases leads to residues containing factors of $F$.
The key point is that after the most singular term has been subtracted,
there is always at least one choice of ordering of momentum sums which
allows the use of the sum-minus-integral identity at each stage.
For most terms in the decomposition there is either a single such choice
or the order is unimportant. However, at $n=4$ we first encounter a
case where there is a significant choice of ordering to be made.
As $n$ increases there are more such cases and we need a convention
for how to deal with them.

We now work through the different contributions to 
$\mathcal K_{3 , L}^{(4,u,u)}$ in some detail, starting from the
most singular and working to the least.
We recall the notation [from Eq.~(\ref{eq:Kndef}) and Fig.~\ref{fig:Ldefs}c] 
that $q_1$ is the leftmost loop momentum and $q_2$ the rightmost.
The most singular term is that shown in Fig.~\ref{fig:K34anal}b,
and gives the contribution
\begin{equation}
\textrm{(b)} = i \K i G i \K i G i \K i G [2 \omega L^3] i \K
\,.
\end{equation}
This term must be left as a sum (which is implicit in our matrix notation).

Contributions with two singular propagators are those of
Fig.~\ref{fig:K34anal}c, its ``reflection'' (not shown) in which the rightmost
diagonal propagator is non-singular, and Fig.~\ref{fig:K34anal}d.
The decomposition of the first of these is also shown in
Fig.~\ref{fig:K34anal}c. We must begin with $q_1$ since the $q_2$
sum runs over two singular propagators.
We first convert the $q_1$ sum into an $F$-insertion plus an integral.
For the $F$-term this is as far as we can go,
since the $q_2$ sum runs over singularities in both $F$ and the
propagator. 
For the integral over $q_1$ we can repeat the $F$ plus integral
decomposition for $q_2$.
Note that in the resulting double integral the order of integration
is important.
The net result is that there are three terms, each with different
levels of singularity. 
The doubly-singular term gives
\begin{equation}
\textrm{(c) [doubly singular]} = 
2 i \Kdfth^{(2,u,s)} \frac{i F}{2\omega L^3} i\K i G [2 \omega L^3] i \K
\,,
\end{equation}
while the term with one singularity contributes as
\begin{equation}
\textrm{(c) [singly singular]} \subset
2 i \Kdfth^{(3,u,s)} {i F} i\K
\,,
\end{equation}
and the non-singular term contributes as
\begin{equation}
\textrm{(c) [non-singular]} \subset
i \Kdfth^{(4,u,u)}
\,.
\end{equation}
The reflected diagram is decomposed similarly.

The decomposition of the remaining term with two singular propagators
is shown in Fig.~\ref{fig:K34anal}d. Here, since the singular
propagators are separated, the sum-integral identity can be applied
to each independently. Thus there are four terms in the decomposition.
The doubly-singular one is
\begin{equation}
\textrm{(d) [doubly singular]} = 
i \K i F 4 i \Kdfth^{(2,s,s)} i F i \K \,.
\end{equation}
Note that here both $(u)$'s have been switched to $(s)$'s.
Each switch comes with a factor of 2, leading to the overall factor of 4.
The singly-singular terms contribute to
$2 i \Kdfth^{(3,u,s)} {i F} i\K$ and $i\K iF 2 i \Kdfth^{(3,s,u)}$,
while the non-singular terms contribute to $i \Kdfth^{(4,u,u)}$.

There are three diagrams containing one singular propagator:
Fig.~\ref{fig:K34anal}e, its reflection, and Fig.~\ref{fig:K34anal}f.
In the first, the sum over $q_2$ can be immediately converted to an
integral, since the summand is non-singular. For $q_1$ we obtain
the usual $F$-term plus integral. The former gives rise to another
contribution to $i\K i F 2 i \Kdfth^{(3,s,u)}$,
while the latter contributes to $i \Kdfth^{(4,u,u)}$.
Analogous results hold for the reflection of Fig.~\ref{fig:K34anal}e.

The diagram of Fig.~\ref{fig:K34anal}f leads to a new effect.
Here we can use the sum-integral identity either on
$\vec q_1$ or $\vec q_2$. Our convention (as above) is to work from
left to right when there is such a choice. This gives
the singly-singular term
\begin{equation}
\textrm{(f) [singly singular]} = 
2 i \Kdfth^{(2,u,s)} \frac{i F}{2 \omega L^3} i \Kdfth^{(2,u,u)} \,,
\end{equation}
where our convention has led to the $(s)$ being on the left side of the
$F$, rather than on the right.
The non-singular term contributes to $i \Kdfth^{(4,u,u)}$.
Here our convention leads to a definite (left to right) ordering of the
$\PV$ integrals.

Another new feature of the $n=4$ analysis is the appearance of singular
contributions in which one of the 
$q_j^0$ integrals does not circle the particle pole.
The corresponding diagrams are Fig.~\ref{fig:K34anal}g and its reflection.
The decomposition exactly follows that of Fig.~\ref{fig:K34anal}e.

Finally, we reach the completely non-singular contributions,
where sums can be immediately converted to integrals.
There are four such diagrams, Fig.~\ref{fig:K34anal}h,
its reflection, Fig.~\ref{fig:K34anal}i and Fig.~\ref{fig:K34anal}j.
These all contribute to $i \Kdfth^{(4,u,u)}$.

Adding all contributions we find the total result
\begin{multline}
 i \KthnL 4 {} = 
i \K i G i \K i G i \K [i G 2 \omega L^3] i \K 
+
i \K i G [2\omega L^3] i \K \frac{i F}{2\omega L^3} 2 i \Kdfth^{(2,s,u)} 
+ 
2 i \Kdfth^{(2,u,s)} \frac{i F}{2 \omega L^3} i \K i G [2 \omega L^3 ] i \K 
\\
+ i \K i F 4 i \Kdfth^{(2,s,s)} i F i \K 
+ 2 i \Kdfth^{(3,u,s)} i F i \K 
+ i\K iF 2 i \Kdfth^{(3,s,u)}
+ 2 i \Kdfth^{(2,u,s)} \frac{i F}{2 \omega L^3} i \Kdfth^{(2,u,u)} 
+ i  \Kdfth^{(4,u,u)} \,,
\label{eq:K44decomp}
\end{multline}
where we have ordered terms in decreasing strength of divergence.
The only aspect of this result not explained above is that
contributions combine properly to give the
quantities $\Kdfth^{(3,u,s)}$ and $\Kdfth^{(3,s,u)}$ in the fifth and
sixth terms, respectively. 
For example, the $\Kdfth^{(3,u,s)}$ term receives 
the required four contributions (see Fig.~\ref{fig:K33anal})
from diagrams (c), (d) and the reflections of (e) and (g).
One can demonstrate that the correct contributions occur in all cases
by observing that (i) the result (\ref{eq:K44decomp}) provides
a complete classification of possible divergence structures and
(ii) that expanding out each term in (\ref{eq:K44decomp}) leads to
a unique set of contributions each of which is necessarily present in the
decomposition of $\KthnL 4 {}$.
Finally, we note that the non-singular term in Eq.~(\ref{eq:K44decomp}),
$\Kdfth^{(4,u,u)}$, is simply defined as the sum of
contributions from all the diagrams in Fig.~\ref{fig:K34anal}
(plus appropriate reflections) that contain only loop {\em integrals}. 

We are now ready to explain the result for general \(i \KthnL n{}\).
What arises are sequences alternating between one of the
${\cal K}$'s,
\begin{gather}
 i \K, i\Kdfth^{(j,u,u)},2i \Kdfth^{(j,s,u)},
2 i \Kdfth^{(j,u,s)}\ {\rm and}\ 4 i \Kdfth^{(j,s,s)},
\end{gather}
and one of
\begin{gather}
\frac{i F}{2\omega L^3} \ {\rm and}\ i G\,.
\end{gather}
All possible combinations should be included, subject to the following rules
\begin{itemize}
\item
The number of switches must add up to $n$. This number is given
by the total number of $F$'s and $G$'s plus the number of switches
in the $\Kdfth$'s.
\item
There must be a $\K$ or $\Kdfth$ on both ends.
\item
Each \(\Kdfth\) must have \(F\) on both sides unless external.
This is because the loop momenta next to a $\Kdfth$ have only
one singular propagator in their summands and so the sum-integral
identity can be used.
This implies, given the rules above, that each \(G\) 
must have a $\K$  (and not a $\Kdfth$) on both sides.
\item
$F$'s must have a $\Kdfth$ on at least one side, or, equivalently,
$F$'s always appear on one side or other of a $\Kdfth$.
This is because one cannot use the sum-integral
identity in the middle of a sequence of singular propagators,
since each loop sum runs over two singularities.
The identity can only be used at the end of the sequence,
and only then if it terminates
with the non-singular part of a propagator.
An example of this rule
is that Fig.~\ref{fig:K34anal}b cannot be decomposed
using the sum-integral identity, whereas Fig.~\ref{fig:K34anal}c
can at the left-hand end.
A consequence of this rule is that the only long subsequences involving
$\K$ have the form $\dots i\K iG i\K iG i\K \dots$.
These correspond to diagrams with sequences of singular propagators.
\item
In a sequence of the form $\dots i\K iG i\K iG i\K \dots$ the
rightmost $G$ is multiplied on the right by $[2\omega L^3]$.
This arises from keeping track of on-shell propagators.
\item
The right-hand superscript of each $\Kdfth$ is $(s)$ unless it is
external, when it is a $(u)$. Examples are the third, fifth and
seventh terms in the expression (\ref{eq:K44decomp}) for
$\KthnL 4 {}$. 
\item
The middle superscript of each $\Kdfth$ is $(s)$ unless it is 
either external or it appears to the right of another $\Kdfth$,
separated by a single $F$, 
in which cases it is a $(u)$. The difference from the previous rule
arises due to our ``left-to-right'' convention of dealing with loop
momenta. An example of the new
exception is given by the penultimate term in Eq.~(\ref{eq:K44decomp}).
\end{itemize}
A simple consequence of these rules is that the most divergent
contribution to \(i \KthnL n{}\) is 
\begin{equation}
\label{eq:divergentsequences}
i\K \left( iG i\K\right)^{n-2} iG [2\omega L^3] i\K
\,,
\end{equation}
Similarly, sequences having this form (but with smaller values of $n$)
can appear both connecting the ends to factors of $\Kdfth$,
or between such factors.

It is simpler to display the full result for the summed quantity
${\cal K}_{3,L}^{(u,u)}=\sum_{n=2}^\infty \KthnL n {}$
than for \(\KthnL n{}\). This removes the constraint of the
first rule, so that the sequences are now composed of
the quantities
\begin{gather}
\Kdfth^{(u,u)} \equiv \sum_{n=2}^\infty \Kdfth^{(n,u,u)} ,\ \
\Kdfth^{(u,s)} \equiv \sum_{n=2}^\infty \Kdfth^{(n,u,s)} ,\ \
\Kdfth^{(s,u)} \equiv \sum_{n=2}^\infty \Kdfth^{(n,s,u)} \ {\rm and}\
\Kdfth^{(s,s)} \equiv \sum_{n=2}^\infty \Kdfth^{(n,s,s)} \,.
\end{gather}
In addition, the sequences of divergent terms of the
form (\ref{eq:divergentsequences}) can be summed, leading to
\begin{equation}
\sum_{n=2}^\infty i\K \left( iG i\K\right)^{n-2} iG [2\omega L^3] i\K
= iT iG [2\omega L^3] i\K
=i\K iG i T [2\omega L^3] \,,
\end{equation}
where
\begin{equation}
iT \equiv \frac{1}{1 - i \K i G} i \K 
= i\K \frac{1}{1 - i G i \K}\,.
\end{equation}
We have used here the result that $\K$ commutes with $[2\omega L^3]$, since both are diagonal.

To show the result in a compact form we collect the $\Kdfth$ into
a two-by-two matrix. Now is a good point to recall that, using the
arguments following Eq.~(\ref{eq:stostilde}), 
we can freely interchange in our formulae 
the superscripts $(s)$ and $(\tilde s)$.
This is allowed because the rules always lead to quantities with
$(s)$ superscripts being adjacent to those with $(u)$ superscripts
(with an intervening factor of $F$).
This allows us, for example,
to replace $2\Kdfth^{(s,u)}$ with $\Kdfth^{(s,u)}+\Kdfth^{(\tilde s,u)}$.
The point of such changes is to move towards a physical quantity
which contains the symmetric combination $(u)+(s)+(\tilde s)$ for
all superscripts.
With this in mind, we introduce the matrix of matrices
\begin{equation}
\Bigg(i\Kdfth\Bigg)
\equiv
\left(\begin{array}{cc}
i\Kdfth^{(u,u)}\ \ \ \ \ & i\Kdfth^{(u,s)}+i\Kdfth^{(u,\tilde s)}\\
i\Kdfth^{(s,u)}+i\Kdfth^{(\tilde s,u)}\ \ \ \ \ \ \ \ &
i\Kdfth^{(s,s)}+i\Kdfth^{(s,\tilde s)} +
i\Kdfth^{(\tilde s,s)}+i\Kdfth^{(\tilde s,\tilde s)} 
\end{array}\right)\,.
\end{equation}
The quantity symmetric under particle exchange is then
\begin{equation}
\label{eq:symmetricKdfth}
i \Kdfth^{[B_2]}
\equiv
\left(\begin{array}{cc} 1 & 1 \end{array}\right)
\Bigg(i\Kdfth\Bigg)
\left(\begin{array}{c} 1 \\ 1 \end{array}\right)\,.
\end{equation}

Using this matrix notation, and implementing the rules described
above, we find
\begin{equation}
i{\cal K}_{3,L}^{(u,u)} = iT iG [2\omega L^3] i\K +
\left(\begin{array}{cc} 1 & iT iF \end{array}\right)
 \Bigg (i\Kdfth\Bigg)
\sum_{j=0}^\infty
\left\{ 
\left(\begin{array}{c} 0 \\ 1 \end{array}\right)
\frac{iF}{2\omega L^3} 
\left(\begin{array}{cc} 1 & iT iF \end{array}\right)
\Bigg(i\Kdfth\Bigg) \right\}^j
\left(\begin{array}{c} 1 \\
\frac{iF}{2\omega L^3} iT 2\omega L^3
\end{array}\right)
\,.
\label{eq:K3Lfinal}
\end{equation}
We have succeeded in pulling out explicit finite-volume factors,
with the infinite-volume quantities being 
$\K$ and the two-by-two matrix $(\Kdfth)$.
The latter, however, does not appear in the symmetric form 
(\ref{eq:symmetricKdfth}).
In particular, our left-to-right convention leads to the
presence of an asymmetric matrix between factors of $(\Kdfth)$.

\bigskip
The final step is to insert the result (\ref{eq:K3Lfinal})
into our expression for $C_L^{[B_2]}$, Eq.~(\ref{eq:CLB2res}),
and simplify. We begin by keeping only the first term in (\ref{eq:K3Lfinal}),
i.e. that which arises from summing the most divergent contributions
to ${\cal K}_{3,L}^{(u,u)}$. We find
\begin{align}
\label{eq:CLB2resa}
C_L^{[B_2]} - C_\infty^{[B_2]} - \delta C_\infty^{[B_2]}
&= A'^{[B_2]} \left[
-\frac23\frac{iF}{2\omega L^3}
+\big[\mathcal A \big] \sum_{n=0}^\infty 
\left( i \mathcal K_{3,L}^{(u,u)} \big[\mathcal A \big]\right)^n
\right] A^{[B_2]}
\\
&=
A'^{[B_2]} iF_3 A^{[B_2]} + \mathcal O(\Kdfth) \,,
\end{align}
where the first line is a restatement of Eq.~(\ref{eq:CLB2res}),
and the second contains the new quantity 
\begin{align}
iF_3 
&= \frac{iF}{2\omega L^3} 
\left[ -\frac23 + \frac1{1-iT iF} \right] \,.
\label{eq:F3defa}
\end{align}
To obtain this form for $F_3$ we have used 
\begin{equation}
\big[\mathcal A \big] \sum_{n=0}^\infty 
\left( iT iG [2\omega L^3] i\K \big[\mathcal A \big]\right)^n
= \frac{iF}{2\omega L^3}\frac1{1-iT iF}\,.
\label{eq:F3defb}
\end{equation}
Here the two sides are different ways of writing the 
sum of sequences in which $F$'s or $G$'s alternate with $\K$'s in 
all possible orders, with the constraint 
that sequences must have $F$'s at both ends.
On the left one sums first over the number of intermediate $F$'s
and then sums over $G$'s,
while on the right the roles of $F$ and $G$ are interchanged.
In the following we will also need two further ways of
writing this quantity
\begin{equation}
\frac{iF}{2\omega L^3}\frac1{1-iT iF}
=
\left\{\sum_{n=0}^\infty 
\left(\big[\mathcal A \big] iT iG [2\omega L^3] i\K \right)^n\right\}
\big[\mathcal A \big]
= 
\frac1{1-\frac{iF}{2\omega L^3} iT [2\omega L^3]}\  \frac{iF}{2\omega L^3}
\,.
\label{eq:F3defc}
\end{equation}

Next we consider terms proportional to \(\Kdfth\).
These are obtained by replacing one of the factors of
$\mathcal K_{3,L}^{(u,u)}$ in the sum over $n$ in Eq.~(\ref{eq:CLB2resa})
with the term linear in $\Kdfth$ from Eq.~(\ref{eq:K3Lfinal}),
with all other $\mathcal K_{3,L}^{(u,u)}$'s replaced by the most divergent
term from Eq.~(\ref{eq:K3Lfinal}). This leads to the contribution
\begin{multline}
C_L^{[B_2]} - C_\infty^{[B_2]} - \delta C_\infty^{[B_2]}
\supset \\
 A'^{[B_2]} \frac{iF}{2\omega L^3}\frac1{1-iT iF}
\left(\begin{array}{cc} 1 & iT iF \end{array}\right)
\Bigg(i\Kdfth\Bigg)
\left(
\begin{array}{c} 1 \\ \frac{iF}{2\omega L^3} iT 2\omega L^3 \end{array}\right)
\frac1{1-\frac{iF}{2\omega L^3} iT [2\omega L^3]}\  \frac{iF}{2\omega L^3}
A^{[B_2]} \,.
\label{eq:CLB2linear}
\end{multline}
We next use the identities
\begin{align}
\frac{iF}{2\omega L^3}\frac1{1-iT iF}
\left(\begin{array}{cc} 1 & iT iF \end{array}\right)
= i F_3 
\left(\begin{array}{cc} 1 & 1 \end{array}\right)
+
\frac{iF}{2\omega L^3}
\left(\begin{array}{cc} \frac23 & -\frac13 \end{array}\right)
\label{eq:identity1}
\\
\left(
\begin{array}{c} 1 \\ \frac{iF}{2\omega L^3} iT 2\omega L^3 \end{array}\right)
\frac1{1-\frac{iF}{2\omega L^3} iT [2\omega L^3]}\  \frac{iF}{2\omega L^3}
=
\left(\begin{array}{c} 1 \\ 1 \end{array}\right) 
i F_3
+
\left(\begin{array}{c} 2/3 \\ -1/3 \end{array}\right) 
\frac{iF}{2\omega L^3}
\label{eq:identity2}
\end{align}
to rewrite the right-hand side of Eq.~(\ref{eq:CLB2linear}) as
\begin{equation}
 A'^{[B_2]} 
\left\{i F_3 
\left(\begin{array}{cc} 1 & 1 \end{array}\right)
+
\frac{iF}{2\omega L^3}
\left(\begin{array}{cc} \frac23 & -\frac13 \end{array}\right)\right\}
\Bigg(i\Kdfth\Bigg)
\left\{
\left(\begin{array}{c} 1 \\ 1 \end{array}\right) 
i F_3
+
\left(\begin{array}{c} 2/3 \\ -1/3 \end{array}\right) 
\frac{iF}{2\omega L^3}
\right\}
A^{[B_2]} \,.
\label{eq:CLB2lineara}
\end{equation}
Here we have separated out the symmetric part of $(\Kdfth)$,
which is multiplied on both sides by $F_3$, from the
asymmetric parts. The latter can be analyzed in the same
way as $\delta C_\infty^{[B_2]}$ 
[see Eq.~(\ref{eq:deltaCinftyanal}) and subsequent text].
This is because the vector $(2,\ -1)$ projects, both from
the left and right, onto a $(u)-(s)$
combination (if we use the freedom to interchange $(\tilde s)$ and
$(s)$ when separated from asymmetric quantity such as 
$A'^{[B_2]}$ by an $F$). For example, using this freedom
one finds
\begin{equation}
 A'^{[B_2]} 
\frac{iF}{2\omega L^3}
\left(\begin{array}{cc} \frac23 & -\frac13 \end{array}\right)
\Bigg(i\Kdfth\Bigg)
\left(
\begin{array}{c} 1 \\ 1 \end{array}\right)
=
 A'^{[B_2]} 
\frac{iF}{2\omega L^3}
\frac23 \left\{
i\Kdfth^{(u,u)}-i\Kdfth^{(s,u)}
+
i\Kdfth^{(u,s)}-i\Kdfth^{(s,s)}
+
i\Kdfth^{(u,\tilde s)}-i\Kdfth^{(s,\tilde s)}\right\}
\,.
\end{equation}
This means that, just as in Eq.~(\ref{eq:deltaCinftyanal}),
$F/L^3$ can be replaced by $\rho$ with the (implicit) sum over
the spectator momentum replaced by an integral.
The same holds for the $F$ on the right of $(\Kdfth)$.
We can therefore rewrite Eq.~(\ref{eq:CLB2lineara}) as
\begin{equation}
 A'^{[B_2]} 
\left\{i F_3 
\left(\begin{array}{cc} 1 & 1 \end{array}\right)
+
\frac{i\rho}{2\omega }
\left(\begin{array}{cc} \frac23 & -\frac13 \end{array}\right)\right\}
\Bigg(i\Kdfth\Bigg)
\left\{
\left(\begin{array}{c} 1 \\ 1 \end{array}\right) 
i F_3
+
\left(\begin{array}{c} 2/3 \\ -1/3 \end{array}\right) 
\frac{i\rho}{2\omega}
\right\}
A^{[B_2]} \,,
\label{eq:CLB2linearb}
\end{equation}
again using the notation with implicit integration for $\rho$ factors that was introduced in Eq.~(\ref{eq:deltaCinftyanal}). The contribution linear in
$\Kdfth$ can thus be broken up into four parts:
(i) a finite-volume term involving symmetric quantities
\begin{equation}
 A'^{[B_2]}  i F_3 i\Kdfth^{[B_2]} iF_3 A^{[B_2]}
\,;
\end{equation}
(ii) a partially asymmetric term with $\rho$ on the left
\begin{equation}
 A'^{[B_2]} 
i F_3 
\frac{i\rho}{2\omega }
\left(\begin{array}{cc} \frac23 & -\frac13 \end{array}\right)
\Bigg(i\Kdfth\Bigg)
\left(\begin{array}{c} 1 \\ 1 \end{array}\right) 
i F_3
A^{[B_2]} \,,
\end{equation}
which can be interpreted as $\delta(A'^{[B_2]}) iF_3 A^{[B_2]}$,
where $\delta(A'^{[B_2]})$ absorbs the infinite-volume integral
involving $\rho$;
(iii)
the reflection of (ii) which gives rise to
$A'^{[B_2]} iF_3 \delta (A^{[B_2]})$;
and (iv) the infinite-volume quantity
\begin{equation}
 A'^{[B_2]} 
\frac{i\rho}{2\omega }
\left(\begin{array}{cc} \frac23 & -\frac13 \end{array}\right)
\Bigg(i\Kdfth\Bigg)
\left(\begin{array}{c} 2/3 \\ -1/3 \end{array}\right) 
\frac{i\rho}{2\omega}
A^{[B_2]} \,,
\end{equation}
which is absorbed by replacing $C_\infty^{[B_2]}$ with the alternative 
infinite-volume quantity defined in Eq.~(\ref{eq:CB2rho}) below.

To see the general pattern we next consider terms 
contributing to $C_L^{[B_2]}$ that are quadratic in $\Kdfth$.
These arise from either a single ${\cal K}_{3,L}^{(u,u)}$
term having two factors of $\Kdfth$ or two 
${\cal K}_{3,L}^{(u,u)}$ terms each containing one such factor.
Adding these, using the identities (\ref{eq:identity1}-\ref{eq:identity2}), 
and replacing $F$ with $\rho$ where allowed, we find
\begin{multline}
 A'^{[B_2]} 
\left\{i F_3 
\left(\begin{array}{cc} 1 & 1 \end{array}\right)
+
\frac{i\rho}{2\omega }
\left(\begin{array}{cc} \frac23 & -\frac13 \end{array}\right)\right\}
\Bigg(i\Kdfth\Bigg)
\left(\begin{array}{c} 1 \\ 1 \end{array}\right) 
\left\{i F_3 
\left(\begin{array}{cc} 1 & 1 \end{array}\right)
+
\frac{i\rho}{2\omega }
\left(\begin{array}{cc} \frac23 & -\frac13 \end{array}\right)\right\}
\Bigg(i\Kdfth\Bigg)
\\
\times
\left\{
\left(\begin{array}{c} 1 \\ 1 \end{array}\right) 
i F_3
+
\left(\begin{array}{c} 2/3 \\ -1/3 \end{array}\right) 
\frac{i\rho}{2\omega}
\right\}
A^{[B_2]} \,.
\label{eq:CLB2quad}
\end{multline}
Extending this analysis, we find that terms of higher order in $\Kdfth$
are obtained by inserting additional factors of the matrix
\begin{equation}
\left(\begin{array}{c} 1 \\ 1 \end{array}\right) 
\left\{i F_3 
\left(\begin{array}{cc} 1 & 1 \end{array}\right)
+
\frac{i\rho}{2\omega }
\left(\begin{array}{cc} \frac23 & -\frac13 \end{array}\right)\right\}
\Bigg(i\Kdfth\Bigg)
\end{equation}
after the final $(\Kdfth)$ in Eq.~(\ref{eq:CLB2quad}).

Our final task is to reorganize the series one last time into
infinite-volume kernels separated by finite-volume quantities.
This is done by generalizing the analysis 
described following Eq.~(\ref{eq:CLB2lineara}).

The following asymmetric quantities are needed
\begin{gather}
i \Kdfth^{x}
=
\left(\begin{array}{cc} \frac23 & -\frac13 \end{array}\right)
\Bigg(i\Kdfth\Bigg)
\left(
\begin{array}{c} 1 \\ 1 \end{array}\right)
\,,\ \
i\Kdfth^{y}
=
\left(\begin{array}{cc} 1 & 1 \end{array}\right)
\Bigg(i\Kdfth\Bigg)
\left(
\begin{array}{c} 2/3 \\ -1/3 \end{array}\right)
\,,\ \
\\
i \Kdfth^{xy}
=
\left(\begin{array}{cc} \frac23 & -\frac13 \end{array}\right)
\Bigg(i\Kdfth\Bigg)
\left(
\begin{array}{c} 2/3 \\ -1/3 \end{array}\right)
\,.
\end{gather}
We find a simple geometric series
\begin{equation}
\label{eq:CLB2resc}
C_L^{[B_2]}  
=
C_\infty^{[B_2,\rho]} + \sum_{n=0}^\infty
A'^{[B_2,\rho]} \left[iF_3  i\Kdfth^{[B_2,\rho]} \right]^n iF_3 
A^{[B_2,\rho]} \,,
\end{equation}
where the redefined infinite-volume quantities are
\begin{align}
i\Kdfth^{[B_2,\rho]} &\equiv \sum_{n=0}^\infty
i\Kdfth^{[B_2]} \left[\frac{i\rho}{2\omega} i\Kdfth^{x}\right]^n\,,
\label{eq:KdfthB2rho}
\\
A'^{[B_2,\rho]} &\equiv \sum_{n=0}^\infty
A'^{[B_2]} \left[\frac{i\rho}{2\omega} i\Kdfth^{x}\right]^n\,,
\label{eq:ApB2rho}
\\
A^{[B_2,\rho]} &\equiv
\left\{1+ i\Kdfth^{y} \frac{i\rho}{2\omega} 
+ i\Kdfth^{[B_2,\rho]}
\frac{i\rho}{2\omega} i\Kdfth^{xy}\frac{i\rho}{2\omega}
\right\} A^{[B_2]}\,,
\label{eq:AB2rho}
\\
C_\infty^{[B_2,\rho]} &\equiv
C_\infty^{[B_2]} + \delta C^{[B_2]}_\infty +
A'^{[B_2,\rho]}
\frac{i\rho}{2\omega} i\Kdfth^{xy} \frac{i\rho}{2\omega}
A^{[B_2]}\,.
\label{eq:CB2rho}
\end{align}
Our notation here is rather compact, with implicit integrals
wherever there is a factor of $\rho$, but we stress that
it is straightforward to rewrite these definitions as integral equations.
We also note that $i\Kdfth^{[B_2,\rho]}$,
$A'^{[B_2,\rho]}$ and $A^{[B_2,\rho]}$ are all symmetric under
external particle interchange. This is because they have the vector $(1, 1)$,
or its transpose, at all ends involving external particles.

We can bring the result of Eq.~(\ref{eq:CLB2resc}) into familiar form
by summing the geometric series, leading to
\begin{equation}
\label{eq:CLB2resd}
C^{[B_2]}_{L} = C^{[B_2,\rho]}_{\infty} 
+ A'^{[B_2,\rho]} 
\frac{1}{1- i F_{3} i \Kdfth^{[B_2,\rho]} } i F_{3} 
A^{[B_2,\rho]} \,.
\end{equation}
This completes the most complicated part of the analysis.

%% file: B3.tex
\subsection{Including three-to-three insertions}

\indent

\label{sec:B3}

In this section we add in all diagrams containing
three-to-three (\(B_{3}\)) kernels, and so complete the derivation.
The new diagrams we are considering are those exemplified
by the first and last lines of Fig.~\ref{fig:fullskelexpansion}.
If there were only $B_3$ kernels, with no $B_2$'s, the
analysis would be a simple generalization of that for two particles.
The complications come from the need to add all possible
$B_2$ kernels between two $B_3$'s 
(or between $\sigma$ and a $B_3$, or a $B_3$ and $\sigma^\dagger$).
A key point here is that the properties of $B_3$
are the same as those of $\sigma$ and $\sigma^\dagger$,
namely that it is symmetric in external momenta (separately on both sides) 
and is a smooth function of these momenta 
(within the range of $E$ that we are considering).
This means that we can piggyback on the previous analysis in which
we added all possible $B_2$'s between $\sigma$ and $\sigma^\dagger$.

In particular, a formula analogous to Eq.~(\ref{eq:CLB2resd}) holds
for each segment of a diagram between two $B_3$'s
(and for that between $\sigma$ and a $B_3$, 
and that between a $B_3$ and $\sigma^\dagger$).
In words, Eq.~(\ref{eq:CLB2resd}) tells us that the finite-volume
correlator can be written as the sum of an infinite-volume part
and a part containing the finite-volume function $F_3$.
The infinite-volume part is obtained in two stages:
first, for each diagram replace all loop sums with $\PV$ integrals
ordered in an appropriate way;
second, add in additional terms involving $\rho$, namely
those of Eqs.~(\ref{eq:deltaCdef}) and (\ref{eq:CB2rho}).
In the second term in Eq.~(\ref{eq:CLB2resd}),
the endcaps $A'^{[B_2,\rho]}$ and $A^{[B_2,\rho]}$
are built up by decorating $\sigma$ and $\sigma^\dagger$, respectively,
with all possible $B_2$ insertions, converting sums to $\PV$ integrals,
and then adding in the ``$\rho$ terms'' of 
Eqs.~(\ref{eq:ApB2rho}) and (\ref{eq:AB2rho}).

Exactly the same analysis holds for segments of
diagrams in which $B_3$'s are playing the role of endcaps.
The $B_3$'s are decorated on both sides with $B_2$'s, and can connect
to an adjacent $B_3$ (or $\sigma/\sigma^\dagger$)
either through infinite-volume loops or through a factor of
\begin{equation}
{\cal Z} = \frac{1}{1- i F_{3} i \Kdfth^{[B_2,\rho]} } i F_{3} \,,
\end{equation}
following decoration analogous to that in $A'^{[B_2,\rho]}$
and $A^{[B_2,\rho]}$.

To present the result, we first introduce ``decoration operators''
$D_C^{[B_2,\rho]}$, $D_{A'}^{[B_2,\rho]}$ and
$D_A^{[B_2,\rho]}$, given by
\begin{equation}
C^{[B_2,\rho]}_{\infty} \equiv \sigma D_C^{[B_2,\rho]}\sigma^\dagger
\,,\ \ 
A'^{[B_2,\rho]} \equiv \sigma D_{A'}^{[B_2,\rho]}
\,,\ \ {\rm and}\ \ 
A^{[B_2,\rho]} \equiv D_{A}^{[B_2,\rho]}\sigma^\dagger\,.
\end{equation}
These are infinite-volume integral operators defined
implicitly by the work of previous subsections.
This allows us to write Eq.~\ref{eq:CLB2resd} as
\begin{align}
\label{eq:CLB2rese}
C^{[B_2]}_{L} = 
\sigma \left\{
D_C^{[B_2,\rho]} + D_{A'}^{[B_2,\rho]} {\cal Z} D_A^{[B_2,\rho]}
\right\} \sigma^\dagger \,.
\end{align}
The reason for using this notation is that it works also
for segments of diagrams involving $B_3$'s at the ends.
Thus, for example, a segment of the finite-volume correlator
between two $B_3$'s can be written
\begin{align}
\label{eq:twoB3s}
\dots B_3 \left\{ D_C^{[B_2,\rho]}
+ D_{A'}^{[B_2,\rho]} {\cal Z} D_A^{[B_2,\rho]} 
\right\} B_3 \cdots \,.
\end{align}
The key point is that the {\em same} decoration operators
appear as in (\ref{eq:CLB2rese}).

We can now write down the result for the full finite-volume correlator
\begin{multline}
C_L = 
\sigma \left\{
D_C^{[B_2,\rho]} + D_{A'}^{[B_2,\rho]} {\cal Z} D_A^{[B_2,\rho]}
\right\} \sigma^\dagger
+
\sigma \left\{
D_C^{[B_2,\rho]} + D_{A'}^{[B_2,\rho]} {\cal Z} D_A^{[B_2,\rho]}
\right\} iB_3
\left\{
D_C^{[B_2,\rho]} + D_{A'}^{[B_2,\rho]} {\cal Z} D_A^{[B_2,\rho]}
\right\} \sigma^\dagger
+
\\
\sigma \left\{
D_C^{[B_2,\rho]} + D_{A'}^{[B_2,\rho]} {\cal Z} D_A^{[B_2,\rho]}
\right\} iB_3
\left\{
D_C^{[B_2,\rho]} + D_{A'}^{[B_2,\rho]} {\cal Z} D_A^{[B_2,\rho]}
\right\} i B_3
\left\{
D_C^{[B_2,\rho]} + D_{A'}^{[B_2,\rho]} {\cal Z} D_A^{[B_2,\rho]}
\right\} \sigma^\dagger
+ \dots
\end{multline}
As in the previous subsection, this can be reorganized into the form
\begin{equation}
\label{eq:CLres}
C_{L} = C_\infty + \sum_{n=0}^\infty A'
\big [ {\cal Z} i B_3^{[B_2,\rho]}  \big ]^n {\cal Z} A 
\end{equation}
where
\begin{align}
iB_3^{[B_2,\rho]} &=
\sum_{n=0}^\infty 
D_A^{[B_2,\rho]} \left[i B_3 D_{C}^{[B_2,\rho]}\right]^n
iB_3 D_{A'}^{[B_2,\rho]}\,,
\\
A' &= 
\sum_{n=0}^\infty \sigma 
\left[D_{C}^{[B_2,\rho]} iB_3\right]^n D_{A'}^{[B_2,\rho]}\,,
\\
A &= 
\sum_{n=0}^\infty
D_{A}^{[B_2,\rho]}
\left[iB_3 D_{C}^{[B_2,\rho]} \right]^n 
\sigma^\dagger\,,
\\
C_\infty &=
\sum_{n=0}^\infty
\sigma
D_{C}^{[B_2,\rho]}
\left[iB_3 D_{C}^{[B_2,\rho]} \right]^n 
\sigma^\dagger\,.
\end{align}
The latter three equations give the final forms of the endcaps
and the infinite-volume correlator, now including all
factors of $B_3$.

We can now sum the geometric series in Eq.~(\ref{eq:CLres})
and perform some simple algebraic manipulations to bring
the result to its final form
\begin{equation}
\label{eq:finresder}
C_{L} = C_\infty  + 
A' \frac{1}{1 + F_3 \Kdfth } i F_3 A\,,
\end{equation}
where
\begin{equation}
\Kdfth \equiv
\Kdfth^{[B_2,\rho]}+ B_3^{[B_2,\rho]}\,,
\end{equation}
is the full divergence-free three-to-three amplitude.
Thus we have obtained our claimed result, Eq.~(\ref{eq:corrresult}),
from which follows the quantization condition Eq.~(\ref{eq:mainres}).

\bigskip

We close our derivation by returning to an issue raised in the introduction to 
this section, namely the possibility of poles in $A$, $A'$ and $C_\infty$.
We argue that, while such poles can be present, they cannot
contribute to the finite-volume spectrum, 
i.e. they do not lead to poles in $C_L$.
Only solutions to the quantization condition (\ref{eq:mainres}) lead to poles in $C_L$.

The intuitive argument for this result is that $A$, $A'$ and $C_\infty$ are
infinite volume quantities. While they are non-standard, being
defined with the $\PV$ prescription and involving the
decoration described above, they have no dependence on $L$. 
Thus, if they did lead to poles in $C_L$, this would imply
states in the finite-volume spectrum whose energies were independent
of $L$ (up to corrections of the form $\exp(-mL)$).
The only plausible state with this property is a single particle,
but this is excluded by our choice of energy range ($m<E^*< 5m$).
Three-particle bound states will have finite-volume corrections
that are exponentially suppressed by $\exp(-\gamma L)$, with
$\gamma \ll m$ the binding momentum, but these should be captured by
our analysis, just as is the case for two-particle bound states~\cite{Beane:2003da}.
Finally, above-threshold ``scattering'' states should have 
energies with power-law dependence on $L$.
This is true in the two-particle case, and we expect it to continue to
hold for three particles. This is confirmed, for example, by
the analysis of three (and more) particles using non-relativistic
quantum mechanics~\cite{Beane:2007qr,Tan:2007}.

For the two-particle analysis
this argument can be made more rigorous,
and it is informative to see how this works.
We have recalled the two-particle quantization condition in Sec.~\ref{sec:noswitch},
and give here the form of the
corresponding two-particle finite-volume correlator:
\begin{equation}
\label{eq:CL2res}
C_{L,2} = C_{\infty,2} + i A'_2 \frac{1}{1 +  F  \mathcal K_2}  F A_2 \,.
\end{equation}
The subscripts ``2'' on $A$, $A'$ and $C$ indicate that these are the
two-particle endcaps and correlator, while $F$ is defined in
Eq.~(\ref{eq:Fdef1}) (although here we drop the spectator momentum argument).

What we now show is that
there are poles in $A_2$, $A'_2$ and $C_{\infty,2}$,
but these cancel in $C_{L,2}$. 
To see this we use the freedom to
arbitrarily choose the interpolating functions $\sigma$ and $\sigma^\dagger$
without affecting the position of poles in $C_{L,2}$.
Specifically, we set both $\sigma$ and $\sigma^\dagger$ equal to the
two-particle Bethe-Salpeter kernel $iB_2$, which, we recall, is a smooth
non-singular function. One then finds that
\begin{equation}
C_{\infty,2} = i\K - iB_2 \ \ {\rm and}\ \
A_2 = A'_2 = i\K\,.
\label{eq:specialsource}
\end{equation}
Inserting these results into Eq.~(\ref{eq:CL2res}) we find that
(for this choice of endcaps)
\begin{equation}
C_{L,2}
= - iB_2 + i\K + i\K \frac1{1-iFi\K}iF i\K
= - iB_2 + \frac{i}{\K^{-1}+F}
\,.
\label{eq:CL2special}
\end{equation}
From Eqs.~(\ref{eq:specialsource}) and (\ref{eq:CL2special})
we draw two conclusions. First, $A_2$, $A'_2$ and $C_{\infty,2}$ 
have poles whenever $\K$ diverges. Such poles occur, for a given
angular momentum, when $\delta_\ell=\pi/2 \ {\rm mod}\,\pi$.
Thus, using the $\PV$ prescription, there are, in general,
poles in $A_2$, $A'_2$ and $C_{\infty,2}$. 
Second, these poles cancel in $C_{L,2}$, as shown by the second
form in Eq.~(\ref{eq:CL2special}), which is clearly finite when $\K$ diverges.

We suspect that a similar result holds for the three-particle analysis,
but have not yet been able to demonstrate this.
Thus, in the three-particle case we must rely for now on the intuitive
argument given above.

%% file: conclusions.tex
\section{Conclusions and outlook}
\label{sec:con}

In this work we have presented and derived a three-particle
quantization condition relating the finite-volume
spectrum to two-to-two and three-to-three infinite-volume
scattering quantities.
This condition separates the
dependence on the volume into kinematic quantities,
as was achieved previously for two particles.

There are two new features of the result compared to the two-particle case.
First, the three-particle scattering
quantity entering the quantization condition has the
physical on-shell divergences removed. The resulting divergence-free
quantity is thus spatially localized. This is crucial for
any practical application of the formalism since it allows
for the partial-wave expansion to be truncated.
Indeed, it is difficult to imagine a quantization condition
involving the three-particle scattering amplitude itself,
given that the latter is divergent for certain physical momenta.

The second feature is that the three-particle scattering quantity
is non-standard---it is not simply related to the (divergence-free part)
of the physical scattering amplitude. This is because it is defined
using the $\PV$ pole prescription, and also because of the ``decorations''
explained in Sec.~\ref{sec:B3}.
We strongly suspect, however, that a relation to the physical amplitude
exists.
In particular, we know from Ref.~\cite{Polejaeva:2012ut}
that the finite-volume spectrum
in a non-relativistic theory can be determined solely in terms of
physical amplitudes, and the same is true in the approximations
adopted in Ref.~\cite{Briceno:2012rv}.
We are actively investigating this issue.

The three-particle quantization condition involves a determinant over
a larger space than that required for two particles.
Nevertheless, as explained in Secs.~\ref{sec:iso}, because the three-particle
quantity that enters has a uniformly convergent partial-wave expansion,
one can make a consistent truncation of the quantization condition
so that it involves only a finite number of parameters.
This opens the way to practical application of the formalism.

We have provided in this paper two mild consistency checks on the 
formalism---that
it correctly reproduces the known results if one particle is non-interacting
(see Sec.~\ref{sec:noswitch}),
and that the number of solutions to the quantization condition in the
isotropic approximation is as expected (see Appendix~\ref{app:iso}).
We have also worked out a more detailed check by comparing our result
close to the three-particle threshold $E^*\approx 3m$ to those obtained
using non-relativistic quantum mechanics~\cite{Beane:2007qr,Tan:2007}. 
Here one has an expansion in powers of $1/L$, and we have checked that
the results agree for the first four non-trivial orders.
This provides, in particular, a non-trivial check of the form of $F_3$, 
Eq.~(\ref{eq:F3def}),
and allows us to relate $\Kdfth$ to physical quantities in the non-relativistic
limit. We will present this analysis separately~\cite{HSinprep}.

Two other issues are deferred to future work.
First, we would like to understand in detail the relation of our formalism
and quantization condition to those obtained in 
Refs.~\cite{Polejaeva:2012ut,Briceno:2012rv}.
Second, we plan to test the formalism using simple models for the scattering 
amplitudes, in order to ascertain how best to use it in practice.

%% file: apps.tex
\appendix

\section{Sum-minus-integral identity}
\label{app:sumintegral}

In this appendix we derive the sum-minus-integral identity
that plays a central role in the main text. This identity
is closely related to that given in Ref.~\cite{Kim:2005}
in the context of the two particle quantization condition.

The identity is 
\begin{equation}
\frac12 \left[\frac1{L^3} \sum_{\vec a} -\PV \int_{\vec a}\right]
\frac{g(\vec k, \vec a) h(\vec k,\vec a) H(\vec k)}
{2\omega_a 2\omega_{ka} (E-\omega_k-\omega_a-\omega_{ka})}
=g^*_{\ell',m'}(\vec k) F_{\ell',m';\ell,m}(\vec k) h^*_{\ell,m}(\vec k)\,,
\label{eq:sumintegral}
\end{equation}
which holds up to (implicit) exponentially-suppressed finite-volume
corrections.
The matrix $F(\vec k)$ is given in the main text but repeated
here for convenience
\begin{align}
F_{\ell', m'; \ell, m}(\vec k) & \equiv 
F^{i \epsilon}_{\ell',m';\ell,m}(\vec k) 
+ \rho_{\ell', m';\ell,m}(\vec k) 
\,,
\label{eq:Fdefapp}
\\
F^{i\epsilon}_{\ell',m';\ell,m}(\vec k) & 
\equiv  \frac{1}{2} \left[ \frac{1}{L^3} \sum_{\vec a} - \int_{\vec a} \right] 
\frac{4 \pi Y_{\ell', m'}(\hat a^*)Y^*_{\ell,m}(\hat a^*) 
H(\vec k) H(\vec a) H(\vec b_{ka})}
{2 \omega_a 2 \omega_{ka}(E - \omega_k - \omega_a - \omega_{ka} + i \epsilon)} 
\left(\frac{a^*}{q_k^*}\right)^{\ell+\ell'} \,.
\label{eq:Fiepsdefapp}
\end{align}
The phase-space quantity $\rho$ and the cutoff function $H$ 
are defined, respectively, in Eqs.~(\ref{eq:rhodef}) and (\ref{eq:Hdef}).
The kinematic notation is that described in Sec.~\ref{sec:res}:
the spectator has fixed four-momentum $(\omega_k,\vec k)$, 
the particle whose momentum is summed/integrated has four-momentum
$(\omega_a,\vec a)$, while the third particle is in general off shell,
with four-momentum $(E_2-\omega_a,\vec b_{ka})$.
The four-momentum of the non-spectator pair is
$P_2=(E_2,\vec P_2)=(E-\omega_k,\vec P-\vec k)$,
and $\vec b_{ka}=\vec P_2-\vec a$.
If the third particle was on shell, it would have energy
$\omega_{ka}$ [defined in Eq.~(\ref{eq:omegadef})], so the on-shell
condition is $E_2=\omega_a+\omega_{ka}$.
This is where the denominator in Eq.~(\ref{eq:Fiepsdefapp}) vanishes.
The boost to the CM-frame of the non-spectator pair sends
$P_2$ to $(E_{2,k}^*,\vec 0)$ and
$(\omega_a,\vec a)$ to $(\omega_a^*,\vec a^*)$.
If all three particles are on shell, then the magnitude
of $\vec a^*$ satisfies $a^*=q_k^*$ 
[with $q_k^*$ given by Eq.~(\ref{eq:qkstardef})].

The two functions in the identity (\ref{eq:sumintegral}),
$g$ and $h$, contain the momentum dependence arising from quantities
respectively on the left and right of the three-particle ``cut''. They could be combined into a single function,
but for our formalism it is advantageous to keep them separate.
We assume that $g$ and $h$ are smooth (infinitely differentiable)
functions of the components of $\vec a$ and that they fall off at
large $|\vec a|$ such that the sum and integral are convergent.
Note that, since $k$ and $a$ are on-shell, and the total momentum is fixed,
the independent quantities are $\vec k$ and $\vec a$, which are thus given
as the arguments of $g$ and $h$. In general, 
the third momentum is off shell, so these functions
involve off-shell amplitudes.
What appears on the right-hand side of the identity, however,
are on-shell projections of these amplitudes after 
decomposition into the angular-momentum
basis in the CM-frame of the non-spectator pair.
This projection is explained around Eq.~(\ref{eq:onshellproj}) for
the case where $g=\sigma$ and $h=\sigma^\dagger$, 
but applies equally well to any functions.

One difference between our identity and that of Ref.~\cite{Kim:2005}
is that, in the three-particle context, the two-particle sub-system
can be arbitrarily far below threshold. The dominant sub-threshold
contribution to $F$ comes from the factor of $\rho$ in
Eq.~(\ref{eq:Fdefapp}), which in turn arises from the 
difference between $\PV$ and $i\epsilon$ pole prescriptions 
[see Eq.~(\ref{eq:PVtildedef})]. This factor is needed so that
the dependence on $\vec k$ is smooth, but is not important for
the derivation of the sum-integral identity. Indeed, we can rewrite
the identity using the $i\epsilon$-prescription and cancel
factors of $\rho$:
\begin{equation}
\frac12 \left[\frac1{L^3} \sum_{\vec a} -\int_{\vec a}\right]
\frac{g(\vec k, \vec a) h(\vec k,\vec a) H(\vec k)}
{2\omega_a 2\omega_{ka} (E_2-\omega_a-\omega_{ka}+i\epsilon)}
=
g^*_{\ell',m'}(\vec k) F^{i\epsilon}_{\ell',m';\ell,m}(\vec k) 
h^*_{\ell,m}(\vec k)\,,
\label{eq:sumintegral2}
\end{equation}
This is now very similar to the identity of Ref.~\cite{Kim:2005},
and we focus on this form henceforth.

To demonstrate (\ref{eq:sumintegral2}) we need simply to subtract
the two sides and show that it is exponentially suppressed.
The difference is proportional to
\begin{equation}
H(\vec k)\left[\frac1{L^3} \sum_{\vec a} -\int_{\vec a}\right]
\frac{
g(\vec k, \vec a) h(\vec k,\vec a)
- g^*_{\ell',m'}(\vec k) 4 \pi Y_{\ell', m'}(\hat a^*)
(a^*/q_k^*)^{\ell'+\ell} Y^*_{\ell,m}(\hat a^*) h^*_{\ell,m}(\vec k) 
H(\vec a) H(\vec b_{ka})
}
{2\omega_a 2\omega_{ka} (E_2-\omega_a-\omega_{ka}+i\epsilon)}
\,,
\label{eq:sumintegral3}
\end{equation}
where we have assumed that the sums over angular-momentum
indices can be interchanged with the $\vec a$ integral.
The overall factor of $H(\vec k)$ serves only to ensure
that the boosts to the two-particle CM frame are well-defined.
We note that the sums and integrals are convergent in the ultraviolet
because of the assumed properties of $g$ and $h$ (in the first term
in the numerator) and the presence of the cutoffs $H$ (in the second term).
The difference (\ref{eq:sumintegral3}) will vanish, 
up to exponentially suppressed corrections, if the
integrand/summand is non-singular and smooth as a function of $\vec a$.
This in turn holds if
(i) all functions appearing in the expression have a smooth 
dependence on $\vec a$, and
(ii) the difference in the numerator cancels that in
the denominator in such a way that the the ratio is smooth.
We address these conditions in turn.

The only non-smooth functions appearing in (\ref{eq:sumintegral3})
are the spherical harmonics, which are ill-defined at $\vec a^*=0$ for 
$\ell>0$. Smoothness is ensured, however,
by the factors of $(a^*)^{\ell'+\ell}$, which turn the spherical harmonics
into polynomials in the components of $\vec a$.
Thus the first condition is satisfied.
For subsequent work, it is useful to understand the
$a^*$ dependence of the coefficients in the
angular-momentum expansion of $g$ and $h$.
Recall that one first changes to $\vec a^*$ as the independent variable,
e.g. $g^*(\vec k,\vec a^*)\equiv g(\vec k, \vec a)$,
and then expands in harmonics:
\begin{equation}
g^*(\vec k,\vec a^*) = 
g^*_{\ell',m'}(\vec k,a^*) Y_{\ell',m'}(\hat a^*)  \sqrt{4\pi}
= 
\frac{g^*_{\ell',m'}(\vec k,a^*)}{a^{*\ell}} 
Y_{\ell',m'}(\hat a^*) a^{*\ell} \sqrt{4\pi}
\,.
\end{equation}
For $a^*\to 0$, the last form is simply a rewriting of the Taylor expansion
in the spherical basis, 
since $Y_{\ell',m'}(\hat a^*) a^{*\ell}$ is a
homogeneous polynomial of order $\ell$ in the components of $\vec a^*$.
This implies that $g^*_{\ell',m'}(\vec k,a^*)/a^{*\ell}$ has a finite
limit as $a^*\to 0$. Furthermore, since $g^*(\vec k,\vec a^*)$ 
is, by assumption, smooth at $a^*=0$, 
$g^*_{\ell',m'}(\vec k,a^*)/a^{*\ell}$ must be a smooth function of
$a^{*2}$ and not $a^*$.
Thus, for small $a^*$,
\begin{equation}
\frac{g^*_{\ell',m'}(\vec k,a^*)}{a^{*\ell}}
=\sum_{n=0}^\infty s_n (a^{*2})^n\,,
\label{eq:gnearzero}
\end{equation}
with $s_n$ the Taylor coefficients.
An analogous result holds for $h$.

We turn now to the second condition, 
that zeroes in the numerator and denominator should cancel. 
To satisfy this we  need the numerator
of (\ref{eq:sumintegral3}) to vanish on shell. 
It does vanish because, when $E_2=\omega_a+\omega_{ka}$, we have
$H(\vec a)=H(\vec b_{ka})=1$, $a^*=q_k^*$, and
\begin{equation}
g^*_{\ell',m'}(\vec k) Y_{\ell',m'}(\hat a^*) \sqrt{4\pi}
=
g^*_{\ell',m'}(\vec k,q_k^*) Y_{\ell',m'}(\hat a^*) \sqrt{4\pi}
=
g^*(\vec k, q_k^* \hat a^*)
= g(\vec k,\vec a)
\label{eq:gstarprop}
\end{equation}
(and similarly for $h$).
In addition, the numerator must vanish fast enough
to cancel the denominator. To see that this is also true
it is convenient to re-express the denominator in terms of
CM variables. Following the arguments of Ref.~\cite{Kim:2005},
we can make the replacement
\begin{equation}
\frac1{2\omega_a 2\omega_{ka} (E_2-\omega_a-\omega_{ka}+i\epsilon)}
\longrightarrow
\frac{\omega_a^*}{2\omega_a E_{2,k}^*(q_k^*+a^*)(q_k^{*}-a^{*}+i\epsilon)}
\label{eq:CMvariables}
\,,
\end{equation}
since the difference is non-singular.
This shows us that the singularity lies in the radial integral over $a^*$.
Now consider the ratio (\ref{eq:sumintegral3}) at fixed angle $\hat a^*$,
so that the spherical harmonics are fixed. Then, from the discussion above,
we know that both terms in the numerator are smooth functions of $a^*$,
which thus have Taylor expansions about $a^*=q_k^*$.
In the difference, the constant term in these Taylor expansions cancels,
and so the ratio with $1/(q_k^*-a*)$ is a smooth function of $a^*$.
In particular we have demonstrated that the numerator does not have 
a non-analytic form such as $\sqrt{q_k^*-a^*}$, which would fail to 
cancel the singularity. 

A special case occurs if $q_k^*\to0$, for then there is a double pole
in $a^*$. In addition, one might be concerned about the factor of
$1/(q_k^*)^{\ell'+\ell}$. These features do not, however, lead to
problems. We know from Eqs.~(\ref{eq:gnearzero}) and (\ref{eq:gstarprop})
that $g^*_{\ell',m'}(\vec k)\propto q_k^{*\ell}$, and similarly for $h$,
so the $1/q_k^*$ factors are canceled.
Furthermore, because of Eq.~(\ref{eq:gnearzero}),
the difference in the numerator of (\ref{eq:sumintegral3}) is
proportional to $a^{*2}$, and thus fully cancels the double pole.
              
\bigskip
This completes the demonstration of the key identity.
We close this section by presenting some further results for
the kinematic functions $F$ and $F^{i\epsilon}$.
First, we give the relation to the
kinematic functions $c^P$ introduced in Ref.~\cite{Kim:2005}.
These replace the product $g\times h$ with a single function,
which is then expanded in a single-set of spherical harmonics. Because of
this, the relation involves Clebsch-Gordon coefficients.
Specifically we find (see also Ref.~\cite{Hansen:2012tf}):\footnote{%
One subtlety in the derivation of Eq.~(\ref{eq:comparetoKim})
is that the powers of $a^*/q_k^*$ do not always match.
This is because we use a double expansion in spherical harmonics
while Ref.~\cite{Kim:2005} use a single expansion. One can show,
however, that the differences always lead to exponentially suppressed
contributions.}
\begin{equation}
F^{i\epsilon}_{\ell',m';\ell,m}(\vec k)
= \frac{i {\rm Re}(q^*_k)}{16 \pi E^*_{2,k}} \delta_{\ell',\ell} \delta_{m',m}+
\sum_{\tilde \ell,\tilde m}
\frac{\sqrt{4\pi} c^P_{\tilde \ell,\tilde m}(q_k^*)}
{4 E_{2,k}^* (q_k^*)^{\tilde\ell}}
\int d\Omega_{a^*}
Y^*_{\tilde\ell,\tilde m}(\hat a^*)             
Y_{\ell',m'}(\hat a^*)
Y^*_{\ell,m}(\hat a^*)
\,.
\label{eq:comparetoKim}
\end{equation}
Because Ref.~\cite{Kim:2005} uses an exponential cut-off
while we use $H(\vec a)H(\vec b_{ka})$,
this result holds only 
up to exponentially suppressed finite-volume corrections.\footnote{%
One might wonder why we use the $H$ functions to provide the cut-off,
since, as far as sum-integral identity is concerned, we could use any 
reasonable cut-off. The reason we use $H$ is that some of the
factors of $F$ arise from insertions of the quantity $G$ 
[Eq.~(\ref{eq:Gdef})]. But $G$ contains, as part of its essential
definition, two factors of $H$, one of which becomes $H(\vec a)$
when we convert the $G$ to an $F$. Thus an $H$ cut-off is forced
upon us for such $F$'s, and we wish to use a uniform definition.
It is then convenient to enforce $a\leftrightarrow b_{ka}$ symmetry
by adding in $H(\vec b_{ka})$.}
%

Finally, we derive a result needed in the main text, namely 
\begin{equation}
F_{\ell',m';\ell,m}(\vec k) = 
F^{i\epsilon}_{\ell',m';\ell,m}(\vec k) = 
0 \ \ {\rm if}\ \ \ell'+\ell={\rm odd}\,,
\label{eq:Foddl}
\end{equation}
up to exponentially suppressed corrections.
It is sufficient to show this result for one of
$F$ and $F^{i\epsilon}$,
since it holds trivially for their difference, $\rho$, which is diagonal in angular momentum. We demonstrate (\ref{eq:Foddl}) for $F^{i\epsilon}$.

The result follows by averaging the original expression (\ref{eq:Fiepsdefapp})
with that obtained by changing variables
$\vec a\to \vec P-\vec a=\vec b_{ka}$.
In this way, the numerator of $F^{i\epsilon}$ is replaced, up to an
overall constant, with
\begin{equation}
\left[Y_{\ell', m'}(\hat a^*)Y^*_{\ell,m}(\hat a^*) (a^*)^{\ell'+\ell}
+Y_{\ell',m'}(\hat b_{ka}^*)Y^*_{\ell,m}(\hat b_{ka}^*)
(b_{ka}^*)^{\ell'+\ell}\right]
H(\vec a) H(\vec b_{ka})
\,.
\label{eq:numeratordiff}
\end{equation}
If all particles are on shell, then from Eq.~(\ref{eq:onshellkin}),
we have that $\vec a^*=-\vec b^*_{ka}$, so the two terms exactly
cancel when the parities of the spherical harmonics are opposite,
i.e. if $\ell+\ell'$ is odd.\footnote{%
A special case occurs when $\vec a=\vec P/2=\vec b_{ka}$, 
for then the two terms in the sum are the same. The derivation remains valid,
however, since in this configuration $\vec a^*=0$, 
implying that the only non-vanishing contributions are from
$\ell=\ell'=0$.}
As we move away from the on-shell condition, the cancellation will
be inexact. However, as we now demonstrate, the residue is
proportional to $E_2-\omega_a-\omega_{ka}$, which is enough to
cancel the pole in $F^{i\epsilon}$, so
that the sum-integral difference of
the residue is exponentially suppressed.
We recall that the boost to the two-particle CM frame transforms
four-vectors as
\begin{equation}
(E_2-\omega_a,\vec b_{ka})\longrightarrow (E_{2,k}^*-\omega_a^*,-\vec a^*)
\ \ {\rm and}\ \ 
(\omega_{ka},\vec b_{ka}) \longrightarrow (\omega_b^*,\vec b_{ka}^*)\,.
\end{equation}
But since
\begin{equation}
(\omega_{ka},\vec b_{ka})
=(E_2-\omega_a,\vec b_{ka})-(E_2-\omega_a-\omega_{ka},\vec 0)\,,
\end{equation}
we see from the linearity of boosts that
\begin{equation}
\vec b_{ka}^*=-\vec a^* + {\cal O}(E_2-\omega_a-\omega_{ka})\,.
\end{equation}
This completes the demonstration.

We note that a similar argument leads to the conclusion that
$c^P_{\ell,m}$ vanishes for odd $\ell$, as first noted in 
Ref.~\cite{Kari:1995}.

\section{Smoothness of $\PV$ pole prescription}

\label{app:PVtilde}

In this appendix we explain why the $\PV$ pole prescription,
defined in Eq.~(\ref{eq:PVtildedef}),
leads to results that are smooth functions of the spectator momentum.
The general integral that appears has the form
\begin{align}
f(\vec k) &= \PV \int_{\vec a} 
\frac{g(\vec k, \vec a)H(\vec k)}
{2 \omega_a 2 \omega_{ka}(E-\omega_k - \omega_a - \omega_{ka})}
\end{align}
The notation is the same as in Eq.~(\ref{eq:sumintegral}),
except that here we have combined the two functions $g$ and
$h$ in the numerator of (\ref{eq:sumintegral}) into 
the single function $g$.
The issue is whether $f(\vec k)$ is a smooth function of $\vec k$.
All quantities appearing in the integrand are smooth functions:
$\omega_k$ and $\omega_{ka}$ manifestly, 
$H(\vec k)$ by construction, and $g(\vec k,\vec a)$ by assumption.\footnote{%
The initial application of the result of this appendix, 
in the discussion following Eq.~(\ref{eq:onshellproj}) in the main text,
has $g$ composed of the product $\sigma \sigma^\dagger$, which is
smooth by construction. Subsequently, one or both of these factors
are replaced by Bethe-Salpeter kernels, which are also smooth because
singularities are far from threshold. The nearest singularity is the
left-hand cut which occurs when $E_{2,k}^*=0$ (corresponding
to $s=u=0$, $t=4m^2$ in Mandlestam variables), but we are protected
from this cut by the cut-off function $H(\vec k)$.
Finally, the factors are replaced
by two particle K-matrices, or decorated end-cap functions.
Here the necessary smoothness is established by the argument of this appendix.
Thus we are using the result of this appendix iteratively.}
We also assume that the behavior of $g$ at large $|\vec a|$ is such
that the integral remains convergent however many derivatives 
of the integrand with respect to the components of $\vec k$ we take.
Then the only source for a lack of smoothness is the pole in the integrand.

It is useful to change variables to $\vec a^*$, 
the momentum in the two-particle CM frame. 
This gives
\begin{align}
f(\vec k) &= \frac{H(\vec k)}{2 E_{2,k}^*} \PV \int \frac{d^3a^*}{(2\pi)^3}
\frac{\tilde g^*(\vec k,\vec a^*)} {(q_k^{*2}-a^{*2})}\,,
\label{eq:app:PVtilde2}
\\
\tilde g^*(\vec k,\vec a^*) &= g(\vec k,\vec a)
\frac{(E-\omega_k-\omega_a+\omega_{ka})(E_{2,k}^*+2\omega_a^*)}
{8\omega_{ka} \omega_a^*}\,.
\label{eq:gtildedef}
\end{align}
The expression multiplying $g$ on the right-hand side of (\ref{eq:gtildedef})
equals unity on shell.
Expanding $\tilde g^*$ in spherical harmonics,
\begin{equation}
\tilde g^*(\vec k,\vec a^*) = \tilde g^*_{\ell,m}(\vec k,a^*)
Y_{\ell,m}(\hat a^*) \sqrt{4\pi}\,,
\end{equation}
we observe that only the $\ell=0$ component
contributes to the integral:
\begin{equation}
f(\vec k) 
= \frac{H(\vec k)}{4 \pi^2 E_{2,k}^*} \PV \int_0^\infty d a^* 
\frac{a^{*2}\tilde g^*_{0,0}(\vec k,a^{*2})} {(q_k^{*2}-a^{*2})}
= \frac{H(\vec k)}{8 \pi^2 E_{2,k}^*} \PV \int_0^\infty d(a^{*2}) 
\frac{\sqrt{a^{*2}}\,\tilde g^*_{0,0}(\vec k,a^{*2})} {(q_k^{*2}-a^{*2})}
\,.
\label{eq:app:PVtilde2a}
\end{equation}
Here we have made explicit that
$\tilde g^*_{0,0}$ is a function of $a^{*2}$,
as follows from the result (\ref{eq:gnearzero}).

In this form, the $\PV$ prescription of Eq.~(\ref{eq:PVtildedef}) becomes
\begin{equation}
f(\vec k) = \frac{H(\vec k)}{8 \pi^2 E_{2,k}^*} \int_0^\infty d(a^{*2}) 
\left[
\frac{\sqrt{a^{*2}}\,\tilde g^*_{0,0}(\vec k,a^{*2})} 
{(q_k^{*2}-a^{*2}+i\epsilon)}
\right]
+ \frac{H(\vec k) \tilde g^*_{0,0}(\vec k, q_k^{*2})}{8\pi E_{2,k}^*}
\times
\begin{cases}
i q_k^* &  \qquad(q_k^{*2}>0)\,,
\\
- |q_k^*| &  \qquad (q_k^{*2}<0)\,.
\end{cases}
\label{eq:app:PVtilde3}
\end{equation}
If $q_k^{*2}>0$, so the non-spectator pair is above threshold,
the $\PV$ and PV prescriptions are the same,
and Eq.~(\ref{eq:app:PVtilde3}) gives the standard relationship between
PV and $i\epsilon$ prescriptions. In particular, the second term
cancels the imaginary part of the $i\epsilon$-regulated integral.
The new feature of the prescription occurs below threshold, i.e. 
for $q_k^{*2}<0$. Here there is no pole to regulate, 
so the $i\epsilon$ prescription is superfluous, 
and the integral is real. Nevertheless, the prescription
adds the second term, also real, which is needed to
avoid a cusp in $f(\vec k)$ at threshold.
We stress that in the second term
$\tilde g^*$ is evaluated on shell, with $a^{*2}=q_k^{*2}$.
If $q_k^{*2}<0$, then $\tilde g^*$ must be evaluated below threshold.
As discussed in the previous appendix, the assumed smoothness of $g$
implies that $\tilde g^*_{0,0}$ is a function of $q_k^{*2}$,
and thus can be straightforwardly evaluated for $q_k^{*2}<0$.

To show that $f(\vec k)$ is smooth, we now extract the essential
features of Eq.~(\ref{eq:app:PVtilde2a}) and consider the integral
\begin{equation}
f(z) = \int_0^\infty dw \frac{\sqrt{w}\, g(w,z)}{z-w} \,.
\label{eq:simpleint}
\end{equation}
Here $w$ and $z$ are playing the roles of $a^{*2}$ and $q_k^{*2}$,
respectively, and $g$, a smooth function of its arguments,
ensures convergence (and includes $H(\vec k)$). The only difference from
Eq.~(\ref{eq:app:PVtilde2a}) is that the dependence on all components
of $\vec k$ in $H$ and $\tilde g^*$ has been simplified to
dependence on $q_k^{*2}$ alone. This simplification is justified
because it is the dependence of the pole term on $\vec k$ that
can lead to a lack of smoothness, and this dependence
is correctly incorporated in Eq.~(\ref{eq:simpleint}).

We treat $z$ as complex, and assume that
$g(w,z)$ can be analytically continued
to complex arguments without encountering singularities.
Then $f(z)$ is well-defined and analytic in the entire complex plane
except along the positive real axis.
As $z$ approaches the positive real axis from above 
or below one obtains the $\pm i\epsilon$-regulated integrals:
\begin{equation}
f(x\pm i\epsilon) 
=
\int_0^\infty dw \frac{\sqrt{w}\, g(w,x)}{x-w\pm i\epsilon} \,.
\end{equation}
These are both complex, with the same real parts 
but differing imaginary parts, $\pm\pi i \sqrt{x} g(x,x)$.
$f(z)$ thus has a cut on the positive real axis.

\begin{figure}[H]
\begin{center}
\includegraphics[scale=.17]{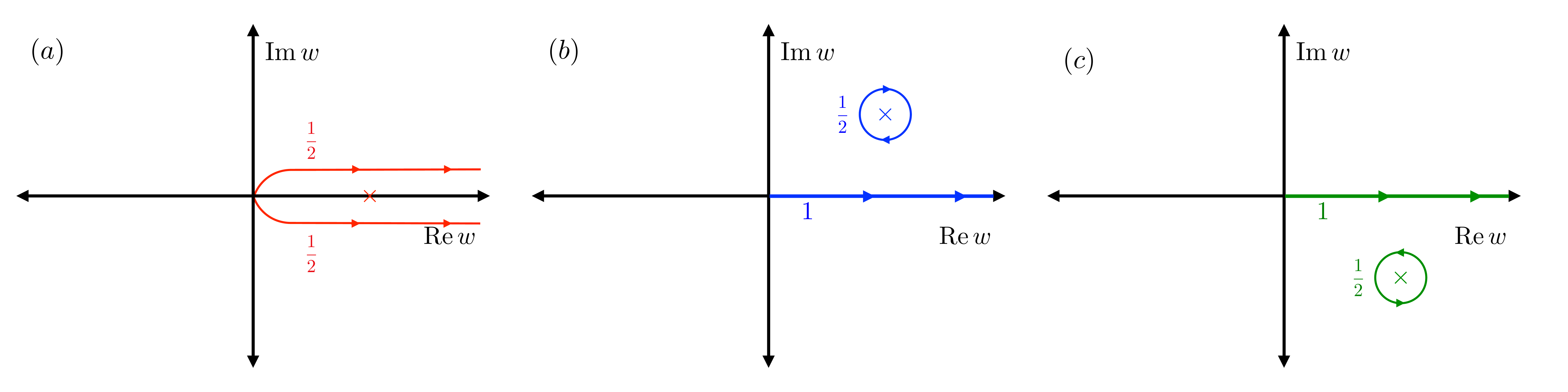} 

\caption{Contours in $w$ complex plane contributing to $f_{\PV}(z)$.
(a) $z$ real and positive;
(b) $z$ above the positive real axis;
(c) $z$ below the positive real axis. In each case the cross indicates the location of the $z=w$ pole, and the numbers indicate the
weights associated with each contour.}
\label{fig:contour}
\end{center}
\end{figure}

The integral of interest, $f(\vec k)$ of Eq.~(\ref{eq:app:PVtilde2a}),
becomes, in our stripped-down version, and for positive $q_k^{*2}$,
\begin{equation}
f_{\PV}(x)= \frac12\left[
f(x + i\epsilon) + f(x- i\epsilon) \right]
\,.
\end{equation}
Here $x$ is real and positive, and we have used the result that
the PV prescription
(which is the same as the $\PV$ prescription for $x>0$)
can be written as the average of the integral with the contour
running above and below the pole (see Fig.~\ref{fig:contour}a).
Our aim is to extend $f_{\PV}$ to a function of $z$, and
study its analyticity properties.

\begin{figure}[H]
\begin{center}
\includegraphics[scale=1.0]{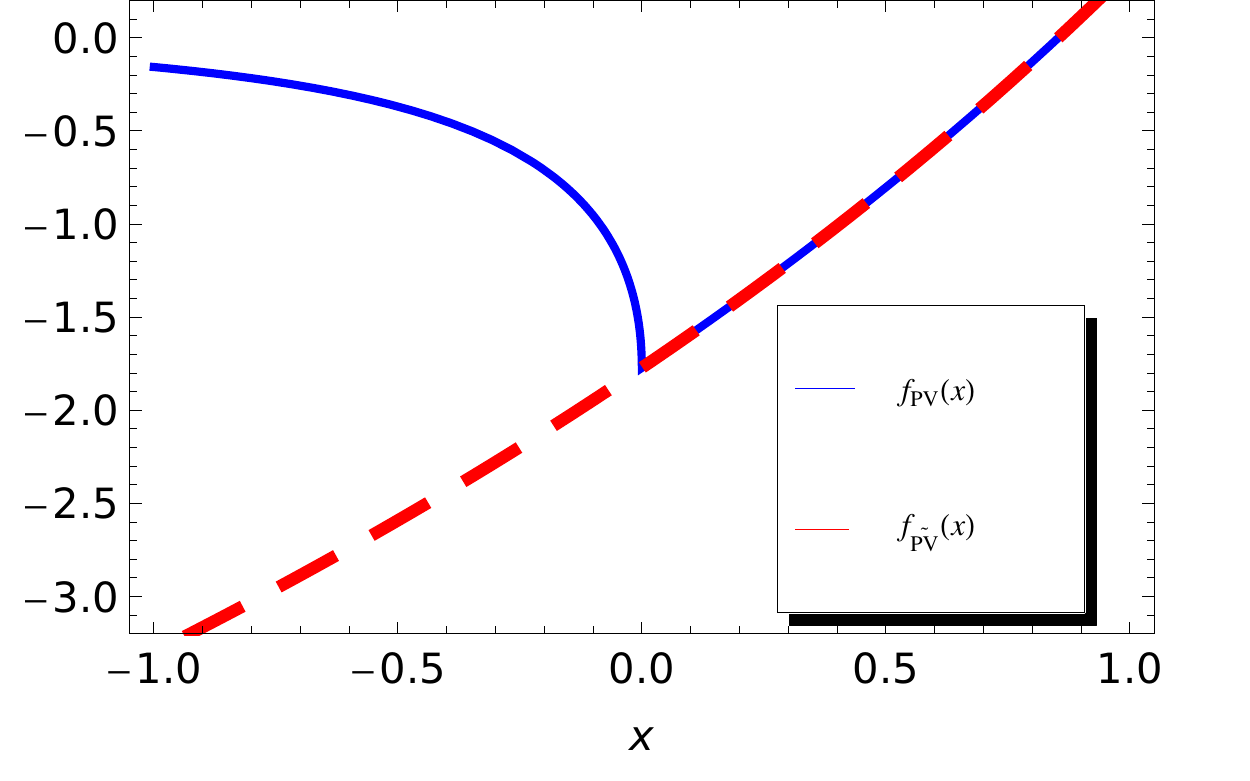}


\caption{$f_{\PV}(x)$ for $g(w,z)=\exp(z-w)$ compared to the
result of using the PV prescription, $f_{\rm PV}(x)$.
The former (dashed [red]) is smooth, while the latter (solid [blue]) has a cusp at $x=0$.
For $x<0$, the difference between the two functions is
the pole term, the second term in Eq.~(\ref{eq:fPVvsf}).}
\label{fig:cusp}
\end{center}
\end{figure}

If $z$ is moved off the positive real axis then,
to avoid non-analytic dependence, the integration contours must be
deformed as shown in Fig.\ref{fig:contour}b 
and Fig.\ref{fig:contour}c. Thus $f_{\PV}(z)$ differs
from $f(z)$ (the integral along the real axis) by pole terms:
\begin{equation}
\label{eq:fPVvsf}
f_{\PV}(z) = f(z) - {\rm sign}({\rm Im} z) i\pi \sqrt{z}\,g(z,z)
= f(z) + \pi \sqrt{-z}\, g(z,z)
\,.
\end{equation}
The sign of ${\rm Im}(z)$ enters because the 
direction of the contour around the pole depends on this sign.
As shown in the second expression, however, the two possibilities
can be combined into a single expression using the properties
of the square root (assuming that the branch cut for $\sqrt{-z}$
is placed along the positive real axis).
We now observe that
there is no discontinuity for negative real $z$,
since both $f(z)$ and $\sqrt{-z}$ are analytic there.
Furthermore, by construction the discontinuities of $f(z)$
and the pole term cancel exactly along the positive real axis
[as they must to yield a real $f_{\PV}(x)$].
Thus we find the key result: $f_{\PV}(z)$
{\em is analytic throughout the complex plane}, i.e. is entire.
Since an entire function is infinitely differentiable,
it follows that $f_{\PV}(x)$ is smooth along the entire real axis.

We now apply this result to our integral of interest,
Eq.~(\ref{eq:app:PVtilde2a}).
The rule is to write the difference between the results of the
$\PV$ and $i\epsilon$ prescriptions, which is standard above threshold,
as an analytic function of $q_k^{*2}$, and then continue to $q_k^{*2}<0$.
Noting that
\begin{equation}
i q_k^* = - \sqrt{-q_k^{*2}} 
\ \ {\rm for}\ \
q_k^{*2}= x + i\epsilon\,,
\end{equation}
we obtain the result quoted in (\ref{eq:app:PVtilde3})
for the below threshold case
\begin{equation}
-\sqrt{-q_k^{*2}} \longrightarrow - |q_k^*| 
\ \ {\rm for}\ \  q_k^{*2} = -x\,.
\end{equation}
Thus our $\PV$ prescription indeed yields a smooth function of $\vec k$.

We have checked this result on an extensive set of examples,
e.g. for $g(w,z) = w^n\exp(-w)$ with $n\ge 0$ an integer,
where the integrals can be done analytically,
and for $g(w,z) = \exp(-w^2)$, where numerical integration is required.
As an illustration,
we show the results for $g(w,z)= \exp(z-w)$ in Fig.~\ref{fig:cusp}.

Finally, we stress that the factor of $\sqrt{w}$ in the
integrand of $f(z)$ is crucial for the smoothness of the $\PV$ prescription.
This factor is present in the original integral,
Eq.~(\ref{eq:app:PVtilde2a}),
because of three-dimensional phase space.
Without this factor, the above- and below-axis pole terms would
not be equal along the negative real axis.
For example, if $\sqrt{w}$ is replaced by an analytic function,
say $x$, then one can easily show, using the arguments above,
that $f_{\PV}(z)$ result has a complex discontinuity
along the negative real axis.

\section{Detailed study of the isotropic approximation}
\label{app:iso}

\indent

In this appendix we study the approach to the isotropic limit
described in Sec.~\ref{sec:iso} in the context of a 
specific choice of parameters.
Our aims are to show how ``lost'' states are recovered,
and also to gain more intuition for the workings of the quantization
condition and, in particular, the impact of the
violation of particle-interchange
symmetry by our coordinates and truncation.
Although our considerations are specific to the chosen parameters,
much of the discussion holds for a general choice of parameters.

We work in the static frame, $\vec P=0$. 
To simplify numerical values we choose a volume such that 
the particle mass satisfies $m=2\pi/L$. While this is artificial,
we stress that none of the general conclusions depend on this choice.
The single-particle momenta are 
$(L/2\pi)\vec k=\vec 0$, $(1,0,0)$, $(1,1,0)$, $(1,1,1)$, etc.,
together with permutations.
We refer to these, respectively, as the $n=0$, $1$, $2$, $3$ shells, etc. 
Given our choice of $m$,
the corresponding single-particle energies are $\omega_n=\sqrt{n+1}\times m$.

We are interested in values of $E$ 
for which there are more than one free three-particle energy levels, 
so that we can see what happens when one of these is replaced by
the solution to Eq.~(\ref{eq:mainiso}).
The minimal case is to have two free levels.
These levels occur at $E=3m$ (all particles at rest),
$E=(1+2\sqrt2)m\approx 3.83 m$
(one particle at rest and two with opposite momenta from the $n=1$ shell),
$E=4.46 m$, etc.
Thus the range of $E$ should extend above $3.83 m$.
We also want to minimize $N$, the size of the matrices appearing
in the quantization condition.
The critical energies above which the cut-off function $H(\vec k)$
is non-vanishing are, for the $n=1-4$ shells,
$(1+\sqrt2)m$, $(\sqrt2+\sqrt3)m\approx 3.15 m$,
$(2+\sqrt3)\approx 3.73m$ and $(2+\sqrt5)m\approx 4.24 m$, respectively.
Thus we are forced to include the $n=3$ shell (in order to attain
$E> 3.83 m$) but if we restrict $E< 4.24 m$ we
do not need the $n=4$ shell.

Thus we end up with the energy range of interest being
$3m < E < 4.24 m$ and $N=1+6+12+8=27$.
Our aim is to find all solutions to the quantization condition,
\begin{equation}
\label{eq:qcapp}
\det(1+F_3^s \Kdfth^s)=0\,,
\end{equation}
within this range.
Here $\Kdfth^s$ is defined in Eq.~(\ref{eq:Kdf3def}),
while $F_3^s$ is simply $F_3$ [Eq.~(\ref{eq:F3def})]
after truncation to $\ell_{\rm max}=0$.
(An explicit expression is given below.)
As already noted, the free states lie at $E=3m$ and $3.83 m$.
There are four such states: one at $3m$ and three at $3.83 m$.
One might have expected six states at $E=3.83 m$, since
two of the particles have momenta in the $n=1$ shell,
but only three are distinct for identical particles.
It is useful to classify the states according to their
transformation properties under
the octahedral symmetry group.
The $E=3m$ state lies in the trivial $A_1$ irrep
(irreducible representation),
while the three $E=3.83 m$ states decompose 
as $A_1+E$, where $E$ is the doublet.
Here we are using standard notation for irreps of
the cubic group, see, e.g., Ref.~\cite{Mandula:1983ut}.
Interactions can lead to mixing between the two $A_1$
states, but not with the doublet.
What we thus expect is that one of the two $A_1$ states is
replaced by the solution to the isotropic quantization condition,
Eq.~(\ref{eq:mainiso}),
while the other remains at its free energy,
as does the doublet.

As a first step in the analysis, 
it is useful to rewrite $F_3^s$ as\footnote{We stress that the matrix $H$ used here has nothing to do with the smooth cutoff function $H(\vec k)$.}
\begin{equation}
F_3^s = \frac1{L^3} \left\{
\frac13 \frac{F^s}{2\omega}
- \frac{F^s}{2\omega} \frac1H
\frac{F^s}{2\omega}\right\}
\,,
\qquad
H =\frac1{2\omega\K^s} + \frac{F^s}{2\omega} + \frac1{2\omega} G^s\,,
\label{eq:F3sdef}
\end{equation}
where $\K^s$, $G^s$ and $F^s$ are defined, respectively,
in Eqs.~(\ref{eq:K2sdef}), (\ref{eq:Gsdef}) and (\ref{eq:Fsdef}).
Recall that all these quantities, as well as $F_3^s$,
are matrices in spectator-momentum space alone, with size $N\times N$.
The form (\ref{eq:F3sdef}) follows by straightforward algebraic 
manipulations from the definition of Eq.~(\ref{eq:F3def}).
One advantage of the new form is that it manifests the symmetry of $F_3^s$,
since $\tilde G^s=\frac1{2\omega}G^s$ is symmetric, 
and all other matrices are diagonal.
Another is that it shows how $F^s$ appears on both ``ends'' of $F_3^s$.

The matrices entering the quantization condition have transformation
properties under the symmetries of the finite box that greatly simplify
their forms. Beginning with $\tilde F_s\equiv F_s/(2\omega)$,
it is clear from its definition, Eq.~(\ref{eq:Fsdef}), that it is invariant
under cubic rotations and parity. Thus its entries, which are all diagonal,
depend only on the class $n$ of the spectator momentum $\vec k$.
For example, all 6 entries for the $n=1$ class are equal.
Thus there are only 4 independent entries in $\tilde F^s$. The same holds for
the other diagonal matrix, $\K^s$.

The situation with $\tilde G^s$ is more complicated---all
entries are non-vanishing.
To simplify $\tilde G^s$ we must decompose the $N$ spectator-momentum
indices into irreps of the finite-box symmetry group, namely
the direct product of the cubic group and parity.
The decomposition is
\begin{align}
n=0 &\longrightarrow A_1^+\,,
\\
n=1 &\longrightarrow A_1^+ + E^+ + T_1^-\,,
\label{eq:n1decomp}
\\
n=2 &\longrightarrow A_1^+ + E^+ + T_1^- + T_2^+ + T_2^-\,,
\\
n=3 &\longrightarrow A_1^+ +       T_1^- + T_2^+ + A_2^-\,,
\end{align}
where the superscript is parity.
$A_2$ is a non-trivial singlet, while $T_1$ and $T_2$ are
three-dimensional irreps.
Off-diagonal elements of $\tilde G^s$ connecting different irreps,
or different elements of the same irrep, vanish.
Thus $\tilde G_s$ is block-diagonal, with a four-dimensional
$A_1^+$ block, a $2\times 2$-dimensional $E^+$ block, etc.

The same block structure holds for $\Kdfth^s$, but here 
additional simplification occurs because of the isotropic approximation,
Eq.~(\ref{eq:isoapprox}). In this approximation, all entries
of $\Kdfth^s$ are equal. Since the $A_1^+$ irreps are obtained by
averaging over their respective momentum shells, while all other
irreps involve differences, 
only the $A_1^+$ block of $\Kdfth^s$ is non-vanishing.
Furthermore, since $\Kdfth^s$ has only a single non-zero eigenvalue, whose eigenvector we call $|1_K\rangle$,
all entries of the $A_1^+$ block are related.
Here we will slightly relax the approximation, so that all
entries which are allowed by symmetries have magnitudes of 
order $\epsilon\ll 1$. Thus we can write
\begin{equation}
\Kdfth^s = |1_K\rangle\,  N \Kdfth^{\rm iso}(E)\, \langle1_K| + 
\left[{\cal O}(\epsilon)\right]
\,,
\label{eq:Kdf3sform}
\end{equation}
where the second term indicates an $N\times N$ matrix
with form consistent with the symmetries 
whose non-zero entries are of ${\cal O}(\epsilon)$
(though unrelated).
If we choose the indices so that the $A_1^+$ block is placed first,
ordered according to the class $n$, the dominant eigenvector is
easily found to be
\begin{equation}
\langle 1_K| = \frac1{\sqrt{N}}(1,\sqrt6,\sqrt{12},\sqrt8,0,\dots)
\,.
\label{eq:dominantevector}
\end{equation}

We now return to the quantization condition (\ref{eq:qcapp}).
Since all matrices which enter are diagonal or block-diagonal,
the condition can be studied block by block.
We begin with the $A_1^+$ block, which is the most interesting
as it contains the dominant eigenvector of $\Kdfth^s$.
Since $\tilde F^s$ and $\K^s$ were diagonal in the original basis,
with no cross terms between different classes $n$,
they remain diagonal in the irrep basis.
In general, there are no relations between the four diagonal elements
of the $A_1^+$ blocks.
As we approach the free-spectrum energies, however,
$\tilde F^s$ does gain further structure. 
This is because it contains poles at these energies 
[from the sum contained in $F^s$, Eq.~(\ref{eq:Fsdef})].
Specifically, one finds that
\begin{equation}
\tilde F_s = {\rm diag}\left(\frac12 B_0 + 3 B_1, B_1,0,0\right)
+ {\rm diag}\left({\cal O}(1)\right)
\,,
\label{eq:Fspole}
\end{equation}
where the pole terms are
\begin{align}
B_0 &= \frac1{L^3} \frac1{(2m)^3} \frac1{E-3m}\,,
\label{eq:B0def}
\\
B_1 &= \frac1{L^3} \frac1{2m(2\omega_1)^2}\frac1{E-m-2\omega_1}
\,.
\label{eq:B1def}
\end{align}
The factor of $3$ multiplying $B_1$ in the first entry
on the right-hand side of Eq.~(\ref{eq:Fspole}) arises 
from the fact that six terms in the sum over $\vec a$ in $F^s$
contribute $B_1/2$.
The second term in $\tilde F^s$ is the non-pole part, arising from
the rest of the sum over $\vec a$ and from the integral.

Pole terms also appear in $\tilde G^s$. Using Eq.~(\ref{eq:Gdef}) we
find the $A_1^+$ block to be
\begin{equation}
\tilde G^s = 
\left(\begin{array}{cccc}
B_0 & \sqrt6 B_1 & {\cal O}(1)& {\cal O}(1) \\
\sqrt6 B_1 & B_1 & {\cal O}(1)& {\cal O}(1) \\
{\cal O}(1)& {\cal O}(1) & {\cal O}(1) & {\cal O}(1) \\
{\cal O}(1)& {\cal O}(1) & {\cal O}(1) & {\cal O}(1) 
\end{array}\right)\,.
\end{equation}
Apart from the fact that the matrix is symmetric, 
there is no relation between the ${\cal O}(1)$ terms.
The combination whose inverse appears in $F_3^s$ 
[Eq.~(\ref{eq:F3sdef})] is thus
\begin{equation}
H \equiv \frac1{2\omega \K^s} + \tilde F^s + \tilde G^s
=
\left(\begin{array}{cccc}
\frac32 B_0 + 3 B_1 & \sqrt6 B_1 & 0& 0 \\
\sqrt6 B_1 & 2 B_1 & 0& 0 \\
0& 0 & 0 & 0 \\
0& 0 & 0 & 0 
\end{array}\right)
+ \left[{\cal O}(1) \right]
\,,
\label{eq:Hres}
\end{equation}
where the ${\cal O}(1)$ symmetric matrix now
contains entries in all positions.

Our task is to combine these forms and insert them in
the quantization condition. Since $B_0$ and $B_1$ become large
for different regions of $E$, we treat these cases one at at time.
The simpler is when $E\approx 3m$, such that $|B_0|\gg 1$,
while $B_1\sim {\cal O}(1)$. Using Raliegh-Schr\"odinger perturbation
theory, one finds that the inverse of $H$ becomes
\begin{equation}
H^{-1} = 
\left(\begin{array}{cc}
\frac2{3B_0} + {\cal O}(1/B_0^2) & {\cal O}(1/B_0) \\
{\cal O}(1/B_0) & {\cal O}(1) 
\end{array}\right) \,.
\label{eq:Hinv}
\end{equation}
Here we are using a block notation in which the first block
has dimension one (the $n=0$ $A_1^+$ subspace)
while the second block has dimension three
(containing the $n=1-3$ $A_1^+$ states).
This result exemplifies two general features of $H^{-1}$
when a one-dimensional subspace of $H$ becomes large.
First, the projection of $H^{-1}$ onto this subspace is,
up to small corrections, simply the inverse of the projection
of $H$ onto the subspace. Thus it is proportional to $1/B_0$.
Second, the off-diagonal elements of the inverse
(those connecting the 1-d subspace to the remainder of the space)
are of ${\cal O}(1/B_0)$.
We use these results again below.

Combining the results above, we find that, when $|B_0|\gg 1$, 
\begin{align}
L^3 F_3^s &= \frac{\tilde F^s}3 - \tilde F_s H^{-1} \tilde F_s
\\
&= 
\left(\begin{array}{cc}
\frac{B_0}6+{\cal O}(1) & 0 \\
0 & {\cal O}(1) 
\end{array}\right)
-
\left(\begin{array}{cc}
\frac{B_0}2+{\cal O}(1) & 0 \\
0 & {\cal O}(1) 
\end{array}\right)
\left(\begin{array}{cc}
\frac2{3B_0} +{\cal O}(1/B_0^2)& {\cal O}(1/B_0) \\
{\cal O}(1/B_0) & {\cal O}(1) 
\end{array}\right)
\left(\begin{array}{cc}
\frac{B_0}2+{\cal O}(1) & 0 \\
0 & {\cal O}(1) 
\end{array}\right)
\\
&= \left[{\cal O}(1)\right] \,,
\label{eq:F3sB0part}
\end{align}
using the same $1+3$ block notation as in Eq.~(\ref{eq:Hinv}).
The key result is that
all terms proportional to positive powers of $B_0$ cancel.
When we combine (\ref{eq:F3sB0part}) with the
result (\ref{eq:Kdf3sform}) for $\Kdfth^s$ and evaluate
the determinant, the quantization condition becomes
\begin{equation}
\label{eq:qcondred1}
\det(1+F_3^s \Kdfth^s)
= 1 + N \Kdfth^{iso}(E) \langle 1_K | F_3^s | 1_K \rangle 
+ {\cal O}(\epsilon)
= 0
\,.
\end{equation}
We see that there is only a single solution in the $A_1^+$ channel,
that given 
essentially by the ``isotropic solution'' of Eq.~(\ref{eq:mainiso}),
aside from small corrections from the $O(\epsilon)$ terms.
There is no possibility of a solution that is $\mathcal O(\epsilon)$ from $E=3m$, 
because there are
no terms of the form $\epsilon B_0$ or $\epsilon B_0^2$, which could have
led to ${\cal O}(1)$ contributions to the quantization condition. 
Such terms are required to cancel the $1$ in Eq.~(\ref{eq:qcondred1}), 
in order to get solutions that are infinitesimally displaced from the the free solution.
\bigskip

The analysis near $E=m+2\omega_1$, when $|B_1|\gg 1$,
is more involved, for in this case the free poles do not cancel.
Using Eq.~(\ref{eq:Hres}), $H$ is now given by
\begin{equation}
H = |B_1\rangle \, 5 B_1 \, \langle B_1| + {\cal O}(1)
\,,
\end{equation}
where the one-dimensional subspace in which $H$ is large is spanned by
\begin{equation}
\langle B_1| = \frac1{\sqrt5} \left(\sqrt3,\sqrt2,0,0\right)
\,.
\end{equation}
Using the results quoted above, we have
\begin{equation}
H^{-1} = 
|B_1\rangle \left[\frac1{5 B_1} +{\cal O}(1/B_1^2)\right] \langle B_1| + 
\sum_{j=2}^4 {\cal O}(1/B_1)
\left(|B_j\rangle \langle B_1| + |B_1\rangle \langle B_j|\right)
+
\sum_{j=2}^4 {\cal O}(1) |B_j\rangle \langle B_j| 
\,,
\end{equation}
where the $|B_j\rangle$, $j=2-4$, are any choice of basis vectors
orthogonal to $|B_1\rangle$. 

To display $F_3^s$ it is better to switch from the ordering
of $A_1^+$ elements according to their momentum class to the
``K-basis''. In this new basis, the vectors are $\langle 1_K |$ together with
\begin{align}
\langle 2_K | &= \frac1{\sqrt7} \left(-\sqrt 6, 1, 0, 0\right)\,,
\\
\langle 3_K | &= 
\frac1{\sqrt{N}} 
\left(-\sqrt{20/7},-\sqrt{120/7},\sqrt{84/20},\sqrt{56/20}
\right)\,,
\\
\langle 4_K | &= \frac1{\sqrt{20}} \left(0, 0, -\sqrt{8}, \sqrt{12}\right)
\,.
\end{align}
This basis makes maximal use of the form of $\Kdfth^s$
[Eq.~(\ref{eq:Kdf3sform})] as well as the fact that the
dominant terms in $\tilde F^s$ lie in the first two entries
on the diagonal.
In addition, we note that $\langle 1_K|$ and $\langle 3_K|$, when contracted
with $\tilde F^s$, have large components only in the subspace
in which $H^{-1}$ is small:
\begin{align}
\langle 1_K | \tilde F^s &= \sqrt{\frac{15}7} B_1 \langle B_1| + {\cal O}(1)
\,,
\\
\langle 3_K | \tilde F^s &= \sqrt{\frac{15}7} B_1 \langle B_1| + {\cal O}(1)
\,.
\end{align}
Also important is that no large terms appear when we contract 
$\langle 4_K |$ with $\tilde F^s$:
\begin{align}
\langle 4_K | \tilde F^s &= {\cal O}(1) 
\,.
\end{align}

The net effect of these results is that the largest contributions
occur when $F_3^s$ is contracted with $\langle 2_K|$.
Specifically, we find
\begin{equation}
F_3^s = 
\left(\begin{array}{cccc}
{\cal O}(1) & {\cal O}(B_1) & {\cal O}(1)& {\cal O}(1) \\
{\cal O}(B_1) & {\cal O}(B_1^2) & {\cal O}(B_1)& {\cal O}(B_1) \\
{\cal O}(1)& {\cal O}(B_1) & {\cal O}(1) & {\cal O}(1) \\
{\cal O}(1)& {\cal O}(B_1) & {\cal O}(1) & {\cal O}(1) 
\end{array}\right)_K
\,,
\label{eq:F3sKbasis}
\end{equation}
where the subscript on the matrix indicates that this result
is in the K-basis.
We see that $F_3^s$ {\em does} contain
single and double $B_1$ pole terms 
(unlike for $B_0$, where they cancel).
There is one cancellation, however,
which leads to the absence of pole terms in the $|1_K\rangle \langle 1_K|$ 
element.
This is important, since we know from Eq.~(\ref{eq:Kdf3sform}) that
the only large entry in $\Kdfth^s$ is in exactly this element:
\begin{equation}
\Kdfth^s =
\left(\begin{array}{cccc}
N \Kdfth^{\rm iso}(E)& {\cal O}(\epsilon)& {\cal O}(\epsilon)& {\cal O}(\epsilon)\\
{\cal O}(\epsilon)& {\cal O}(\epsilon)& {\cal O}(\epsilon)& {\cal O}(\epsilon)\\
{\cal O}(\epsilon)& {\cal O}(\epsilon)& {\cal O}(\epsilon)& {\cal O}(\epsilon)\\
{\cal O}(\epsilon)& {\cal O}(\epsilon)& {\cal O}(\epsilon)& {\cal O}(\epsilon)
\end{array}\right)_K
\,.
\end{equation}

The final step of the $B_1$ analysis is to insert these results into the
quantization condition (\ref{eq:qcapp}) and evaluate the determinant.
We are looking for solutions which occur when $B_1$ is large, so that
they are almost at the free-particle energy. 
By explicit evaluation, we find that the dominant contributions
to the determinant involving $B_1$ are of ${\cal O}(\epsilon B_1^2)$.
Thus the appropriate scaling of $B_1$ relative to $\epsilon$
is such that $\epsilon B_1^2={\cal O}(1)$.
Using this scaling, it turns out that only the upper-left
$2\times 2$ blocks in the K-basis are relevant for solutions to the
quantization condition. Other blocks lead to contributions
proportional to $\epsilon B_1$, which remains small. 
Thus, for the purpose of finding solutions to the quantization
condition we can make the replacements
\begin{equation}
F_3^s \longrightarrow
\left(\begin{array}{cccc}
{\cal O}(1) & {\cal O}(B_1) & 0& 0\\
{\cal O}(B_1) & {\cal O}(B_1^2) & 0& 0\\
0& 0& 0& 0\\
0& 0& 0& 0
\end{array}\right)
\ \ {\rm and}\ \ 
\Kdfth^s \longrightarrow
\left(\begin{array}{cccc}
N \Kdfth^{\rm iso}(E)& {\cal O}(\epsilon)& 0& 0\\
{\cal O}(\epsilon)& {\cal O}(\epsilon)& 0& 0\\
0& 0& 0& 0\\
0& 0& 0& 0
\end{array}\right)
\,.
\end{equation}
This shows that, in the $A_1^+$ block, the quantization
condition involves only two states, and not four:
\begin{equation}
\det
\left(\begin{array}{cc}
1 + N \Kdfth^{\rm iso}(E)\langle 1_K|F^s|1_K\rangle +{\cal O}(\epsilon B_1) & 
{\cal O}(\epsilon B_1)\\
{\cal O}(B_1)& 1+ {\cal O}(\epsilon B_1^2)
\end{array}\right) = 0
\end{equation}
When $E$ is far from $m+2\omega_1$,
so that $B_1={\cal O}(1)$, the $B_1$ terms are small,
and one finds only the isotropic solution, Eq.~(\ref{eq:mainiso}).
But now, when $\epsilon B_1^2={\cal O}(1)$, there is the possibility
of a second solution. To demonstrate the existence and to find
position of this solution, however, appears to require knowledge 
of the subdominant parts of $F^s$, $\K^s$ and $\Kdfth^s$.
Nevertheless, what is clear is that any solution will lie
very close to the free-particle energy, since it will require
$|B_1|\gg 1$.

\bigskip
We now turn to blocks of the matrices in other irreps.
These can only lead to solutions close to free-particle energies since
$\Kdfth^s$ is of ${\cal O}(\epsilon)$ throughout these blocks.
Such solutions require factors of $B_1$ to counterbalance those
of $\epsilon$. $B_1$ appears in $\tilde F^s$ in all
diagonal elements with spectator momentum of class $n=1$,
and thus [see Eq.~(\ref{eq:n1decomp})]
appears in both $E^+$ and $T_1^-$ blocks.
The same can be seen to hold for $\tilde G^s$.

We consider first the $E^+$ block. This has dimension four,
since there are $E^+$ irreps in both $n=1$ and $n=2$ classes,
while the $E^+$ irrep itself is a doublet.
The structure within each $E^+$ irrep is, however, always
proportional to the identity matrix. Thus we display the
blocks as $2\times 2$ matrices, each element of which is implicitly
proportional to the identity matrix.
The matrices have the form
\begin{equation}
\tilde F^s = \left(\begin{array}{cc}
B_1 +{\cal O}(1)& 0 \\ 0 & {\cal O}(1) 
\end{array}\right)\,,\ \
H = \left(\begin{array}{cc}
2B_1 + {\cal O}(1)& {\cal O}(1) \\ {\cal O}(1) & {\cal O}(1)
\end{array}\right)\,, \ \
\Kdfth^s = \left(\begin{array}{cc}
{\cal O}(\epsilon)& {\cal O}(\epsilon) \\ {\cal O}(\epsilon) & {\cal O}(\epsilon)
\end{array}\right)\,, 
\end{equation}
from which it follows that
\begin{equation}
F_3^s = \left(\begin{array}{cc}
-\frac{B_1}6 +{\cal O}(1)& {\cal O}(1) \\ {\cal O}(1) & {\cal O}(1) 
\end{array}\right)\,,\ \
1 + F_3^s \Kdfth^s
=
\left(\begin{array}{cc}
1+{\cal O}(\epsilon B_1)& {\cal O}(\epsilon B_1) 
\\ 
{\cal O}(\epsilon) & 1+ {\cal O}(\epsilon) 
\end{array}\right)\,.
\end{equation}
Thus the determinant is $1 + {\cal O}(\epsilon B_1)$,
and can vanish if $\epsilon B_1 ={\cal O}(1)$. 
In this case we know that such a solution will exist,
irrespective of the overall sign of the $\epsilon B_1$ term,
since $B_1$ can take either sign, depending on whether $E$ is
above or below $m+2\omega_1$.
Thus we conclude that the $E^+$ irrep yields a solution with
$E\approx m+2\omega_1$.
Recalling the implicit $2\times 2$ identity matrix in each entry,
this will be a degenerate doublet.

\bigskip

At this stage we have uncovered all the solutions we want---four in all
(assuming that the $A_1^+$ quantization condition does have an almost-free
solution). But there remains the $T_1^-$ block in which both 
$\tilde F^s$ and $\tilde G^s$ have entries of $B_1$.
If these end up multiplying factors of $\epsilon$, as in the $E^+$ block,
then there is potential for unwanted solutions to the quantization
condition, corresponding to states which violate particle-interchange
symmetry. The way in which the formalism avoids this is through
the particle-interchange symmetry that has been carefully
maintained in $\Kdfth$. The issue is subtle, however, because
our coordinates, and, in particular, the truncation we are using,
is not particle-interchange symmetric.

The $T_1^-$ block has contributions from classes $n=1$, $2$ and $3$,
each of which is three-dimensional, so the overall block dimension is nine.
Entries within each $3\times 3$ sub-block are, however,
proportional to the identity matrix, so we leave this implicit and
display only the $3\times 3$ matrix indexed by momentum class.
We find
\begin{equation}
\tilde F^s = \left(\begin{array}{ccc}
B_1 +{\cal O}(1)& 0 & 0\\ 
0 & {\cal O}(1) & 0 \\
0 & 0 & {\cal O}(1)
\end{array}\right)\,,\ \
\tilde G^s = \left(\begin{array}{ccc}
-B_1 +{\cal O}(1)& {\cal O}(1) & {\cal O}(1)\\ 
{\cal O}(1) & {\cal O}(1) & {\cal O}(1) \\
{\cal O}(1) & {\cal O}(1) & {\cal O}(1)
\end{array}\right)\,,
\label{eq:FsGsT1}
\end{equation}
where the minus sign on $B_1$ in $\tilde G^s$ arises
from the negative parity of the $T_1^-$ irrep and
the fact that the non-zero elements of $\tilde G^s$ in the
original basis are those connecting an $n=1$ momentum to its
parity conjugate.
It follows from (\ref{eq:FsGsT1})
that $B_1$ cancels from $H$, so that $H^{-1}$ is
a general symmetric ${\cal O}(1)$ matrix with no small elements. 
We then find
\begin{equation}
F_3^s = \left(\begin{array}{ccc}
{\cal O}(B_1^2)& {\cal O}(B_1)& {\cal O}(B_1) \\ 
{\cal O}(B_1) & {\cal O}(1) & {\cal O}(1) \\
{\cal O}(B_1) & {\cal O}(1) & {\cal O}(1) 
\end{array}\right)\,,
\label{eq:F3sT1}
\end{equation}
so that, as in the $E^+$ block, $F_3^s$ contains single
and double poles.

If $\Kdfth^s$ was simply a symmetric matrix containing
terms of ${\cal O}(\epsilon)$, then an analysis similar
to that for the $E^+$ block would imply the presence of
almost-free solutions to the quantization condition in the 
$T_1^-$ block. These would be unexpected, and indicate
that our formalism was violating particle-interchange symmetry
in a fundamental way.
We are saved from this conclusion by the presence of
additional structure in $\Kdfth^s$, following,
not surprisingly, from particle-interchange symmetry. 
We recall that, before truncation, $\Kdfth$ is, by construction,
exactly symmetric under particle interchange.
We argue below that a consequence of this symmetry
is that, if $E=m+2\omega_1$ 
(so that the free three-particle state is exactly on shell), then
\begin{equation}
\left[\Kdfth^s\right]_{(1,0,0);k'} = \left[\Kdfth^s\right]_{(-1,0,0);k'}\,,
\label{eq:Kdfthparity}
\end{equation}
(and similarly for the permutations of the left-hand momentum index).
Here the right-hand index indicates an arbitrary momentum.
The key feature of this result is that only the 
first index is parity-inverted---the second is unchanged.
This implies that the projection on the left-hand index onto
the irrep $T_1^-$, which involves taking the difference between
the two sides of Eq.~(\ref{eq:Kdfthparity}), vanishes identically.
As we move away from $E=m+2\omega_1$, the two sides start to differ,
but we expect this difference to grow at least linearly
in $E-(m+2\omega_1)\propto 1/B_1$.

The upshot is that particle-interchange symmetry leads to the
following form for $\Kdfth^s$:
\begin{equation}
\Kdfth^s = \left(\begin{array}{ccc}
{\cal O}(\epsilon/B_1^2)& {\cal O}(\epsilon/B_1)& {\cal O}(\epsilon/B_1) \\ 
{\cal O}(\epsilon/B_1) & {\cal O}(\epsilon) & {\cal O}(\epsilon) \\
{\cal O}(\epsilon/B_1) & {\cal O}(\epsilon) & {\cal O}(\epsilon) 
\end{array}\right)\,.
\end{equation}
The top-left element is doubly suppressed because it 
involves a cancellation of the type just described for both
left and right-hand indices.
Combined with the result for $F_3^s$, Eq.~(\ref{eq:F3sT1}),
this implies that
$\det(1 + F_3^s \Kdfth^s)=1 + {\cal O}(\epsilon)$.
Thus there are no solutions to the quantization condition,
and no unwanted states.\footnote{%
There is another source of suppression arising from particle-interchange
symmetry, arising from the endcaps $A$ and $A'$ [see Eq.~(\ref{eq:CLB2resd})].
These have vanishing coupling to the symmetry-violating
states when $E=m+2\omega_1$. However, this alone would not be enough to
remove these states from the spectrum if their energies were shifted
slightly from the free-particle value.}

We now demonstrate Eq.~(\ref{eq:Kdfthparity}). 
On the left-hand side
the spectator momentum is $\vec k=(1,0,0)$,
so the total momentum of the other two particles is $-\vec k$.
By assumption (given our truncation) the amplitude in the CM frame
of the other two particles is independent of angle.
For one choice of angle the other two momenta are $\vec 0$ and $-\vec k$
(since this gives the correct energy $E$).
Thus the amplitude on the left-hand side of (\ref{eq:Kdfthparity})
is equal to the original $\Kdfth$ (with no superscript $s$)
when the three outgoing momenta are $\vec k$, $\vec 0$ and $-\vec k$.
By exactly the same argument, the amplitude on the right-hand side
equals $\Kdfth$ for outgoing momenta $-\vec k$, $\vec 0$ and $\vec k$.
But since $\Kdfth$ is symmetric under incoming particle exchange,
the amplitudes on the two sides are equal.

%% file: 3pi.bbl
\begin{thebibliography}{30}%
\makeatletter
\providecommand \@ifxundefined [1]{%
 \@ifx{#1\undefined}
}%
\providecommand \@ifnum [1]{%
 \ifnum #1\expandafter \@firstoftwo
 \else \expandafter \@secondoftwo
 \fi
}%
\providecommand \@ifx [1]{%
 \ifx #1\expandafter \@firstoftwo
 \else \expandafter \@secondoftwo
 \fi
}%
\providecommand \natexlab [1]{#1}%
\providecommand \enquote  [1]{``#1''}%
\providecommand \bibnamefont  [1]{#1}%
\providecommand \bibfnamefont [1]{#1}%
\providecommand \citenamefont [1]{#1}%
\providecommand \href@noop [0]{\@secondoftwo}%
\providecommand \href [0]{\begingroup \@sanitize@url \@href}%
\providecommand \@href[1]{\@@startlink{#1}\@@href}%
\providecommand \@@href[1]{\endgroup#1\@@endlink}%
\providecommand \@sanitize@url [0]{\catcode `\\12\catcode `\$12\catcode
  `\&12\catcode `\#12\catcode `\^12\catcode `\_12\catcode `\%12\relax}%
\providecommand \@@startlink[1]{}%
\providecommand \@@endlink[0]{}%
\providecommand \url  [0]{\begingroup\@sanitize@url \@url }%
\providecommand \@url [1]{\endgroup\@href {#1}{\urlprefix }}%
\providecommand \urlprefix  [0]{URL }%
\providecommand \Eprint [0]{\href }%
\providecommand \doibase [0]{http://dx.doi.org/}%
\providecommand \selectlanguage [0]{\@gobble}%
\providecommand \bibinfo  [0]{\@secondoftwo}%
\providecommand \bibfield  [0]{\@secondoftwo}%
\providecommand \translation [1]{[#1]}%
\providecommand \BibitemOpen [0]{}%
\providecommand \bibitemStop [0]{}%
\providecommand \bibitemNoStop [0]{.\EOS\space}%
\providecommand \EOS [0]{\spacefactor3000\relax}%
\providecommand \BibitemShut  [1]{\csname bibitem#1\endcsname}%
\let\auto@bib@innerbib\@empty
\bibitem [{\citenamefont {Prelovsek}(2013)}]{Prelovsek:2013cta}%
  \BibitemOpen
  \bibfield  {author} {\bibinfo {author} {\bibfnamefont {S.}~\bibnamefont
  {Prelovsek}},\ }\href@noop {} {\  (\bibinfo {year} {2013})},\ \Eprint
  {http://arxiv.org/abs/1310.4354} {arXiv:1310.4354 [hep-lat]} \BibitemShut
  {NoStop}%
\bibitem [{\citenamefont {Thomas}(2014)}]{Thomas:2014dpa}%
  \BibitemOpen
  \bibfield  {author} {\bibinfo {author} {\bibfnamefont {C.}~\bibnamefont
  {Thomas}},\ }\href@noop {} {\bibfield  {journal} {\bibinfo  {journal} {PoS}\
  }\textbf {\bibinfo {volume} {LATTICE2013}},\ \bibinfo {pages} {003} (\bibinfo
  {year} {2014})}\BibitemShut {NoStop}%
\bibitem [{\citenamefont {Doering}(2014)}]{Doering:2014fpa}%
  \BibitemOpen
  \bibfield  {author} {\bibinfo {author} {\bibfnamefont {M.}~\bibnamefont
  {D\"oring}},\ }\href@noop {} {\bibfield  {journal} {\bibinfo  {journal} {PoS}\
  }\textbf {\bibinfo {volume} {LATTICE2013}},\ \bibinfo {pages} {006} (\bibinfo
  {year} {2014})}\BibitemShut {NoStop}%
\bibitem [{\citenamefont {Luscher}(1986)}]{Luscher:1986n2}%
  \BibitemOpen
  \bibfield  {author} {\bibinfo {author} {\bibfnamefont {M.}~\bibnamefont
  {L\" uscher}},\ }\href {\doibase 10.1007/BF01211097} {\bibfield  {journal}
  {\bibinfo  {journal} {Commun.Math.Phys.}\ }\textbf {\bibinfo {volume}
  {105}},\ \bibinfo {pages} {153} (\bibinfo {year} {1986})}\BibitemShut
  {NoStop}%
\bibitem [{\citenamefont {Luscher}(1991{\natexlab{a}})}]{Luscher:1991n1}%
  \BibitemOpen
  \bibfield  {author} {\bibinfo {author} {\bibfnamefont {M.}~\bibnamefont
  {L\" uscher}},\ }\href {\doibase 10.1016/0550-3213(91)90366-6} {\bibfield
  {journal} {\bibinfo  {journal} {Nucl.Phys.}\ }\textbf {\bibinfo {volume}
  {B354}},\ \bibinfo {pages} {531} (\bibinfo {year}
  {1991}{\natexlab{a}})}\BibitemShut {NoStop}%
\bibitem [{\citenamefont {Luscher}(1991{\natexlab{b}})}]{Luscher:1991n2}%
  \BibitemOpen
  \bibfield  {author} {\bibinfo {author} {\bibfnamefont {M.}~\bibnamefont
  {L\" uscher}},\ }\href {\doibase 10.1016/0550-3213(91)90584-K} {\bibfield
  {journal} {\bibinfo  {journal} {Nucl.Phys.}\ }\textbf {\bibinfo {volume}
  {B364}},\ \bibinfo {pages} {237} (\bibinfo {year}
  {1991}{\natexlab{b}})}\BibitemShut {NoStop}%
\bibitem [{\citenamefont {Lage}\ \emph {et~al.}(2009)\citenamefont {Lage},
  \citenamefont {Meissner},\ and\ \citenamefont {Rusetsky}}]{Lage:2009}%
  \BibitemOpen
  \bibfield  {author} {\bibinfo {author} {\bibfnamefont {M.}~\bibnamefont
  {Lage}}, \bibinfo {author} {\bibfnamefont {U.-G.}\ \bibnamefont {Meissner}},
  \ and\ \bibinfo {author} {\bibfnamefont {A.}~\bibnamefont {Rusetsky}},\
  }\href {\doibase 10.1016/j.physletb.2009.10.055} {\bibfield  {journal}
  {\bibinfo  {journal} {Phys.Lett.}\ }\textbf {\bibinfo {volume} {B681}},\
  \bibinfo {pages} {439} (\bibinfo {year} {2009})},\ \Eprint
  {http://arxiv.org/abs/0905.0069} {arXiv:0905.0069 [hep-lat]} \BibitemShut
  {NoStop}%
\bibitem [{\citenamefont {Bernard}\ \emph {et~al.}(2011)\citenamefont
  {Bernard}, \citenamefont {Lage}, \citenamefont {Meissner},\ and\
  \citenamefont {Rusetsky}}]{Bernard:2010}%
  \BibitemOpen
  \bibfield  {author} {\bibinfo {author} {\bibfnamefont {V.}~\bibnamefont
  {Bernard}}, \bibinfo {author} {\bibfnamefont {M.}~\bibnamefont {Lage}},
  \bibinfo {author} {\bibfnamefont {U.-G.}\ \bibnamefont {Meissner}}, \ and\
  \bibinfo {author} {\bibfnamefont {A.}~\bibnamefont {Rusetsky}},\ }\href
  {\doibase 10.1007/JHEP01(2011)019} {\bibfield  {journal} {\bibinfo  {journal}
  {JHEP}\ }\textbf {\bibinfo {volume} {1101}},\ \bibinfo {pages} {019}
  (\bibinfo {year} {2011})},\ \Eprint {http://arxiv.org/abs/1010.6018}
  {arXiv:1010.6018 [hep-lat]} \BibitemShut {NoStop}%
\bibitem [{\citenamefont {Doring}\ \emph {et~al.}(2011)\citenamefont {Doring},
  \citenamefont {Meissner}, \citenamefont {Oset},\ and\ \citenamefont
  {Rusetsky}}]{Doering:2011}%
  \BibitemOpen
  \bibfield  {author} {\bibinfo {author} {\bibfnamefont {M.}~\bibnamefont
  {D\"oring}}, \bibinfo {author} {\bibfnamefont {U.-G.}\ \bibnamefont
  {Meissner}}, \bibinfo {author} {\bibfnamefont {E.}~\bibnamefont {Oset}}, \
  and\ \bibinfo {author} {\bibfnamefont {A.}~\bibnamefont {Rusetsky}},\ }\href
  {\doibase 10.1140/epja/i2011-11139-7} {\bibfield  {journal} {\bibinfo
  {journal} {Eur.Phys.J.}\ }\textbf {\bibinfo {volume} {A47}},\ \bibinfo
  {pages} {139} (\bibinfo {year} {2011})},\ \Eprint
  {http://arxiv.org/abs/1107.3988} {arXiv:1107.3988 [hep-lat]} \BibitemShut
  {NoStop}%
\bibitem [{\citenamefont {Hansen}\ and\ \citenamefont
  {Sharpe}(2012)}]{Hansen:2012tf}%
  \BibitemOpen
  \bibfield  {author} {\bibinfo {author} {\bibfnamefont {M.~T.}\ \bibnamefont
  {Hansen}}\ and\ \bibinfo {author} {\bibfnamefont {S.~R.}\ \bibnamefont
  {Sharpe}},\ }\href {\doibase 10.1103/PhysRevD.86.016007} {\bibfield
  {journal} {\bibinfo  {journal} {Phys.Rev.}\ }\textbf {\bibinfo {volume}
  {D86}},\ \bibinfo {pages} {016007} (\bibinfo {year} {2012})},\ \Eprint
  {http://arxiv.org/abs/1204.0826} {arXiv:1204.0826 [hep-lat]} \BibitemShut
  {NoStop}%
\bibitem [{\citenamefont {Briceno}\ and\ \citenamefont
  {Davoudi}(2013{\natexlab{a}})}]{Briceno:2012yi}%
  \BibitemOpen
  \bibfield  {author} {\bibinfo {author} {\bibfnamefont {R.~A.}\ \bibnamefont
  {Brice\~no}}\ and\ \bibinfo {author} {\bibfnamefont {Z.}~\bibnamefont
  {Davoudi}},\ }\href {\doibase 10.1103/PhysRevD.88.094507} {\bibfield
  {journal} {\bibinfo  {journal} {Phys.Rev.}\ }\textbf {\bibinfo {volume}
  {D88}},\ \bibinfo {pages} {094507} (\bibinfo {year} {2013}{\natexlab{a}})},\
  \Eprint {http://arxiv.org/abs/1204.1110} {arXiv:1204.1110 [hep-lat]}
  \BibitemShut {NoStop}%
\bibitem [{\citenamefont {Dudek}\ \emph {et~al.}(2014)\citenamefont {Dudek},
  \citenamefont {Edwards}, \citenamefont {Thomas},\ and\ \citenamefont
  {Wilson}}]{Dudek:2014qha}%
  \BibitemOpen
  \bibfield  {author} {\bibinfo {author} {\bibfnamefont {J.~J.}\ \bibnamefont
  {Dudek}}, \bibinfo {author} {\bibfnamefont {R.~G.}\ \bibnamefont {Edwards}},
  \bibinfo {author} {\bibfnamefont {C.~E.}\ \bibnamefont {Thomas}}, \ and\
  \bibinfo {author} {\bibfnamefont {D.~J.}\ \bibnamefont {Wilson}},\
  }\href@noop {} {\  (\bibinfo {year} {2014})},\ \Eprint
  {http://arxiv.org/abs/1406.4158} {arXiv:1406.4158 [hep-ph]} \BibitemShut
  {NoStop}%
\bibitem [{\citenamefont {Polejaeva}\ and\ \citenamefont
  {Rusetsky}(2012)}]{Polejaeva:2012ut}%
  \BibitemOpen
  \bibfield  {author} {\bibinfo {author} {\bibfnamefont {K.}~\bibnamefont
  {Polejaeva}}\ and\ \bibinfo {author} {\bibfnamefont {A.}~\bibnamefont
  {Rusetsky}},\ }\href {\doibase 10.1140/epja/i2012-12067-8} {\bibfield
  {journal} {\bibinfo  {journal} {Eur.Phys.J.}\ }\textbf {\bibinfo {volume}
  {A48}},\ \bibinfo {pages} {67} (\bibinfo {year} {2012})},\ \Eprint
  {http://arxiv.org/abs/1203.1241} {arXiv:1203.1241 [hep-lat]} \BibitemShut
  {NoStop}%
\bibitem [{\citenamefont {Briceno}\ and\ \citenamefont
  {Davoudi}(2013{\natexlab{b}})}]{Briceno:2012rv}%
  \BibitemOpen
  \bibfield  {author} {\bibinfo {author} {\bibfnamefont {R.~A.}\ \bibnamefont
  {Brice\~no}}\ and\ \bibinfo {author} {\bibfnamefont {Z.}~\bibnamefont
  {Davoudi}},\ }\href {\doibase 10.1103/PhysRevD.87.094507} {\bibfield
  {journal} {\bibinfo  {journal} {Phys.Rev.}\ }\textbf {\bibinfo {volume}
  {D87}},\ \bibinfo {pages} {094507} (\bibinfo {year} {2013}{\natexlab{b}})},\
  \Eprint {http://arxiv.org/abs/1212.3398} {arXiv:1212.3398 [hep-lat]}
  \BibitemShut {NoStop}%
\bibitem [{\citenamefont {Guo}(2013)}]{Guo:2013qla}%
  \BibitemOpen
  \bibfield  {author} {\bibinfo {author} {\bibfnamefont {P.}~\bibnamefont
  {Guo}},\ }\href@noop {} {\  (\bibinfo {year} {2013})},\ \Eprint
  {http://arxiv.org/abs/1303.3349} {arXiv:1303.3349 [hep-lat]} \BibitemShut
  {NoStop}%
\bibitem [{\citenamefont {Aoki}\ \emph {et~al.}(2013)\citenamefont {Aoki},
  \citenamefont {Ishii}, \citenamefont {Doi}, \citenamefont {Ikeda},\ and\
  \citenamefont {Inoue}}]{Aoki:2013cra}%
  \BibitemOpen
  \bibfield  {author} {\bibinfo {author} {\bibfnamefont {S.}~\bibnamefont
  {Aoki}}, \bibinfo {author} {\bibfnamefont {N.}~\bibnamefont {Ishii}},
  \bibinfo {author} {\bibfnamefont {T.}~\bibnamefont {Doi}}, \bibinfo {author}
  {\bibfnamefont {Y.}~\bibnamefont {Ikeda}}, \ and\ \bibinfo {author}
  {\bibfnamefont {T.}~\bibnamefont {Inoue}},\ }\href {\doibase
  10.1103/PhysRevD.88.014036} {\bibfield  {journal} {\bibinfo  {journal}
  {Phys.Rev.}\ }\textbf {\bibinfo {volume} {D88}},\ \bibinfo {pages} {014036}
  (\bibinfo {year} {2013})},\ \Eprint {http://arxiv.org/abs/1303.2210}
  {arXiv:1303.2210 [hep-lat]} \BibitemShut {NoStop}%
\bibitem [{\citenamefont {Kim}\ \emph {et~al.}(2005)\citenamefont {Kim},
  \citenamefont {Sachrajda},\ and\ \citenamefont {Sharpe}}]{Kim:2005}%
  \BibitemOpen
  \bibfield  {author} {\bibinfo {author} {\bibfnamefont {C.}~\bibnamefont
  {Kim}}, \bibinfo {author} {\bibfnamefont {C.}~\bibnamefont {Sachrajda}}, \
  and\ \bibinfo {author} {\bibfnamefont {S.~R.}\ \bibnamefont {Sharpe}},\
  }\href {\doibase 10.1016/j.nuclphysb.2005.08.029} {\bibfield  {journal}
  {\bibinfo  {journal} {Nucl.Phys.}\ }\textbf {\bibinfo {volume} {B727}},\
  \bibinfo {pages} {218} (\bibinfo {year} {2005})},\ \Eprint
  {http://arxiv.org/abs/hep-lat/0507006} {arXiv:hep-lat/0507006 [hep-lat]}
  \BibitemShut {NoStop}%
\bibitem [{\citenamefont {Hansen}\ and\ \citenamefont
  {Sharpe}(2014)}]{Hansen:2013dla}%
  \BibitemOpen
  \bibfield  {author} {\bibinfo {author} {\bibfnamefont {M.~T.}\ \bibnamefont
  {Hansen}}\ and\ \bibinfo {author} {\bibfnamefont {S.~R.}\ \bibnamefont
  {Sharpe}},\ }\href@noop {} {\bibfield  {journal} {\bibinfo  {journal} {PoS}\
  }\textbf {\bibinfo {volume} {LATTICE2013}},\ \bibinfo {pages} {221} (\bibinfo
  {year} {2014})},\ \Eprint {http://arxiv.org/abs/1311.4848} {arXiv:1311.4848
  [hep-lat]} \BibitemShut {NoStop}%
\bibitem [{\citenamefont {Rubin}\ \emph {et~al.}(1966)\citenamefont {Rubin},
  \citenamefont {Sugar},\ and\ \citenamefont {Tiktopoulos}}]{Rubin:1966zz}%
  \BibitemOpen
  \bibfield  {author} {\bibinfo {author} {\bibfnamefont {M.}~\bibnamefont
  {Rubin}}, \bibinfo {author} {\bibfnamefont {R.}~\bibnamefont {Sugar}}, \ and\
  \bibinfo {author} {\bibfnamefont {G.}~\bibnamefont {Tiktopoulos}},\ }\href
  {\doibase 10.1103/PhysRev.146.1130} {\bibfield  {journal} {\bibinfo
  {journal} {Phys.Rev.}\ }\textbf {\bibinfo {volume} {146}},\ \bibinfo {pages}
  {1130} (\bibinfo {year} {1966})}\BibitemShut {NoStop}%
\bibitem [{\citenamefont {Brayshaw}(1968)}]{Brayshaw:1969ab}%
  \BibitemOpen
  \bibfield  {author} {\bibinfo {author} {\bibfnamefont {D.}~\bibnamefont
  {Brayshaw}},\ }\href {\doibase 10.1103/PhysRev.176.1855} {\bibfield
  {journal} {\bibinfo  {journal} {Phys.Rev.}\ }\textbf {\bibinfo {volume}
  {176}},\ \bibinfo {pages} {1855} (\bibinfo {year} {1968})}\BibitemShut
  {NoStop}%
\bibitem [{\citenamefont {Potapov}\ and\ \citenamefont
  {Taylor}(1977{\natexlab{a}})}]{Taylor:1977A}%
  \BibitemOpen
  \bibfield  {author} {\bibinfo {author} {\bibfnamefont {V.}~\bibnamefont
  {Potapov}}\ and\ \bibinfo {author} {\bibfnamefont {J.}~\bibnamefont
  {Taylor}},\ }\href@noop {} {\bibfield  {journal} {\bibinfo  {journal}
  {Phys.Rev.}\ }\textbf {\bibinfo {volume} {A16}},\ \bibinfo {pages} {2264}
  (\bibinfo {year} {1977}{\natexlab{a}})}\BibitemShut {NoStop}%
\bibitem [{\citenamefont {Potapov}\ and\ \citenamefont
  {Taylor}(1977{\natexlab{b}})}]{Taylor:1977B}%
  \BibitemOpen
  \bibfield  {author} {\bibinfo {author} {\bibfnamefont {V.}~\bibnamefont
  {Potapov}}\ and\ \bibinfo {author} {\bibfnamefont {J.}~\bibnamefont
  {Taylor}},\ }\href@noop {} {\bibfield  {journal} {\bibinfo  {journal}
  {Phys.Rev.}\ }\textbf {\bibinfo {volume} {A16}},\ \bibinfo {pages} {2276}
  (\bibinfo {year} {1977}{\natexlab{b}})}\BibitemShut {NoStop}%
\bibitem [{\citenamefont {Beane}\ \emph {et~al.}(2004)\citenamefont {Beane},
  \citenamefont {Bedaque}, \citenamefont {Parreno},\ and\ \citenamefont
  {Savage}}]{Beane:2003da}%
  \BibitemOpen
  \bibfield  {author} {\bibinfo {author} {\bibfnamefont {S.}~\bibnamefont
  {Beane}}, \bibinfo {author} {\bibfnamefont {P.}~\bibnamefont {Bedaque}},
  \bibinfo {author} {\bibfnamefont {A.}~\bibnamefont {Parreno}}, \ and\
  \bibinfo {author} {\bibfnamefont {M.}~\bibnamefont {Savage}},\ }\href
  {\doibase 10.1016/j.physletb.2004.02.007} {\bibfield  {journal} {\bibinfo
  {journal} {Phys.Lett.}\ }\textbf {\bibinfo {volume} {B585}},\ \bibinfo
  {pages} {106} (\bibinfo {year} {2004})},\ \Eprint
  {http://arxiv.org/abs/hep-lat/0312004} {arXiv:hep-lat/0312004} \BibitemShut
  {NoStop}%
\bibitem [{\citenamefont {Sasaki}\ and\ \citenamefont
  {Yamazaki}(2006)}]{Sasaki:2006jn}%
  \BibitemOpen
  \bibfield  {author} {\bibinfo {author} {\bibfnamefont {S.}~\bibnamefont
  {Sasaki}}\ and\ \bibinfo {author} {\bibfnamefont {T.}~\bibnamefont
  {Yamazaki}},\ }\href {\doibase 10.1103/PhysRevD.74.114507} {\bibfield
  {journal} {\bibinfo  {journal} {Phys.Rev.}\ }\textbf {\bibinfo {volume}
  {D74}},\ \bibinfo {pages} {114507} (\bibinfo {year} {2006})},\ \Eprint
  {http://arxiv.org/abs/hep-lat/0610081} {arXiv:hep-lat/0610081} \BibitemShut
  {NoStop}%
\bibitem [{\citenamefont {Rummukainen}\ and\ \citenamefont
  {Gottlieb}(1995)}]{Kari:1995}%
  \BibitemOpen
  \bibfield  {author} {\bibinfo {author} {\bibfnamefont {K.}~\bibnamefont
  {Rummukainen}}\ and\ \bibinfo {author} {\bibfnamefont {S.~A.}\ \bibnamefont
  {Gottlieb}},\ }\href {\doibase 10.1016/0550-3213(95)00313-H} {\bibfield
  {journal} {\bibinfo  {journal} {Nucl.Phys.}\ }\textbf {\bibinfo {volume}
  {B450}},\ \bibinfo {pages} {397} (\bibinfo {year} {1995})},\ \Eprint
  {http://arxiv.org/abs/hep-lat/9503028} {arXiv:hep-lat/9503028} \BibitemShut
  {NoStop}%
\bibitem [{\citenamefont {Beane}\ \emph {et~al.}(2007)\citenamefont {Beane},
  \citenamefont {Detmold},\ and\ \citenamefont {Savage}}]{Beane:2007qr}%
  \BibitemOpen
  \bibfield  {author} {\bibinfo {author} {\bibfnamefont {S.~R.}\ \bibnamefont
  {Beane}}, \bibinfo {author} {\bibfnamefont {W.}~\bibnamefont {Detmold}}, \
  and\ \bibinfo {author} {\bibfnamefont {M.~J.}\ \bibnamefont {Savage}},\
  }\href {\doibase 10.1103/PhysRevD.76.074507} {\bibfield  {journal} {\bibinfo
  {journal} {Phys.Rev.}\ }\textbf {\bibinfo {volume} {D76}},\ \bibinfo {pages}
  {074507} (\bibinfo {year} {2007})},\ \Eprint {http://arxiv.org/abs/0707.1670}
  {arXiv:0707.1670 [hep-lat]} \BibitemShut {NoStop}%
\bibitem [{\citenamefont {Tan}(2008)}]{Tan:2007}%
  \BibitemOpen
  \bibfield  {author} {\bibinfo {author} {\bibfnamefont {S.}~\bibnamefont
  {Tan}},\ }\href {\doibase 10.1103/PhysRevA.78.013636} {\bibfield  {journal}
  {\bibinfo  {journal} {Phys.Rev.}\ }\textbf {\bibinfo {volume} {A78}},\
  \bibinfo {pages} {013636} (\bibinfo {year} {2008})},\ \Eprint
  {http://arxiv.org/abs/0709.2530} {arXiv:0709.2530 [cond-mat.stat-mech]}
  \BibitemShut {NoStop}%
\bibitem [{\citenamefont {Hansen}\ and\ \citenamefont
  {Sharpe}(tion)}]{HSinprep}%
  \BibitemOpen
  \bibfield  {author} {\bibinfo {author} {\bibfnamefont {M.~T.}\ \bibnamefont
  {Hansen}}\ and\ \bibinfo {author} {\bibfnamefont {S.~R.}\ \bibnamefont
  {Sharpe}},\ }\href@noop {} {\  (\bibinfo {year} {in
  preparation})}\BibitemShut {NoStop}%
\bibitem [{\citenamefont {Sterman}(1993)}]{Stermantext}%
  \BibitemOpen
  \bibfield  {author} {\bibinfo {author} {\bibfnamefont {G.}~\bibnamefont
  {Sterman}},\ }\href@noop {} {\emph {\bibinfo {title} {An Introduction to
  Quantum Field Theory}}}\ (\bibinfo  {publisher} {Cambridge University
  Press},\ \bibinfo {year} {1993})\BibitemShut {NoStop}%
\bibitem [{\citenamefont {Mandula}\ \emph {et~al.}(1983)\citenamefont
  {Mandula}, \citenamefont {Zweig},\ and\ \citenamefont
  {Govaerts}}]{Mandula:1983ut}%
  \BibitemOpen
  \bibfield  {author} {\bibinfo {author} {\bibfnamefont {J.~E.}\ \bibnamefont
  {Mandula}}, \bibinfo {author} {\bibfnamefont {G.}~\bibnamefont {Zweig}}, \
  and\ \bibinfo {author} {\bibfnamefont {J.}~\bibnamefont {Govaerts}},\ }\href
  {\doibase 10.1016/0550-3213(83)90399-1} {\bibfield  {journal} {\bibinfo
  {journal} {Nucl.Phys.}\ }\textbf {\bibinfo {volume} {B228}},\ \bibinfo
  {pages} {91} (\bibinfo {year} {1983})}\BibitemShut {NoStop}%
\end{thebibliography}%
